\documentclass[twocolumn]{aastex631}

\newcommand{\kms}{km s$^{-1}$}
\newcommand{\dego}{$^\circ$}
\newcommand{\msun}{M$_\odot$}

\shorttitle{Mira Ceti, atypical archetype}
\shortauthors{Nhung et al. 2021}

\begin{document}

\title{Mira Ceti, atypical archetype}

\correspondingauthor{Pham T. Nhung}
\email{pttnhung@vnsc.org.vn}

\author[0000-0002-0311-0809]{Pham T. Nhung}
\affiliation{Department of Astrophysics, Vietnam National Space Center, Vietnam Academy of Science and Technology, \\
18, Hoang Quoc Viet, Nghia Do, Cau Giay, Ha Noi, Vietnam}

\author[0000-0002-3816-4735]{Do T. Hoai}
\affiliation{Department of Astrophysics, Vietnam National Space Center, Vietnam Academy of Science and Technology, \\
18, Hoang Quoc Viet, Nghia Do, Cau Giay, Ha Noi, Vietnam}

\author[0000-0002-3773-1435]{Pham Tuan-Anh}
\affiliation{Department of Astrophysics, Vietnam National Space Center, Vietnam Academy of Science and Technology, \\
18, Hoang Quoc Viet, Nghia Do, Cau Giay, Ha Noi, Vietnam}

\author[0000-0002-8979-6898]{Pierre Darriulat}
\affiliation{Department of Astrophysics, Vietnam National Space Center, Vietnam Academy of Science and Technology, \\
  18, Hoang Quoc Viet, Nghia Do, Cau Giay, Ha Noi, Vietnam}

\author[0000-0002-2808-0888]{Pham N. Diep}
\affiliation{Department of Astrophysics, Vietnam National Space Center, Vietnam Academy of Science and Technology, \\
  18, Hoang Quoc Viet, Nghia Do, Cau Giay, Ha Noi, Vietnam}

\author[0000-0002-5913-5554]{Nguyen B. Ngoc}
\affiliation{Department of Astrophysics, Vietnam National Space Center, Vietnam Academy of Science and Technology, \\
  18, Hoang Quoc Viet, Nghia Do, Cau Giay, Ha Noi, Vietnam}

\author[0000-0002-8408-4816]{Tran T. Thai}
\affiliation{Department of Astrophysics, Vietnam National Space Center, Vietnam Academy of Science and Technology, \\
18, Hoang Quoc Viet, Nghia Do, Cau Giay, Ha Noi, Vietnam}

\begin{abstract}
  With the aim of unraveling the complexity of the morpho-kinematics of the circumstellar envelope (CSE) of Mira Ceti, we review, extend and in some cases revisit ALMA millimeter observations of the emission of the SiO(5-4) and CO(3-2) molecular lines. In addition, we present a detailed analysis of the optically thin $^{13}$CO(3-2) emission, which provides several important new results. In agreement with observations at infrared and visible wavelengths, we give evidence for confinement and probably rotation of a dense gas volume within $\sim$50 au from the star and for a large SiO line-width within $\sim$15 au. We show that the mass loss process is episodic and takes the form of clumps having a very low SiO/CO abundance ratio compared with similar oxygen-rich long period variables, probably a result of depletion on dust grains and photo-dissociation. We evaluate the  mass loss rate associated with the main clumps and compare it with values obtained from single dish observations. We argue that the SiO emission observed in the south-western quadrant is not related to the mechanism of generation of the nascent wind but  to a mass ejection that occurred eleven years before the observations. We remark that Mira Ceti is not a good archetype in terms of its wind: models aiming at describing  the very complex gas-dust chemistry in action in the CSE of oxygen-rich AGB stars may find it difficult to account for its peculiar features, small variations in the parameters deciding when and where mass loss can proceed significantly. 
\end{abstract}


\section{Introduction}

The present article aims at reviewing and improving our knowledge of the mechanisms governing the generation of the nascent wind of Mira Ceti (or omicron Ceti, or Mira A), one of the most observed AGB stars, remarkable for the large amplitude of its variability ($\sim$8 mag), and known to be accompanied by a distant companion (Mira B, $\sim$70 to 80 au away). It is a long period variable of spectral type M5-9IIIe+DA \citep{Skiff2014} with a period of 333 d \citep{Templeton2009}, a temperature of 2900 to 3200 K \citep{Woodruff2004} and a stellar radius of 21 mas \citep{Khouri2018}. The values commonly quoted for the mass loss rate, 2 to 3$\times$10$^{-7}$ \msun yr$^{-1}$ \citep{Ryde2001, Heras2005} are obtained from single dish observations assuming that the wind is stable with time and spherically symmetric, which is far from being the case. In the present article we adopt a distance of 100 pc, 10 mas spanning 1 au \citep{Haniff1995, vanLeeuwen2007}. While \citet{Vanture1991} have detected technetium in the Mira A spectrum, suggesting that the star has already experienced the third dredge up, conflicting negative results have been obtained by \citet{Kipper1992}; moreover \citet{Hinkle2016} and \citet{Hoai2020}, hereafter referred to as ``HTN20'', have measured  $^{12}$CO/$^{13}$CO abundance ratios of 10$\pm$3 and 12$\pm$2, respectively,  rather low values for a star having already experienced the third dredge up. The time spent by Mira A on the AGB is therefore uncertain; it is at least 30,000 years, as testified by the turbulent wake detected by GALEX \citep{Martin2007}, which traces its interaction with the ISM.  

\subsection{Binarity related features and past history}

Binarity is expected to have little impact on the mechanism currently governing the formation of the nascent wind of Mira A namely in the first phase of its existence, but influences its future evolution.    

The first image showing a clear separation between Mira A and its companion was obtained  with the HST \citep{Karovska1997} and the pair has been resolved later at many other wavelengths. A clear bridge linking the two stars has been observed in X-ray by Chandra \citep{Karovska2005}, in UV by the HST \citep{Wood2006} and in the infrared by the VLT \citep{Matthews2006}. Mira B is generally considered to be a White Dwarf \citep{Wood2002, Sokoloski2010} surrounded by a thin accretion disc rotating at high velocity \citep{Reimers1985} and its highly variable UV emission has been claimed to dominate over that of Mira A. Both statements have been disputed, however: \citet{Ireland2007} have argued that Mira B is a Main Sequence star and \citet{Montez2017} have questioned the supposed dominance of the UV emission of Mira B over Mira A. As Mira is moving at a velocity of 130 \kms\ through the interstellar medium it leaves a turbulent wake behind it that has been detected in the UV with evidence for a bow shock \citep{Martin2007}. The astropause has been detected in the FIR \citep{Ueta2008}. Closer to the star, observations in the FUV and H$\alpha$ have revealed the possible presence of a fast bipolar outflow \citep{Meaburn2009}. The Mira pair is thought to display occasional nova-like eruptions and a soft X ray outburst, probably associated with mass ejection, was observed in 2003 \citep{Karovska2005}.

At millimeter wavelengths, the companion is detected separately from the AGB star \citep{Vlemmings2015}, with a clear bridge linking the two stars, and its presence is seen to influence the morpho-kinematics of the wind, in particular by focusing the wind of Mira A \citep{Nhung2016}. Yet, the impact of binarity on the morpho-kinematics of the CSE has been shown to be small \citep[][ HTN20]{Vlemmings2015, Nhung2016}: it is limited to Mira B focusing the wind of Mira A that blows toward it, part of which being accreted, but is unrelated to the overall morpho-kinematics of the nascent wind.  In particular a southern arc, which was described by \citet{Ramstedt2014} as the birth of a spiral, has been shown to have no simple relation with Mira B and to wind in the wrong direction to be interpreted as a Wind Roche Lobe Overflow spiral (HTN20).

The orbit has a semi-major axis of 70 to 80 au (0.7-0.8 arcsec) with a period at 500 yr scale and is inclined by some 60\dego\ with respect to the plane of the sky \citep{Prieur2002, Vlemmings2015, Planesas2016}. From the observation of the blue-shifted wind blowing from Mira A to Mira B \citep[][HTN20]{Nhung2016}, we know that Mira B is closer to us than Mira A by some 60 au. This result has since been confirmed \citep{Saberi2018} by the observation of CI emission from the neighborhood of Mira B, blue-shifted by $\sim$4 \kms. The mass of Mira A is $\sim$2 solar masses, that of Mira B $\sim$0.7 solar masses \citep{Planesas2016}. The escape velocity at 25 au from a 2 solar masses star is $\sim$12 \kms. 

\subsection{Shocks and dust formation in the close neighborhood of Mira A}

From infrared to millimeter wavelengths, the latter probing the proximity of the stellar surface with vibrationally excited molecular lines or continuum emission, all studies show the importance of pulsation- and shock-induced dynamics in levitating the molecular atmosphere \citep{Chandler2007, Matthews2015, Vlemmings2015, Kaminski2016, Planesas2016, Khouri2016, Khouri2018, Vlemmings2019, Perrin2020}. They reveal a complex process of the formation of dust grains, influenced by variability, displaying significant inhomogeneity and suggesting the presence of shocks related to star pulsations \citep{Wittkowski2016}. Evidence is found for short time variability (at month scale), occasional gas in-fall, hot spots covering a small fraction of the stellar disc, radial velocities at the 10 \kms\ scale, all revealing the impact of shocks associated with pulsations and convective cell ejections.  A consequence on the millimeter observation of the emission of molecular lines is the presence of large Doppler velocity wings near the line of sight crossing the stellar disc in its center (HTN20).

Another feature revealed by most studies is an abrupt decline of the gas emission of SiO and other molecules at distances from the star in excess of some  6 au \citep{Wong2016, Khouri2018}. Dust is observed to cluster near the edge of this region. No convincing interpretation has been proposed, whether in terms of shock wave, of recent increase of mass loss or of dust condensation. The gas phase of Al-bearing molecules is seen to deplete very close to the star, within 3 to 4 au, revealing their condensation in dust grains, in contrast with Ti-bearing molecules, which are thought to play essentially no role in the formation of dust \citep{Kaminski2017}. Evidence for the presence of  dust grains at 0.1 $\mu$m scale, within a very few au from the radiosphere, is obtained from the study of continuum emission as well as from the observation of polarized light \citep{Khouri2019}.

VLA observations of SiO masers \citep{Cotton2004,Cotton2006} suggest the presence of rotation at some 7 au from the center of the star and recurrent OH maser flaring events have been observed close to Mira A \citep{Etoka2017}, apparently unrelated to Mira B location.

Finally, we note that there have been speculations about a possible magnetic origin of hot spots and/or rotation in Mira A \citep{Thirumalai2013}. 

\subsection{Morpho-kinematics of the CSE beyond 20 au from Mira A}

ALMA observations of the CO(3-2) line emission have been analyzed  by \citet{Ramstedt2014} with an angular resolution of  $\sim$0.5 arcsec (FWHM), and by  \citet{Nhung2016} and HTN20 with an angular resolution of 0.32 arcsec (FWHM). A very complex morpho-kinematics is observed, witness of several past episodes of enhanced mass loss. However, by limiting the range of Doppler velocities to $|V_z|$$<$4 \kms, one can reasonably well select most of the more recent emission. In particular it excludes the contribution, covering up to ~10 arcsec, of a blue-shifted bubble, or ring, emitted some 2,000 years ago, together with red-shifted fragments probably ejected at the same epoch. At shorter distances from Mira A, within some 2.5 arcsec, two broad outflows are seen in the south-western and north-eastern quadrants. They display important inhomogeneity, emission of the north-eastern outflow being strongly depressed at distances between $\sim$1 and 2 arcsec. ALMA observations of the SiO(5-4) line, with an angular resolution of $\sim$50 mas (FWHM) have been analyzed by HTN20;  the emission is confined to the south-western outflow for distances from the star in excess of some 0.5 arcsec. Several interpretations have been suggested to explain these observed features, none of which, however, can be considered as well established. In HTN20, the authors remarked that the CO(3-2) emission seems to be enhanced in the orbital plane of the AB pair; however, the long orbital period, at 500 yr scale, makes it difficult to conceive a sensible scenario. They also suggested a mechanism for the SiO and CO emissions in the south-western quadrant that relates to the 2003 mass ejection associated with the detection of a soft X ray outburst, but again lacking a solid basis. 

\subsection{Outline of the article}

The aim of the article is to contribute to unraveling the complexity of the observed morpho-kinematics of the wind of Mira Ceti. Such complexity, first explicitly revealed by \citet{Ramstedt2014}, made it difficult to draw a simple picture. The subsequent studies of \citet{Nhung2016} and HTN20, which are the only additional studies of the CSE beyond a few 100 mas projected distance from the star, offer a fragmentary description that lacks unity. Section 2 gives a brief reminder of the essentials of observations and data reduction. Section 3 revisits an analysis of the SiO line width within 15 au from the star and Section 4  presents a detailed analysis of the emission of the $^{13}$CO(3-2) line, on which many of the results presented in the article, in particular an evaluation of the mass loss rate, are based. Section 5 addresses the peculiarities of the observed SiO emission:  the confinement of high gas density within some 50 au from the star, particularly low SiO/CO abundance ratio and complex, unexpected and different patterns displayed by CO and SiO emission in the south-western quadrant. Section 6 summarizes the results and concludes. 

\section{OBSERVATIONS AND DATA REDUCTION}

As observations and data reduction are described in detail in HTN20, we refer the reader to it and simply recall that the data are retrieved from ALMA archives and correspond to two different sets of observations. Project ADS/JAO.ALMA\#2011.0.00014.SV is used for the analysis of the SiO($\nu$=0, $J$=5-4) line emission at distances from the star exceeding $\sim$0.5 arcsec; the same data have been analyzed by \citet{Wong2016} for projected distances from the star well below 0.5 arcsec. Project ADS/JAO.ALMA\#2013.1.00047.S is used for the analysis of the $^{12,13}$CO($\nu$=0, $J$=3-2) line emissions; the same data have been analyzed earlier by \citet{Planesas2016} to study continuum emission and by \citet{Nhung2016} and HTN20 to study molecular line emission. Issues related to continuum subtraction and to maximal recoverable scale have been addressed in detail in HTN20, from which we copy below (Table \ref{tab1}) a summary of the main parameters of relevance to observations and data reduction. The maximal recoverable scales listed in Table \ref{tab1}, 11.3 arcsec for SiO emission and $\sim$4 arcsec  for CO emission, provide a measure of the radial dependence of the missing flux. The reliability of the imaging process is however confined to shorter distances \citep{Hoai2021}. However, the present paper focusses on projected distances from Mira A not exceeding 2-3 arcsec and we have checked that the impact of the short spacing problem on the present analyses can be safely neglected.

We use orthonormal coordinates with the $x$ axis pointing east and the $y$ axis pointing north. The origin is taken to match the center of continuum emission of Mira A as observed by \citet{Planesas2016} and \citet{Wong2016}. The projected angular distance of pixel ($x,y$) to the origin is $R$$=$$\sqrt{x^2+y^2}$ and its position angle, measured counter-clockwise from north, is $\omega$$=$$\tan^{-1}(x/y)$. The $z$ axis is parallel to the line of sight pointing away from us. As origin of velocity coordinate, we use a systemic velocity of 47.7 \kms\ with respect to the local standard of rest (LSR) as obtained by \citet{Khouri2018} from a fit of the CO($\nu$=1, $J$=3-2) line emission in the close environment of the star. This is 1.0 \kms\ larger than obtained by \citet{Wong2016} from a fit of the SiO($\nu$=0,$J$=5-4) line over a large field of view: the uncertainty on this number may be as large as 1 \kms.

\begin{deluxetable*}{cccc}
  \tablenum{1}
  \tablecaption{Parameters of relevance to observations and data reduction. Data reduction and imaging were done with GILDAS.\label{tab1}}
  \tablehead{\colhead{}&\colhead{SiO(5-4)}&\colhead{$^{12}$CO(3-2)}&\colhead{$^{13}$CO(3-2)}}
  \startdata
  ALMA project (ADS/JAO.ALMA\#)&2011.0.00014.SV &\multicolumn{2}{c}{2013.1.00047.S}\\
    PI&ALMA&\multicolumn{2}{c}{P. Planesas}\\
    Date of observation& 1st November 2014 &\multicolumn{2}{c}{12-15 June 2014}  \\
    Number of antennas&39&\multicolumn{2}{c}{34/36}\\
    Maximal baseline (km)&15.24&\multicolumn{2}{c}{0.65}\\
    Maximal recoverable scale (arcsec)&11.3&\multicolumn{2}{c}{$\sim$4}\\
    Frequency (GHz)&217.1049&345.7960&330.5880\\
    Beam size (mas$^2$)&60$\times$30&390$\times$360&400$\times$370\\
    Spectral resolution (\kms)&0.4&0.4&0.4\\
    Pixel size (arcsec$^2$)&0.01$\times$0.01&0.06$\times$0.06&0.06$\times$0.06\\
    Noise level (mJy\,beam$^{-1}$)&0.66&14&5\\
 \enddata
\end{deluxetable*}

\section{SiO(5-4) line width in the 5 to 15 au range}

The suggested evidence for   the presence of important shocks induced by pulsations and convection cell ejections in the inner layer of the CSE, mentioned in the preceding section, is a common feature of many O-rich AGB stars that have been observed with high angular and spectral resolutions and its description by 3-D hydrodynamical models has recently made significant progress \citep[e.g.][and references therein]{Freytag2019}. At millimeter wavelengths, its impact on the line profile of single dish observations, in the form of broad Doppler velocity wings, had been noticed early and properly interpreted as suggesting that the emission occurs close to the star  where the SiO gas phase is not yet significantly depleted and is somehow related to stellar pulsations \citep{Winters2003}. \citet{deVicente2016} observed their presence in the SiO(1-0) emission of five oxygen-rich AGB stars, including Mira A.   However, when the presence of high Doppler velocity wings close to the line of sight crossing the star in its center was first noticed on R Dor \citep{Decin2018}, it was not immediately understood.

\begin{figure*}
  \centering
  \includegraphics[height=5.5cm,trim=0.cm 1.5cm 0cm 0.5cm,clip]{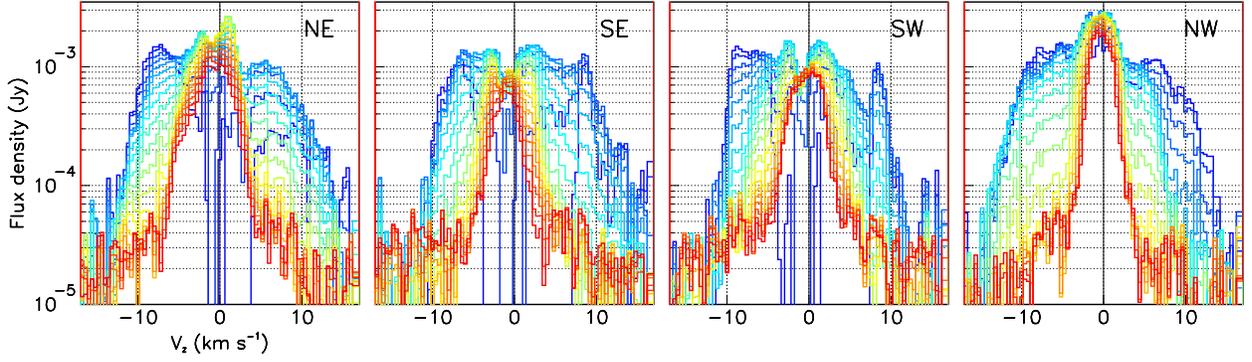}
  \caption{SiO(5-4) emission of Mira A. Doppler velocity spectra averaged in quadrants of 3 au broad annular rings centered on Mira A with mean radii increasing from 2.5 au (blue) to 21.5 au (red) in steps of 1 au (1 au$\sim$10 mas). (Reproduced from Figure A2 of HTN20 with permission of the RAS)}
 \label{fig1}
\end{figure*}

\begin{figure*}
  \centering
  \includegraphics[height=3.9cm,trim=.5cm 1.5cm 1.5cm 1.8cm,clip]{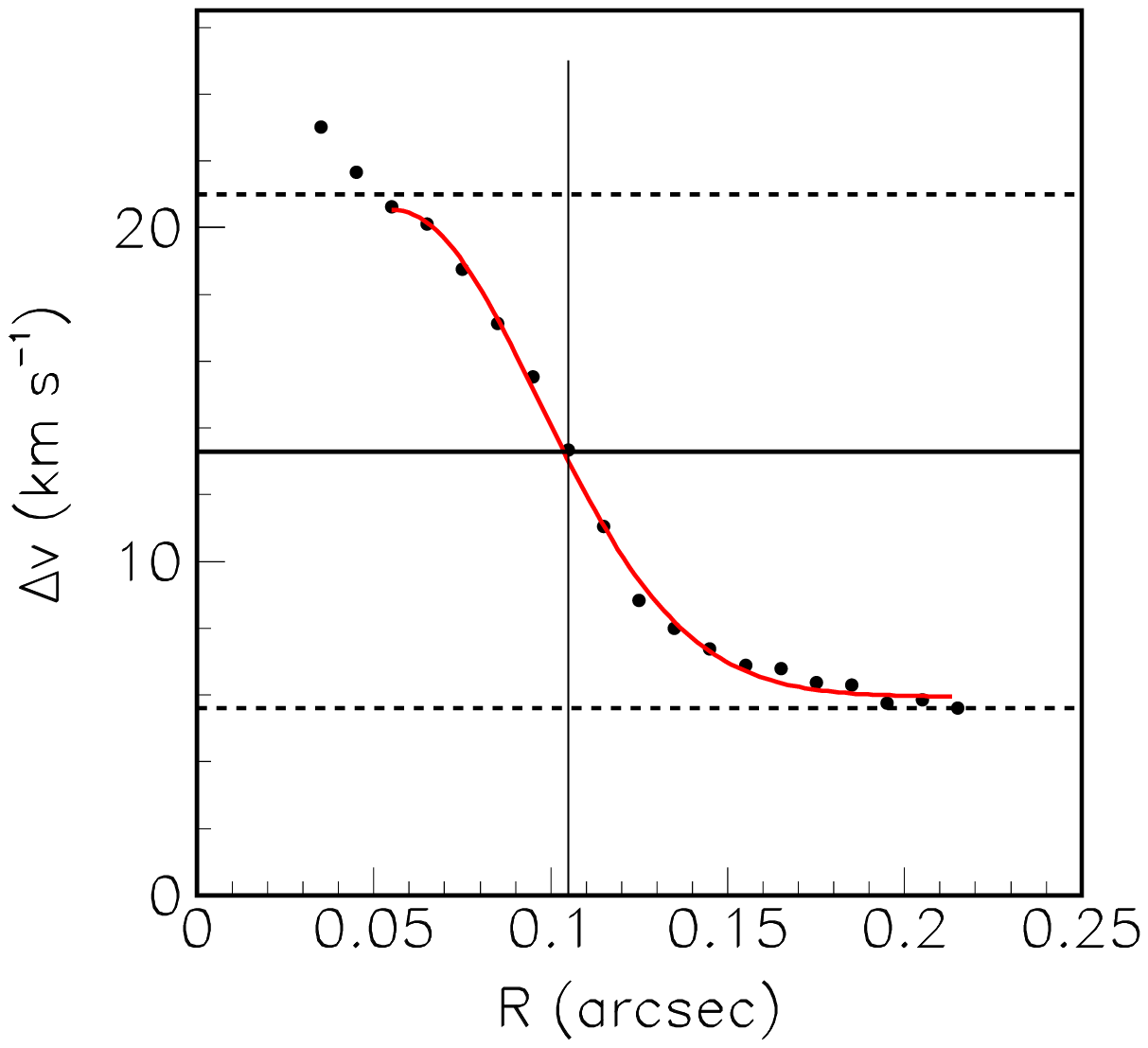}
  \includegraphics[height=3.9cm,trim=.5cm 1.5cm 1.5cm 1.8cm,clip]{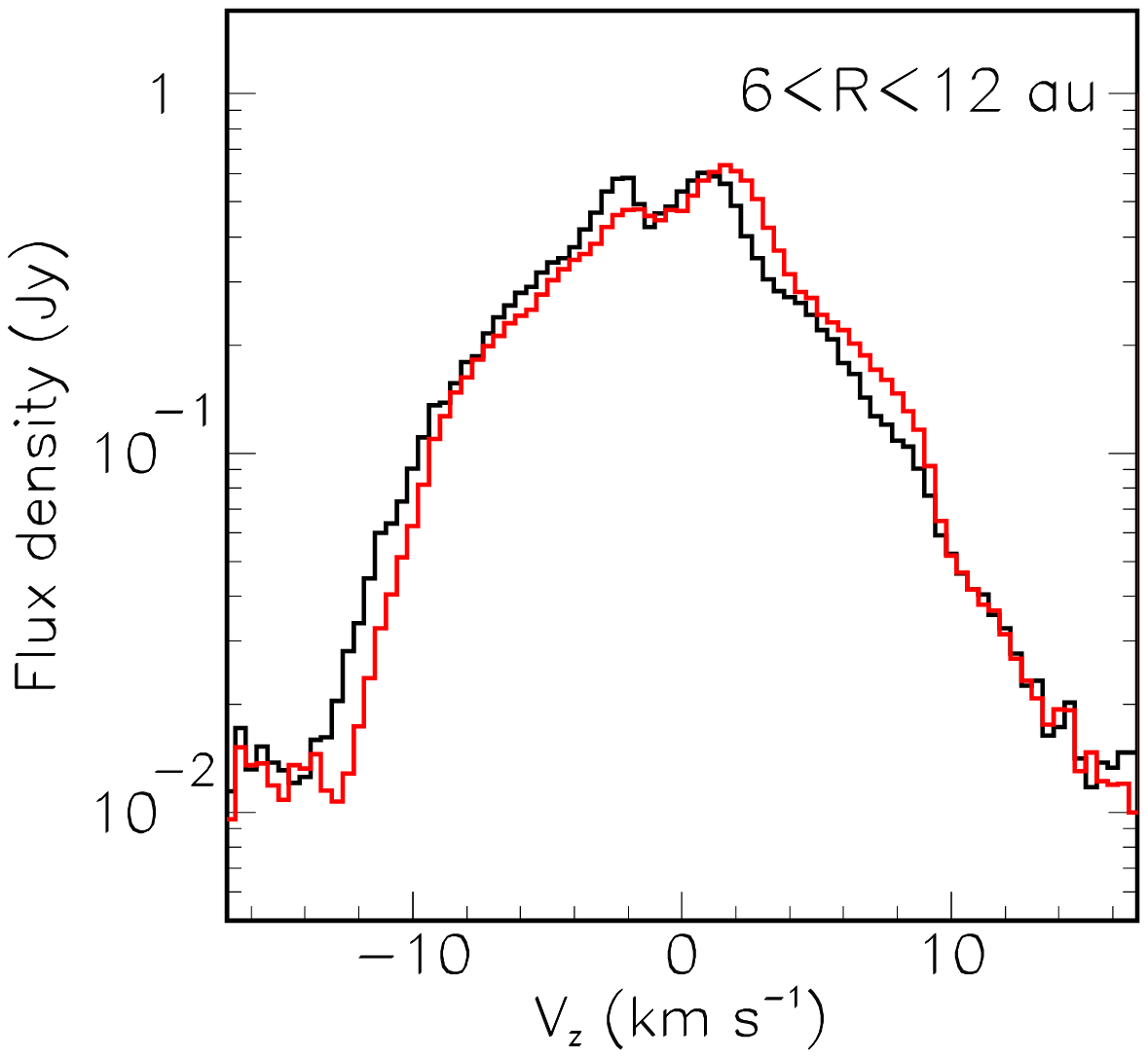}
  \includegraphics[height=3.9cm,trim=.5cm 1.5cm 1.5cm 1.8cm,clip]{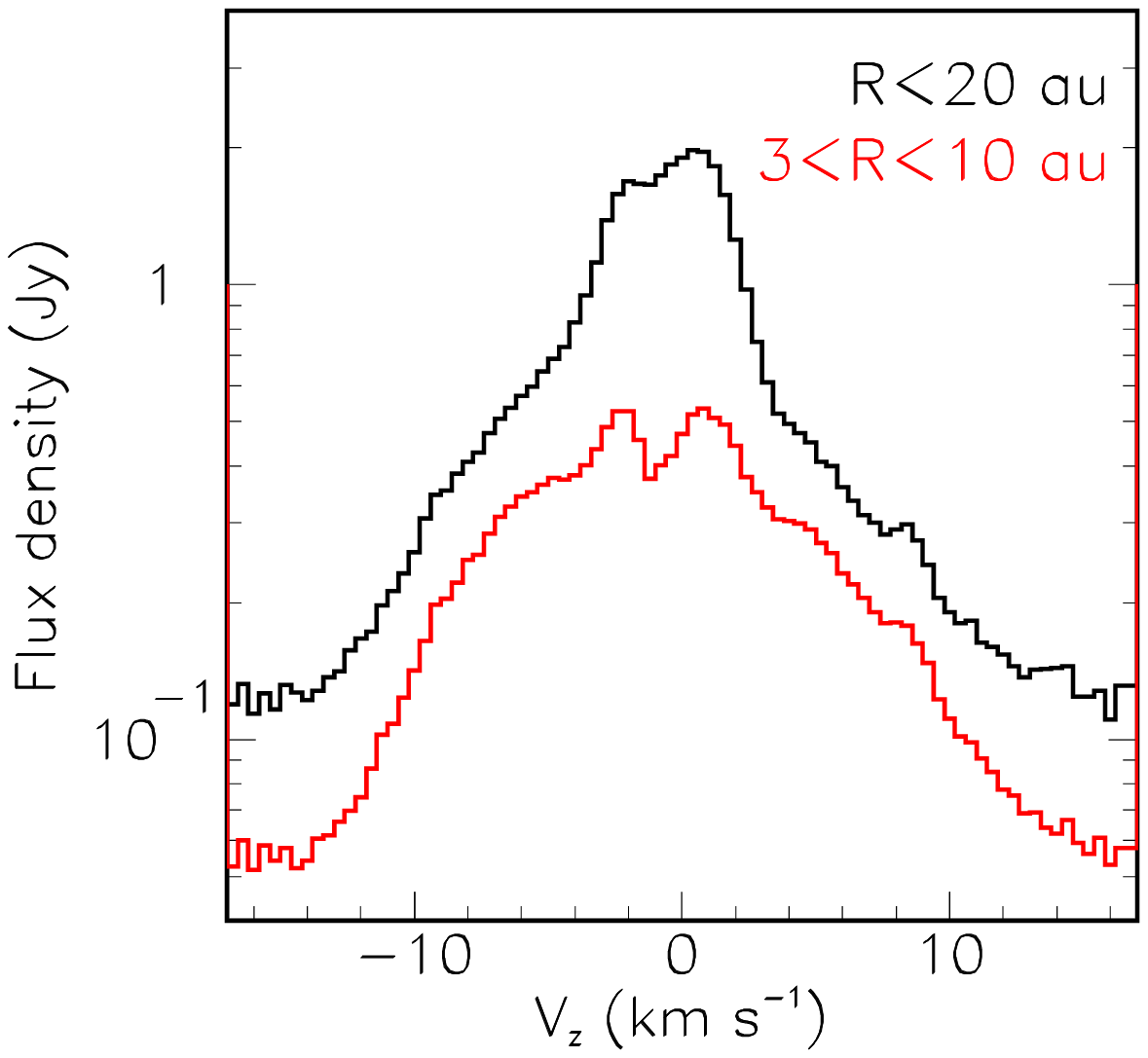}
  \includegraphics[height=3.9cm,trim=.5cm 1.5cm 1.5cm 1.8cm,clip]{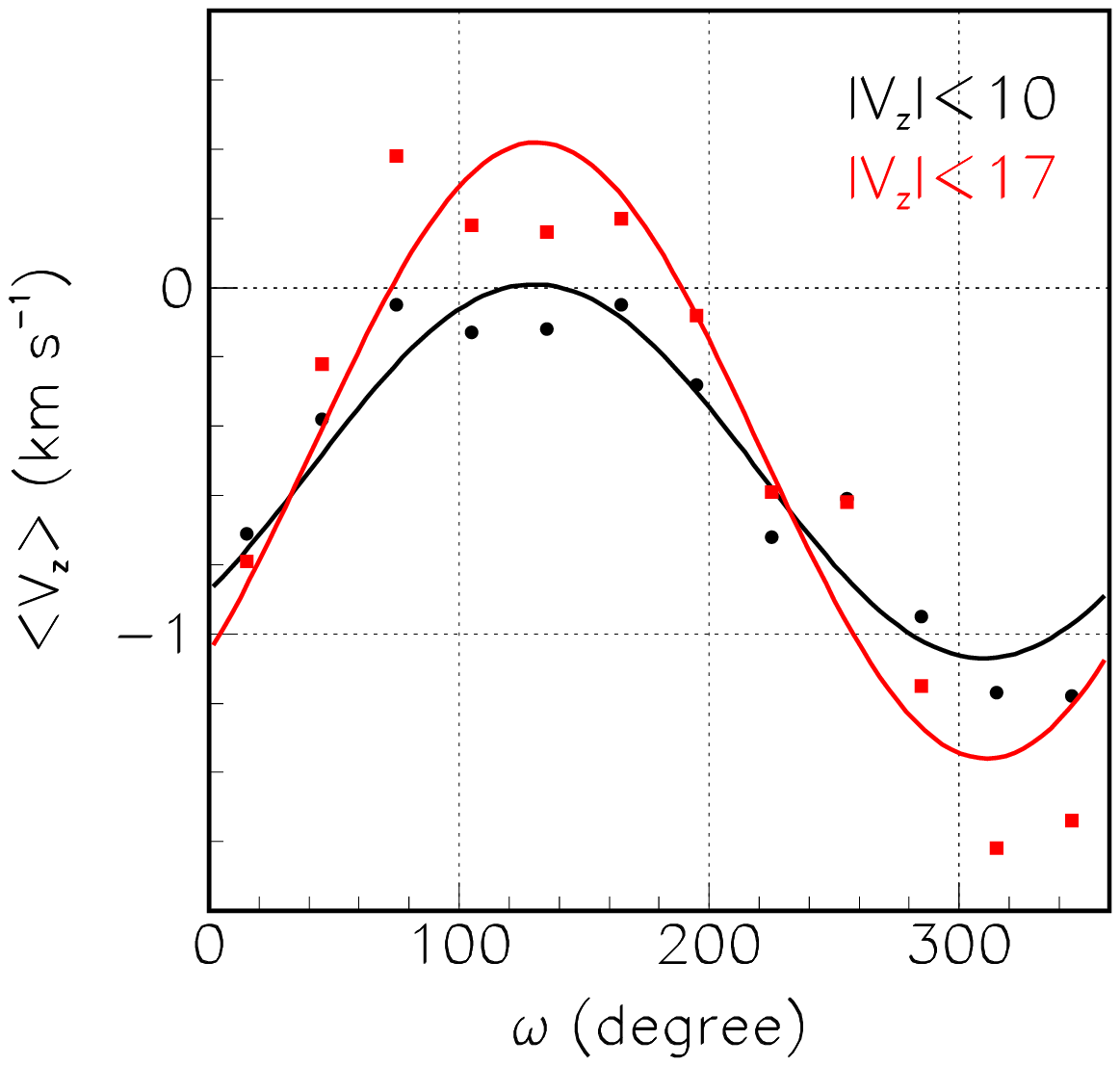}
  \caption{SiO(5-4) emission of Mira A. Left: Dependence on $R$ of the full-width at 1/5 maximum of the spectra displayed in Figure \ref{fig1} (all four quadrants together, reproduced from Figure 14d of HTN20). Middle-left: Doppler velocity spectrum observed in the ring 6 au$<$$R$$<$12 au before (black) and after (red) re-centering (1 au$\sim$10 mas). Middle-right: Doppler velocity spectra evaluated in the circle $R$$<$20 au and in the ring 3 au$<$$R$$<$10 au. Right: dependence of the mean Doppler velocity on position angle $\omega$ in the ring 3 au$<$$R$$<$10 au. The lines are sine wave best fits. }
 \label{fig2}
\end{figure*}

On Mira A, their presence was first revealed (HTN20) in the emission of the SiO(5-4) molecular line. In the present section we revisit and refine this analysis, in order, in particular, to account for a possible effect of rotation.     Figure \ref{fig1} displays, in each of four quadrants, Doppler velocity spectra averaged in 30 mas broad annular rings centered on Mira A with mean radii increasing from 25 mas (blue) to 215 mas (red) in steps of 10 mas. A cut at 3$\sigma$ is applied to the data. The dependence on $R$ (projected distance from the center of Mira A) of  the full-width at 1/5 maximum, where the profile is well defined, is displayed in Figure \ref{fig2} left; it shows that the line width decreases from 21 to 7 \kms\  when $R$ spans from $\sim$50 to $\sim$150 mas. In principle, part of the line width might be due to a velocity gradient within a quadrant, such as produced by rotation. In order to assess the importance of such a contribution, we repeat the analysis by re-centering, in each 30$\times$30 mas$^2$ pixel, the observed Doppler velocity spectrum on its mean value. The effect is minimal, as illustrated in Figure \ref{fig2} middle-left with the example of the ring 60 mas$<$$R$$<$120 mas. The decrease of the rms deviation with respect to the mean never exceeds 0.1 \kms\ and, on average, is only 0.05 \kms.

\begin{figure*}
  \centering
   \includegraphics[height=4.2cm,trim=0.2cm 1.2cm 0.2cm 1cm,clip]{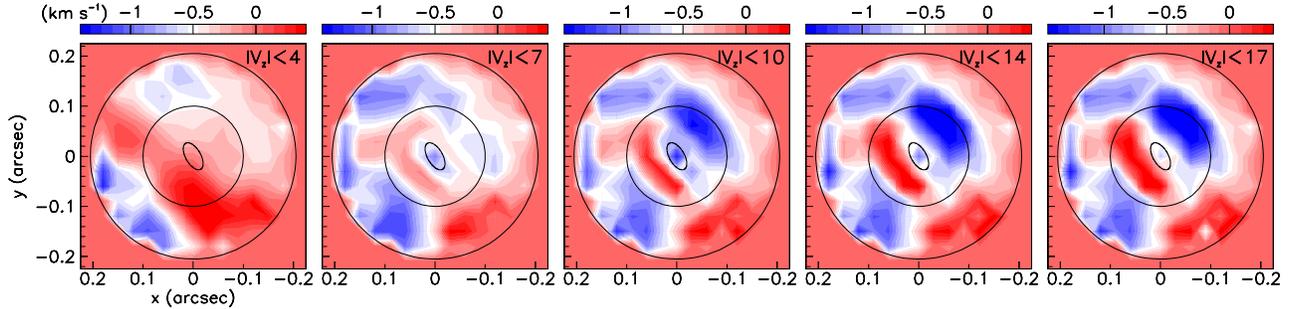}
  \caption{Maps of $<$$V_z$$>$ evaluated over different intervals of $V_z$ as indicated in the upper right corner of each panel. The circles in the center of each panel have a radius of 10 au. The ellipse in the center of each panel shows the beam.}
 \label{fig3}
\end{figure*}

This result strengthens the conclusion that, with a resolution of $\sim$30 mas, the line width becomes very broad at projected distances smaller than $\sim$150 mas, reaching over 20 \kms\ at $R$$<$$\sim$50 mas. At the same time, it makes it difficult to detect a small velocity gradient, in particular the possible presence of rotation, known to be present in some other long period variables, as for example R Dor \citep{Vlemmings2018, Homan2018, Nhung2021}. Such rotation would cause the mean Doppler velocity, $<$$V_z$$>$, to have a sine wave dependence on position angle $\omega$  at distances from the star where rotation is significant. We illustrate it using a ring covering 30 mas$<$$R$$<$100 mas: Figure \ref{fig2} middle-right shows that the spectrum in such a ring is  broad, at the scale of 20 \kms.  Moreover, it is rather flat-topped in comparison with the spectrum in the broader circle $R$$<$200 mas: if one evaluates $<$$V_z$$>$ over too short an interval of $V_z$, one is not sensitive to a possible shift; one needs therefore to calculate it over a broader interval, implying a large uncertainty on the result. This is illustrated in Figure \ref{fig3}, which displays maps of $<$$V_z$$>$ calculated over different intervals of $V_z$. For intervals broader than $\sim$$\pm$10 \kms, the map displays a pattern typical of rotation about an axis projecting in the north-east/south-west direction; but for intervals narrower than $\sim$$\pm$7 \kms, the evidence is much weaker. Figure \ref{fig2} right shows the dependence on $\omega$ of $<$$V_z$$>$ evaluated in intervals of $\pm$17 \kms\ and $\pm$10 \kms, respectively. The best sine wave fits are $<V_z>=-0.53+0.54\sin(\omega-40$\dego) and $-0.47+0.89\sin(\omega-41$\dego) \kms, respectively. Uncertainties accounting for the systematic effects of choosing different intervals of $R$ and/or $V_z$ are estimated as $\sim$$\pm$0.2 \kms\ on the projected velocity and $\sim$$\pm$15\dego\ on the position angle of the projected rotation axis. The offsets may reveal an overestimate of the systemic velocity, which, as remarked in Section 2, is not accurately known.   

The sine wave fits suggest the presence of rotation within some 15 au from the center of the star with a projected velocity of 0.7$\pm$0.2 \kms\ about an axis projecting 40\dego$\pm$15\dego\ east of north. It is in agreement with the values quoted by \citet{Cotton2004,Cotton2006}: 1.5$\pm$0.9 \kms\ and 17$\pm$36\dego, respectively. In principle, the fits could also be interpreted in terms of radial outflows, but such a morphology would be very unusual for the inner layer of such an AGB star. Moreover, no trace of such outflows is seen at larger projected distances from the star in the emission of the CO lines (see Section 4). Instead, rotation is commonly interpreted as evidence for the presence of a companion. In the present case, however, such a possible rotation, if confirmed, could not have anything to do with Mira B, which is much farther away (HTN20).

\section{ $^{13}$CO(3-2) emission}

As mentioned in Section 1.3, the emission of the $^{12}$CO(3-2) line has been studied by \citet{Ramstedt2014}, \citet{Nhung2016} and  HTN20. In addition, the authors of the latter study have reduced observations of the $^{13}$CO(3-2) emission, made within the same ALMA project, in order to evaluate the $^{12}$CO/$^{13}$CO ratio. They used a simple radiative transfer calculation to evaluate the opacity of the $^{12}$CO line at the level of $\sim$37\% averaged over regions of large column density where the mean brightness ratio between $^{12}$CO and $^{13}$CO emissions is $\sim$7.3. Correcting for the different opacity in $^{12}$CO and $^{13}$CO data boosts this ratio to $\sim$10.5 and accounting for the different excitation energies and frequencies translates into a $^{12}$CO/$^{13}$CO abundance ratio of 12$\pm$2.

In the present section, we take advantage of the smaller optical depth of the $^{13}$CO(3-2) observations to study the gas morphology in finer details than was possible using $^{12}$CO(3-2) observations. Imaging was done using GILDAS with natural weighting, giving a nearly circular beam  of $\sim$0.38 arcsec HPBW. The brightness distribution is displayed in Figure \ref{fig4} left, showing a noise rms of 5 mJy\,beam$^{-1}$. Channel maps of the  $^{13}$CO(3-2) emission are shown in the Appendix. Table \ref{tab2} lists some parameters of relevance to the comparison between $^{12}$CO(3-2) and $^{13}$CO(3-2) emissions, including size of the beam, frequency, Einstein coefficient and energy of the upper level. Their values are similar, with the result that the ratio between the two emissions is nearly independent from temperature and, in the optically thin limit (which however is not the case here), would provide a direct measure of the isotopic ratio. 

\subsection{Comparing $^{13}$CO(3-2) and $^{12}$CO(3-2) emissions}

As the $^{13}$CO(3-2) detected emission is typically one order of magnitude smaller than the $^{12}$CO(3-2) emission, we limit the present study (in addition to the restriction to the Doppler velocity interval of $\pm$4 \kms) to the central region where the intensity of the $^{12}$CO(3-2) emission exceeds 10 Jy arcsec$^{-2}$ \kms. Its map is displayed in the left panel of Figure \ref{fig5}. It is elongated in the north-east/south-west direction and covers fragments that were identified in HTN20 and were given names of south-western outflow (SWO), north-eastern outflow (NEO) and north-eastern arc (NEA), respectively. In what follows we refer to it as region C (for central), implying in addition that the inequality $|V_z|$$<$4 \kms\ is satisfied.  

\begin{figure*}
  \centering
  \includegraphics[height=4.5cm,trim=.5cm 1.5cm 1.2cm 1.5cm,clip]{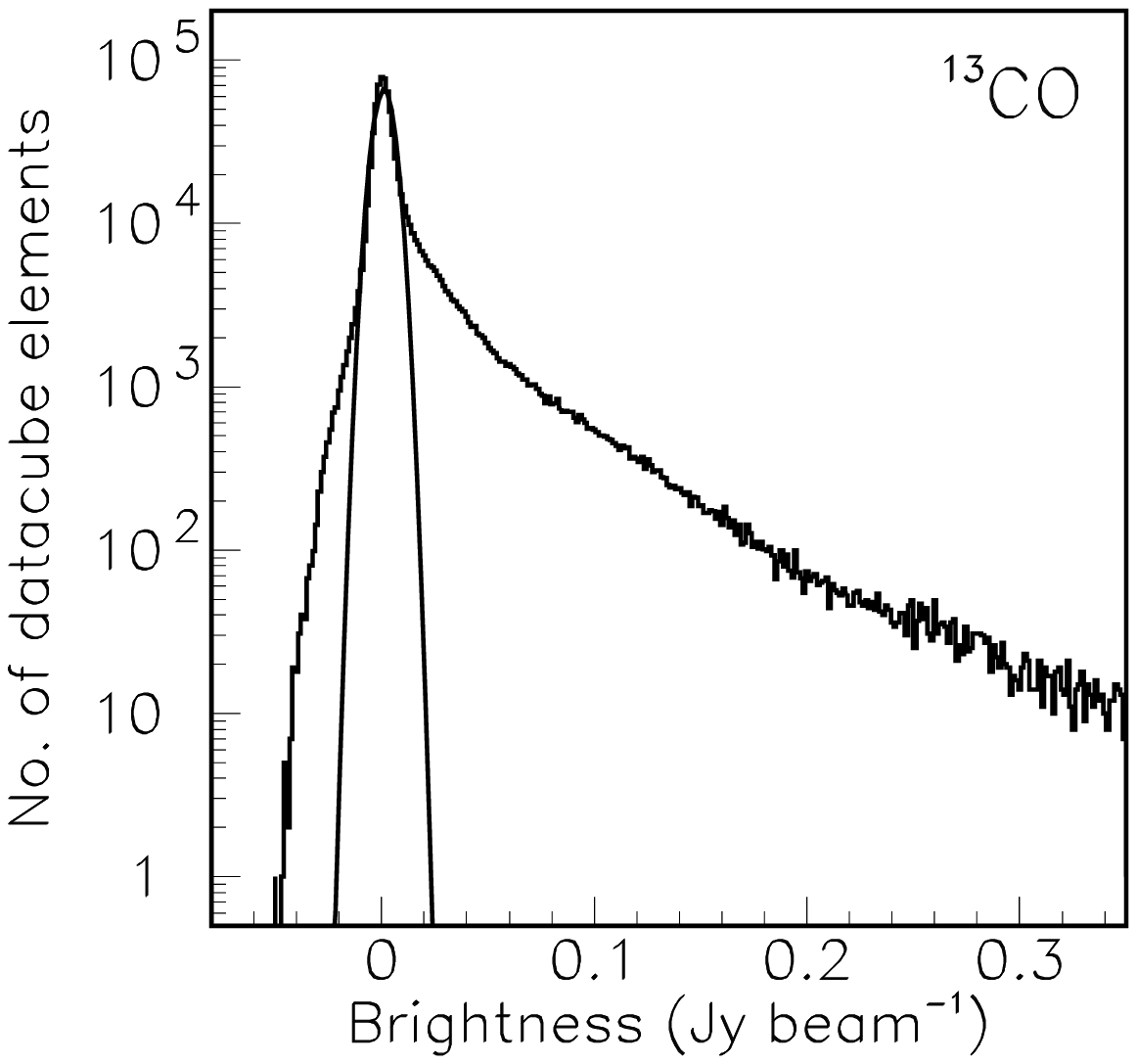}
  \includegraphics[height=4.5cm,trim=.5cm 1.5cm 1.2cm 1.5cm,clip]{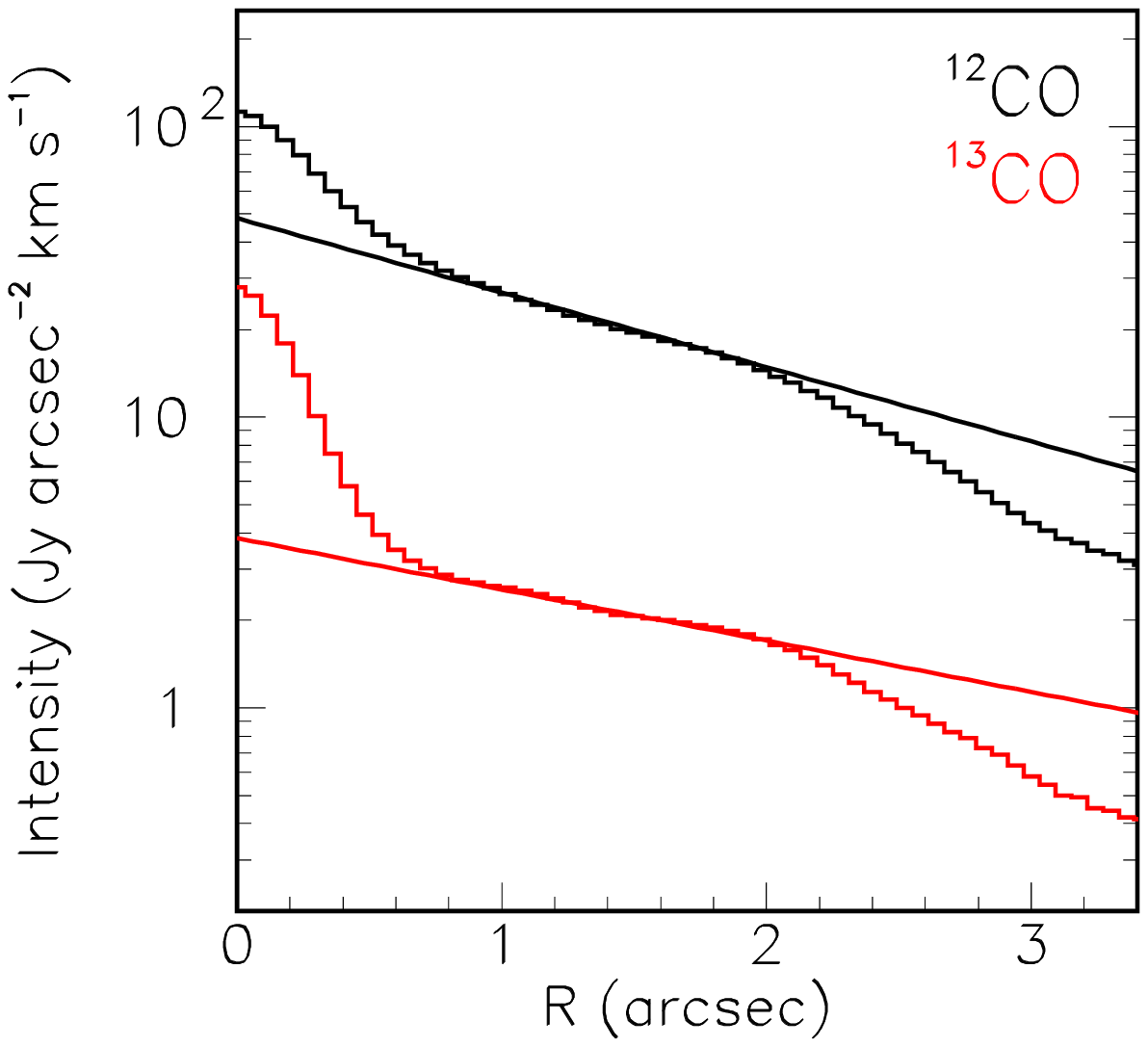}
  \includegraphics[height=4.5cm,trim=.5cm 1.5cm 1.2cm 1.5cm,clip]{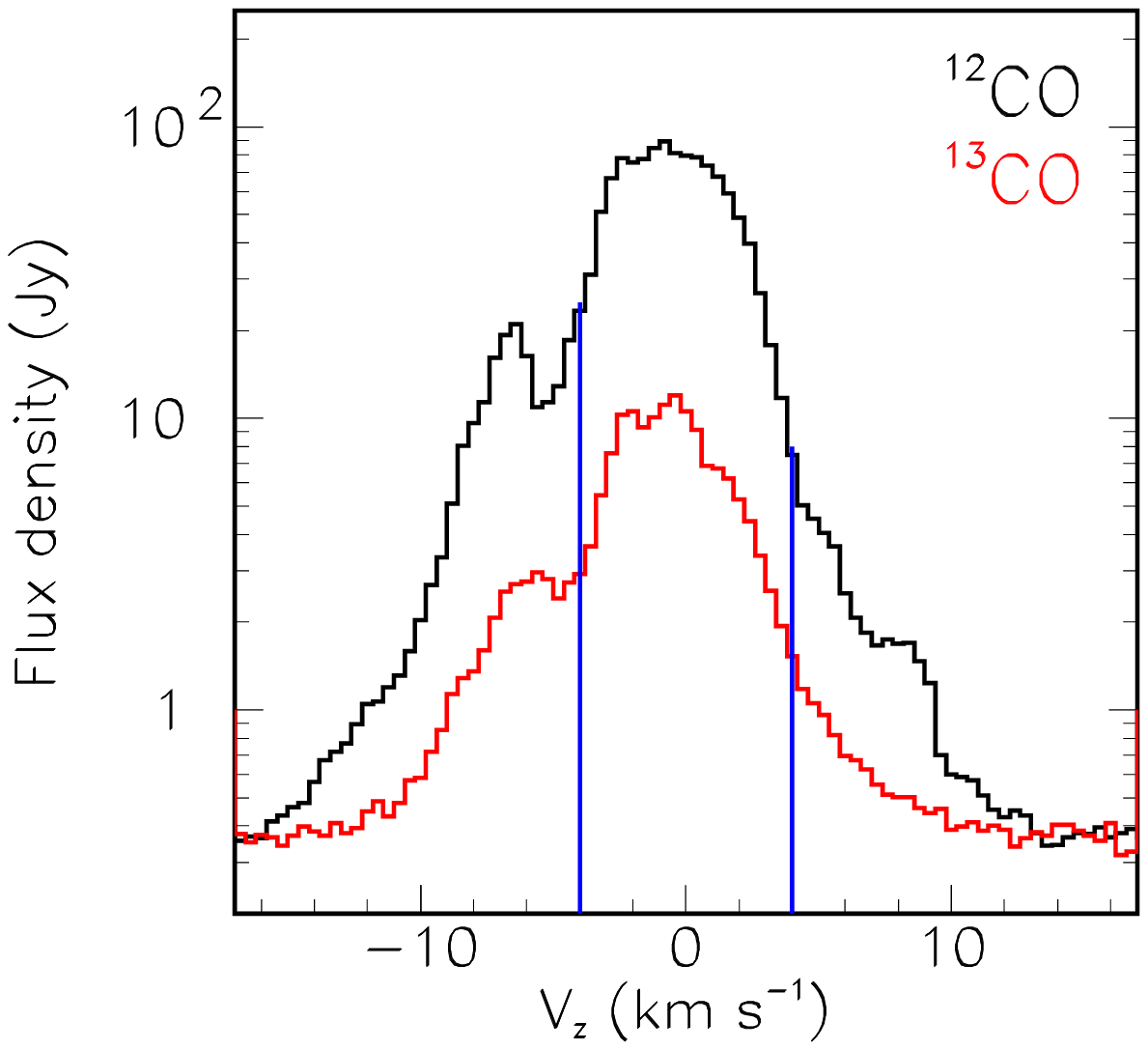}
    \caption{Left: brightness distribution of the $^{13}$CO(3-2) emission measured over a data cube covering $\pm$3.7 arcsec in $x$ and $y$ and $\pm$10 \kms\ in $V_z$. The curve is a Gaussian fit with a $\sigma$ of 5 mJy beam$^{-1}$. Center: radial distributions (after continuum subtraction) of the $^{12}$CO(3-2) (black) and $^{13}$CO(3-2) (red) emissions integrated over position angles and over $|V_z|$$<$4 \kms ; the lines are exponential fits to the interval 1$<$$R$$<$2 arcsec. Right: Doppler velocity spectra of the $^{12}$CO(3-2) (black) and $^{13}$CO(3-2) (red) line emissions integrated over $R$$<$3 arcsec. The blue vertical lines show the interval considered in the present article, $|V_z|$$<$4 \kms.}
 \label{fig4}
\end{figure*}

\begin{figure*}
  \centering
  \includegraphics[height=4.5cm,trim=1.cm 1.5cm 1.2cm 0.5cm,clip]{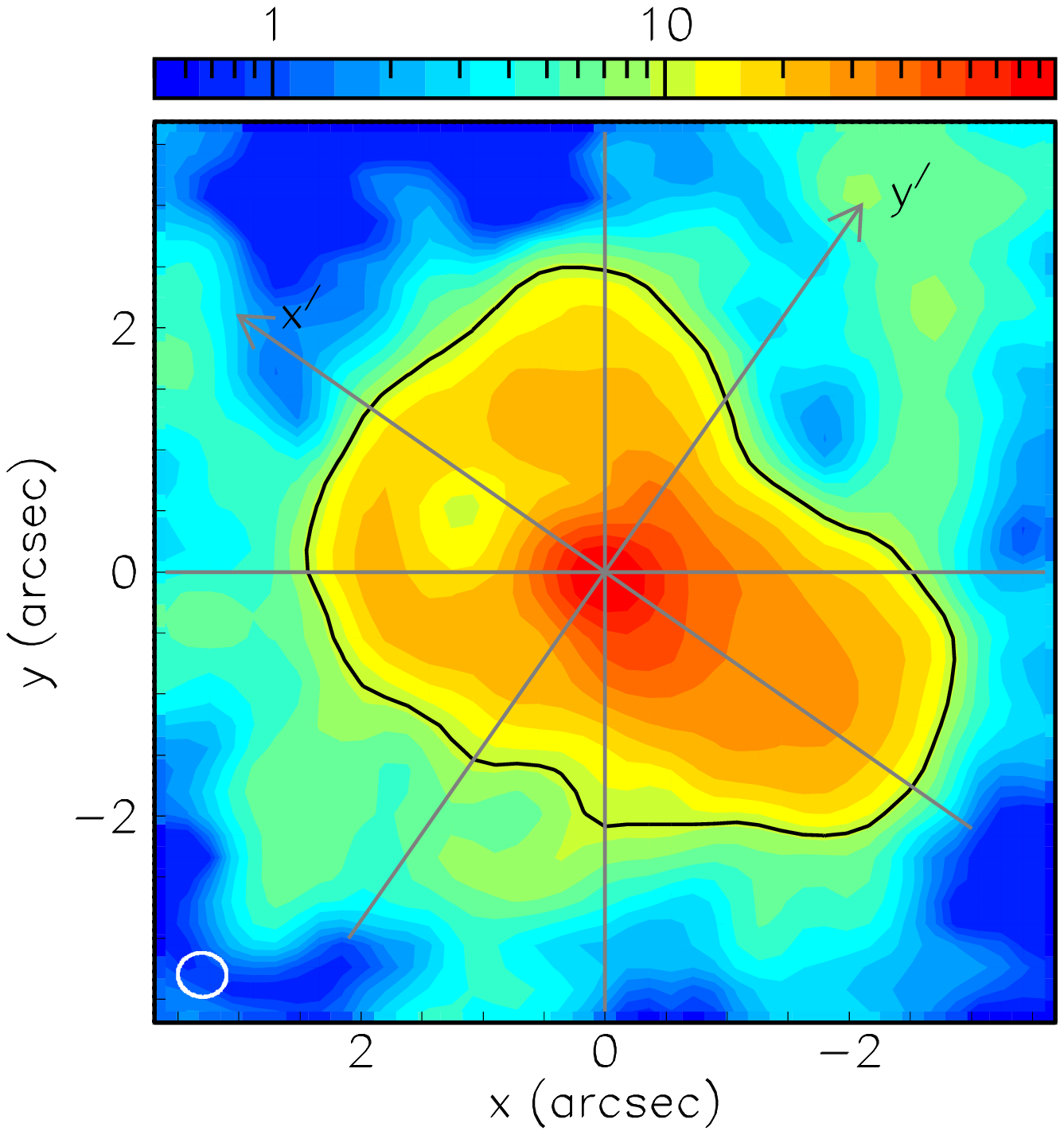}
  \includegraphics[height=4.5cm,trim=1.cm 1.5cm 1.2cm 0.5cm,clip]{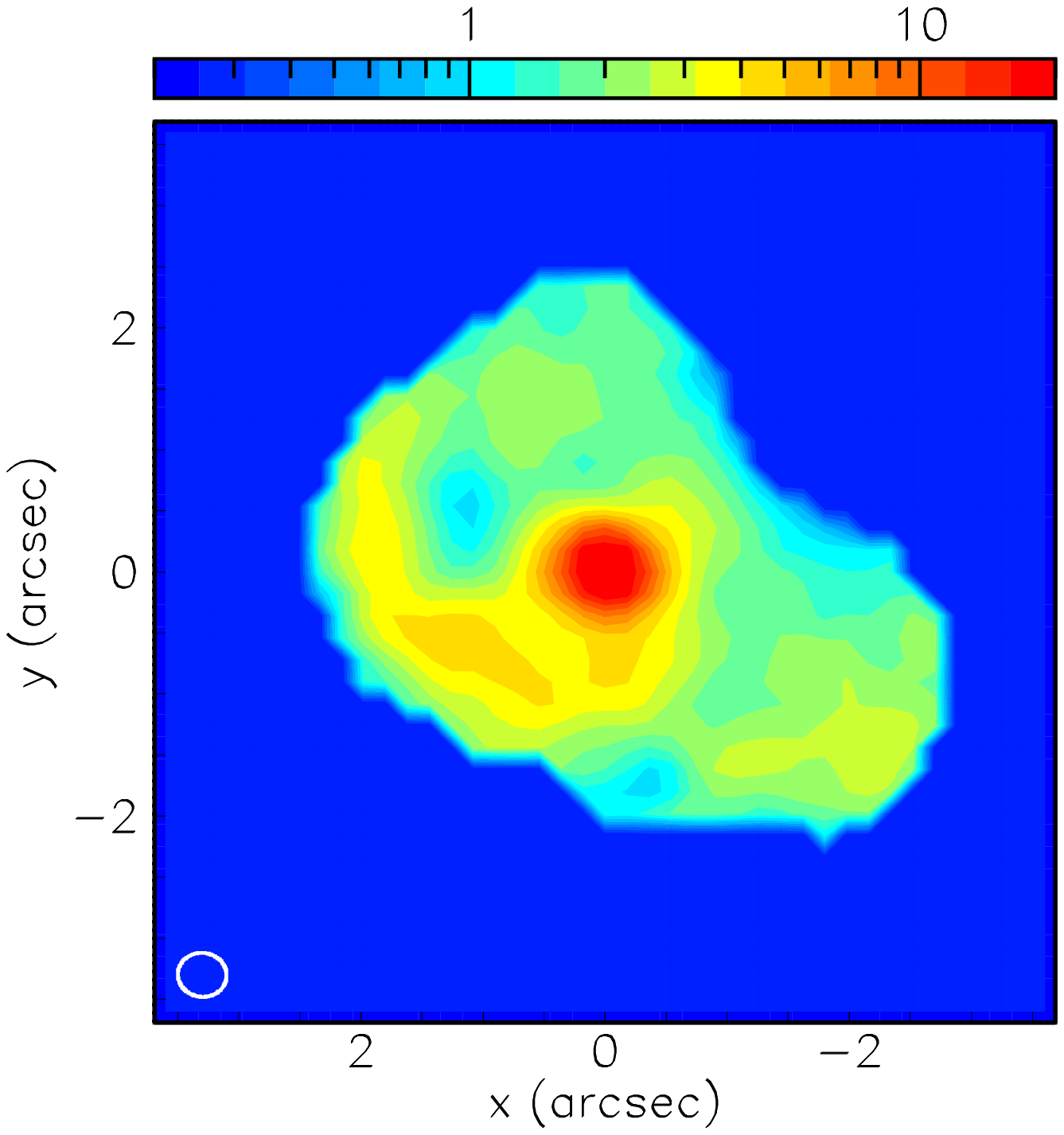}
  \includegraphics[height=4.5cm,trim=1.cm 1.5cm 1.2cm 0.5cm,clip]{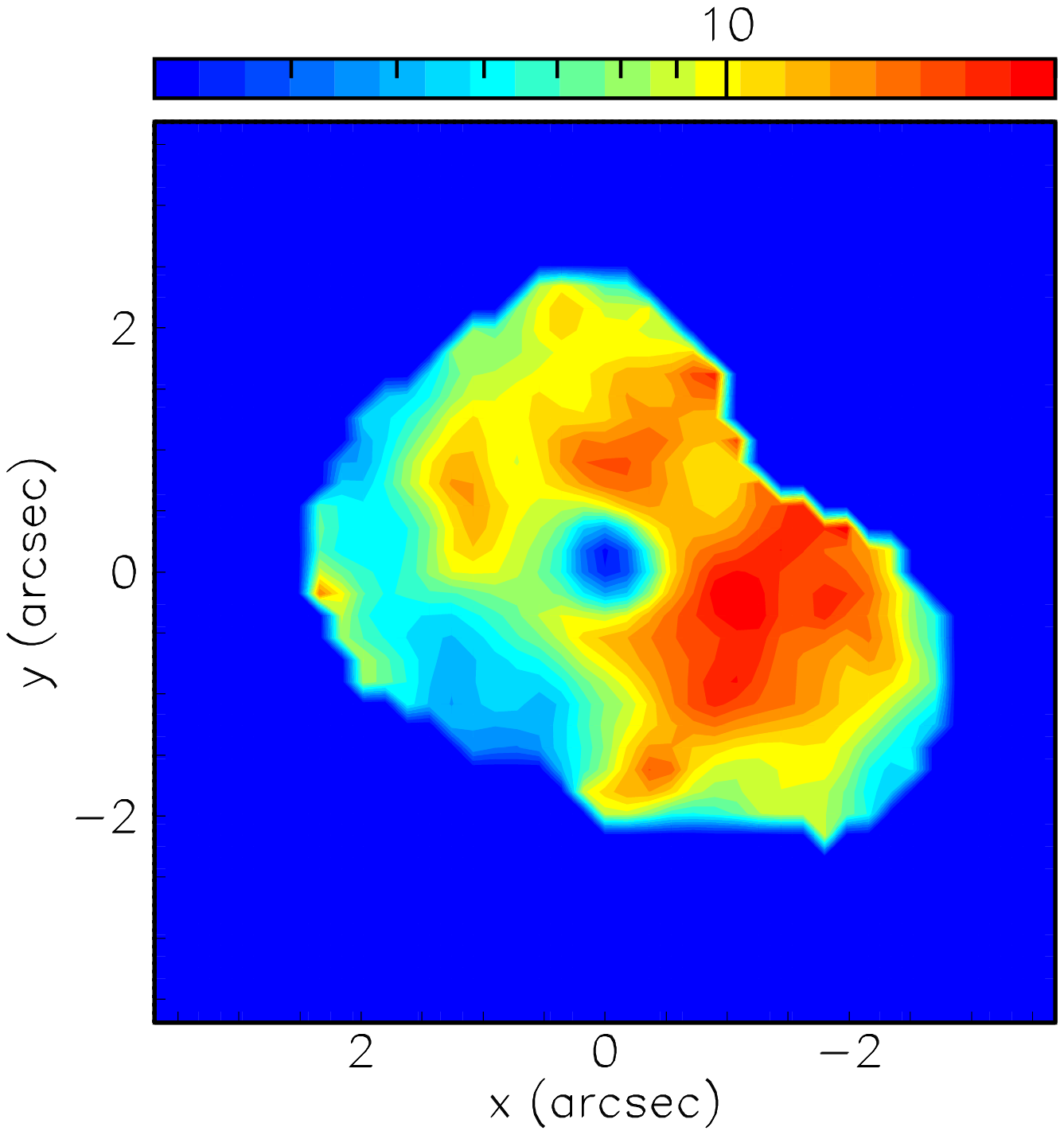}
  \caption{Left: map of the $^{12}$CO(3-2) intensity (Jy \kms\ arcsec$^{-2}$) integrated over $|V_z|$$<$4 \kms. The black contour defines region C retained for the comparison between $^{12}$CO and $^{13}$CO emissions. The $x'$ and $y'$ axes are shown. Center: map of the $^{13}$CO(3-2) intensity (Jy \kms arcsec$^{-2}$) in region C. Right: map of the $^{12}$CO(3-2)/$^{13}$CO(3-2) intensity ratio in region C.}
 \label{fig5}
\end{figure*}

 The radial dependence of the emission in region C of the $^{12}$CO(3-2) and $^{13}$CO(3-2) lines is illustrated in the central panel of Figure \ref{fig4}. Both line emissions give evidence for an important concentration of CO molecules in the close neighborhood of the star. Extrapolation to $R$=0 of the intensity observed in the ring 1$<$$R$$<$2 arcsec gives excesses of 28.9 and 6.4 Jy \kms\ for $^{12}$CO(3-2) and $^{13}$CO(3-2) respectively, excluding a contribution of $\sim$2.7 Jy \kms\ of the unresolved continuum emission. Evidence for such confinement had been previously obtained by \citet{Khouri2018} from high angular resolution ALMA observations of both the $^{12}$CO($\nu$=1,$J$=3-2) and $^{13}$CO($\nu$=0,$J$=3-2) line emissions; the beam was $\sim$30$\times$20 mas$^2$ but the maximal recoverable scale was only $\sim$0.4 arcsec: the present analysis provides a useful complement to their work, extending the reach to over 3 arcsec. Doppler velocity spectra are shown in the right panel of Figure \ref{fig4}; from earlier studies of the $^{12}$CO(3-2) emission, the features seen on the blue-shifted and red-shifted wings are associated with fragments emitted some few 1,000 years ago and the cut $|V_z|$$<$ 4 \kms\ is effective at selecting the central, more recent emission.

\begin{deluxetable}{ccc}
\tablenum{2}
\tablecaption{Parameters of relevance to the comparison between the $^{13}$CO(3-2) and $^{12}$CO(3-2) emissions. The values are taken from Leiden Atomic and Molecular Database (LAMDA; \citet{Schoier2005}). \label{tab2}}
\tablewidth{0pt}
\tablehead{  \colhead{Line}&\colhead{$^{12}$CO(3-2)}&\colhead{$^{13}$CO(3-2)}}
\startdata
Frequency (GHz)&
345.7960 &
330.5880 \\
Einstein coefficient, $A_{ji}$ (s$^{-1}$)&
2.50$\times$10$^{-6}$ &
2.18$\times$10$^{-6}$\\
Upper level energy, $E_u$ (K)&
33.2 &
31.7 \\
\enddata
\end{deluxetable}

\begin{figure*}
  \centering
  \includegraphics[height=5.5cm,trim=1.cm 1.5cm 1.2cm 0.5cm,clip]{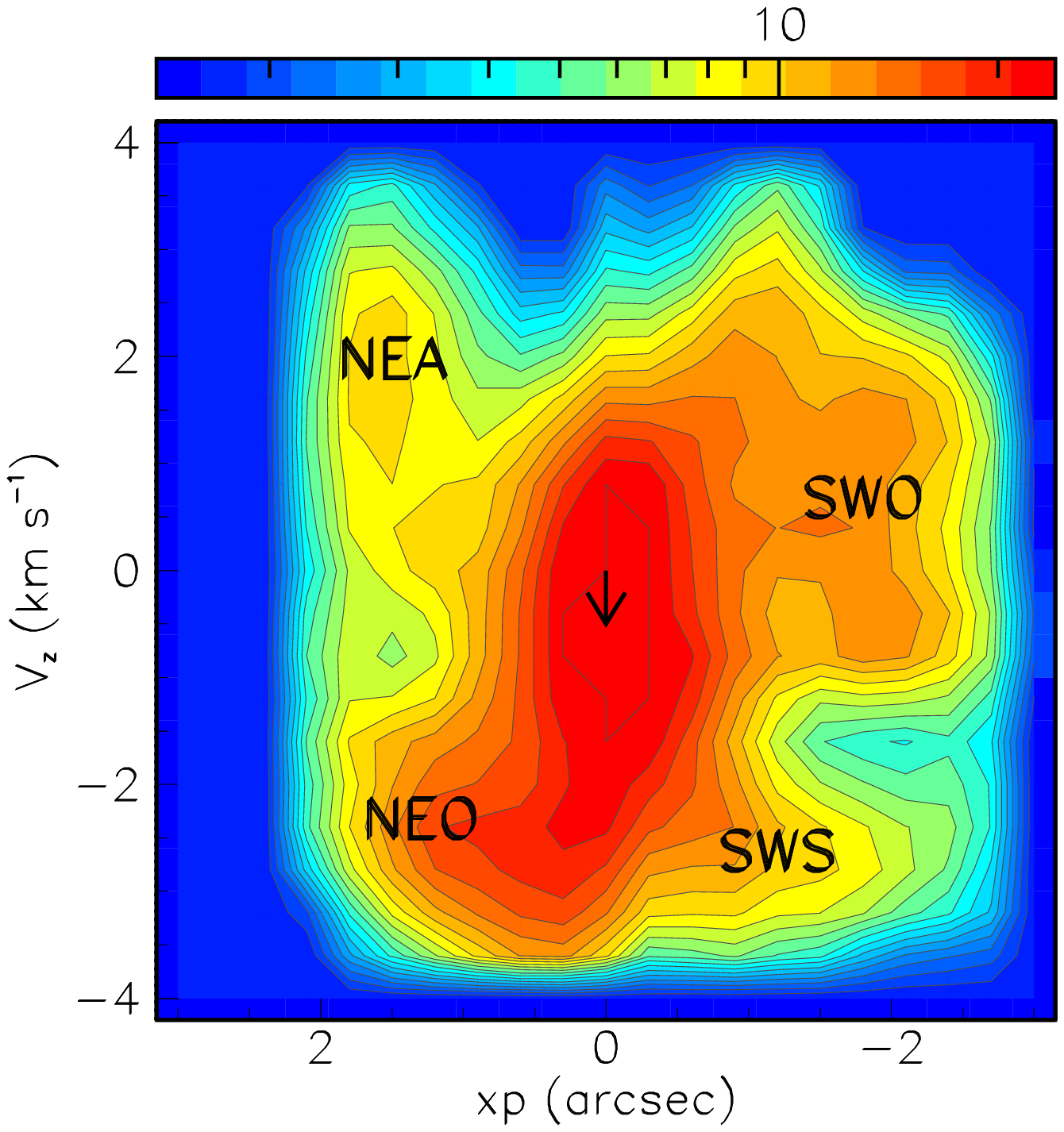}
  \includegraphics[height=5.5cm,trim=1.cm 1.5cm 1.2cm 0.5cm,clip]{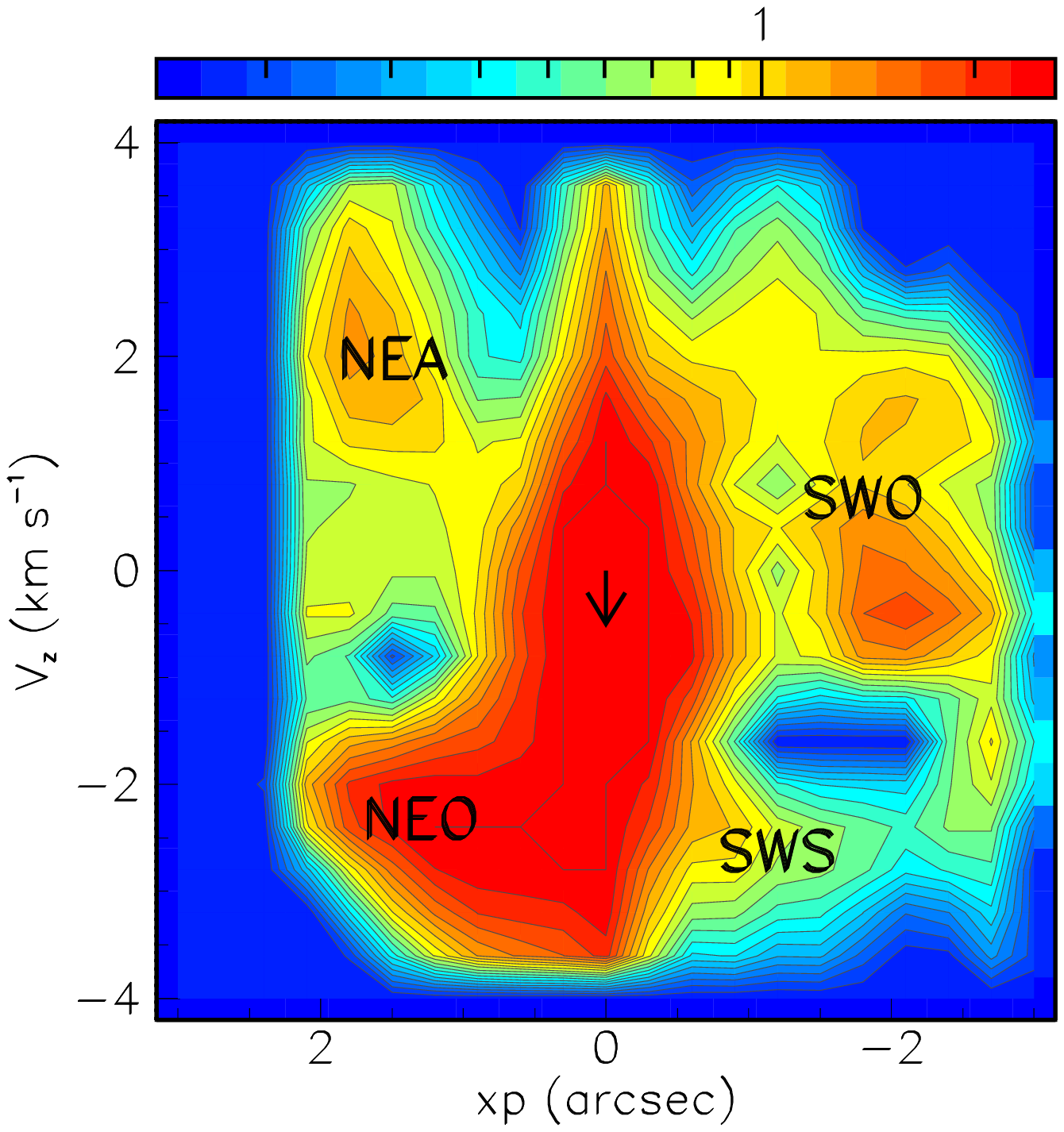}\\
  \includegraphics[height=5.5cm,trim=1.cm 1.5cm 1.2cm 0.5cm,clip]{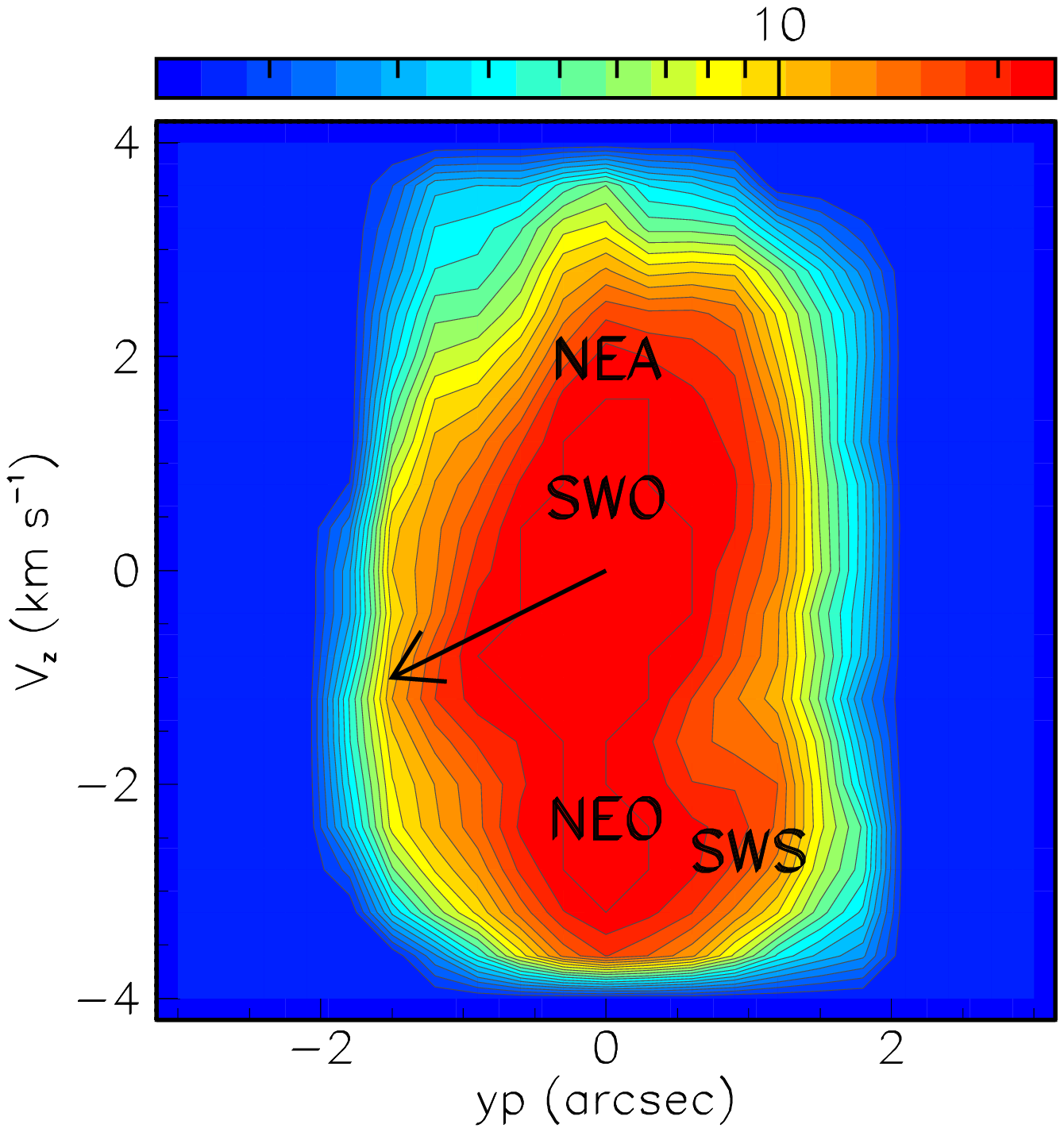}
  \includegraphics[height=5.5cm,trim=1.cm 1.5cm 1.2cm 0.5cm,clip]{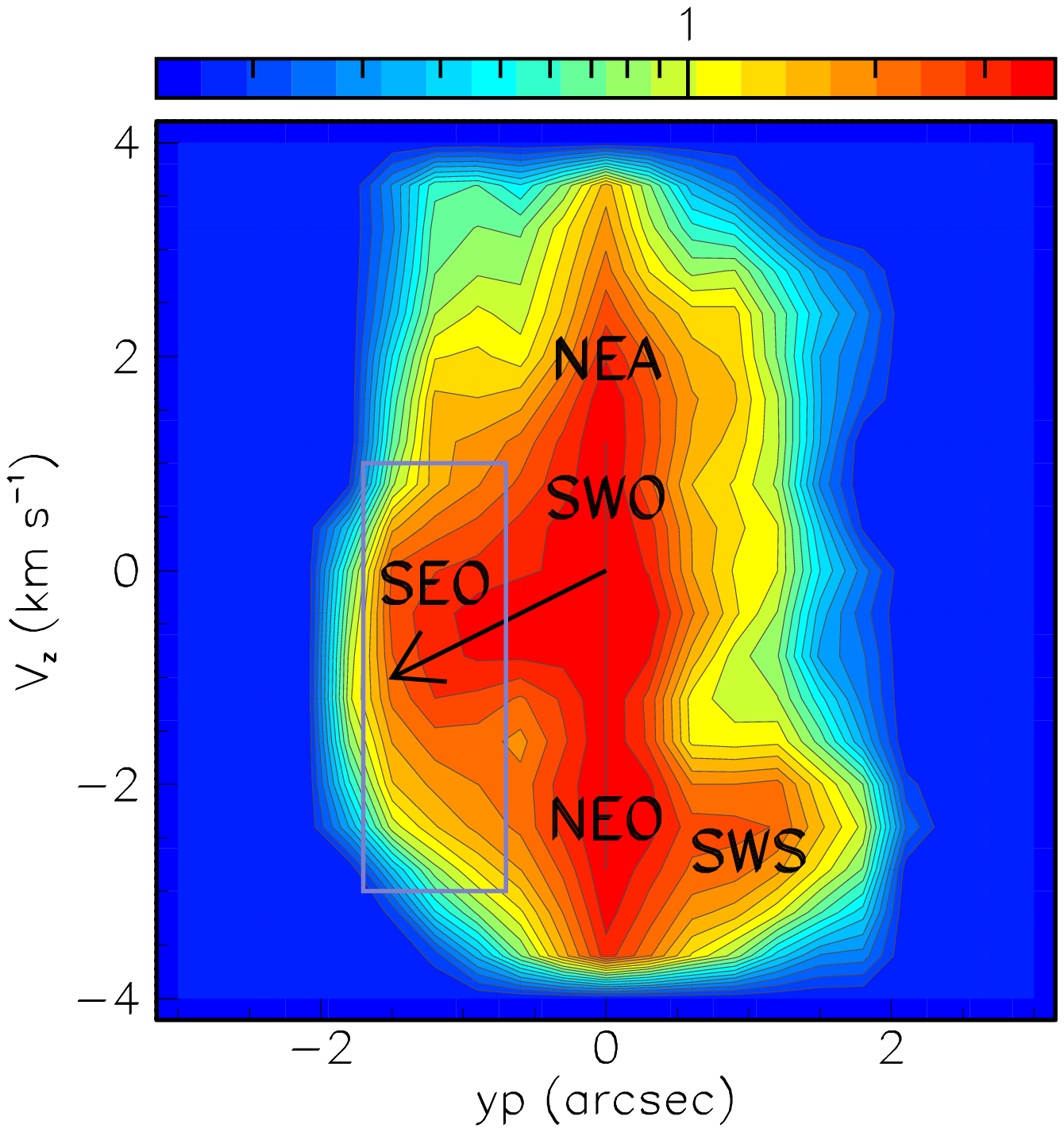}
  \caption{Comparing $^{12}$CO(3-2) and $^{13}$CO(3-2) emissions in region C: PV maps $V_z$ vs $x'$ (upper row) and vs $y'$ (lower row).  From left to right, $^{12}$CO(3-2), $^{13}$CO(3-2).  The labels refer to the fragments defined in Table \ref{tab3} of HTN20 and SWS stands for the newly identified south-western stream. NEA is for north-eastern arm, SWO for south-western outflow and NEO for north-eastern outflow; the arrow and the blue lines refer to the south-eastern outflow (SEO) dominated by the wind blowing from Mira A to Mira B. The color scales are in units of Jy arcsec$^{-1}$. }
 \label{fig7}
\end{figure*}

The maps of the intensity in region C of the $^{13}$CO(3-2) emission  and of its ratio to the $^{12}$CO(3-2) emission are displayed in the central and right panels of Figure \ref{fig5}.  To ease the comparison, we take advantage of the approximate symmetry displayed by region C about an axis $x'$ making an angle of $\sim$35\dego\ north of east: we use coordinates ($x',y'$) rotated clockwise by 35\dego\ from ($x,y$) to explore the morpho-kinematics. This is done in Figure \ref{fig7}, which uses $x'$ (pointing north-east) as axis of abscissa and $y'$, normal to it and pointing north-west, as axis of ordinate.  PV maps of the Doppler velocity $V_z$ vs $x'$ and $y'$ respectively, restricted to region C, are displayed for the $^{12}$CO(3-2) and $^{13}$CO(3-2) emissions. Such PV maps, rather than being confined to a narrow slit as usually done in PV diagrams, are integrated over the third coordinate of the data cube not used in the map, namely $y'$ for $V_z$ vs $x'$ maps and $x'$ for $V_z$ vs $y'$ maps. They are used in several places in the present article. The same approach to data-cube projection is adopted for the familiar $x$ vs $y$ intensity maps integrated over $V_z$. The lower opacity of the latter allows for a finer definition of the morphology of the emission. This representation makes it easier to identify the different fragments revealed in HTN20: a north-eastern arc (NEA) and a pair of outflows, north-eastern (NEO) and south-western (SWO). It also shows clearly an outflow that blows in the south-eastern direction, along $y'$$<$0, and is slightly blue-shifted: we refer to it as the south-eastern outflow (SEO). Most of it is the wind of Mira A focused by Mira B that has been studied in detail in HTN20; we shall comment on it in Section 4.2 when discussing the mass loss rate. In addition, the representation using rotated coordinates suggests separating a blue-shifted stream from the south-western outflow; it was clearly visible as a blob in the $V_z$ vs $\omega$ map but had been grouped together with the south-western outflow in HTN20. We prefer to give it a proper identity and we refer to it as south-western stream (SWS).

\begin{figure*}
  \centering
  \includegraphics[height=4.5cm,trim=1.cm 1.5cm 1.2cm 1.5cm,clip]{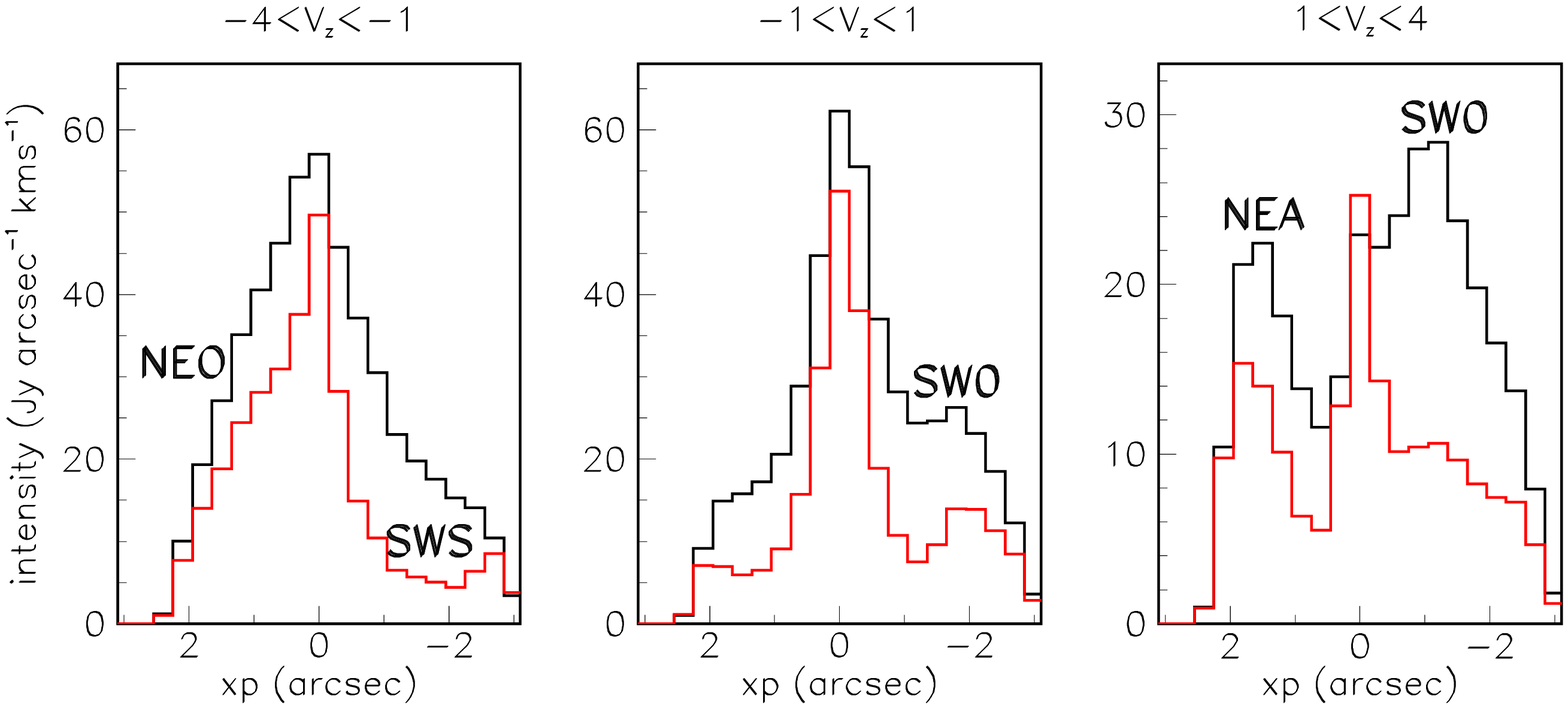}
  \includegraphics[height=4.5cm,trim=1.cm 1.5cm 1.2cm 1.5cm,clip]{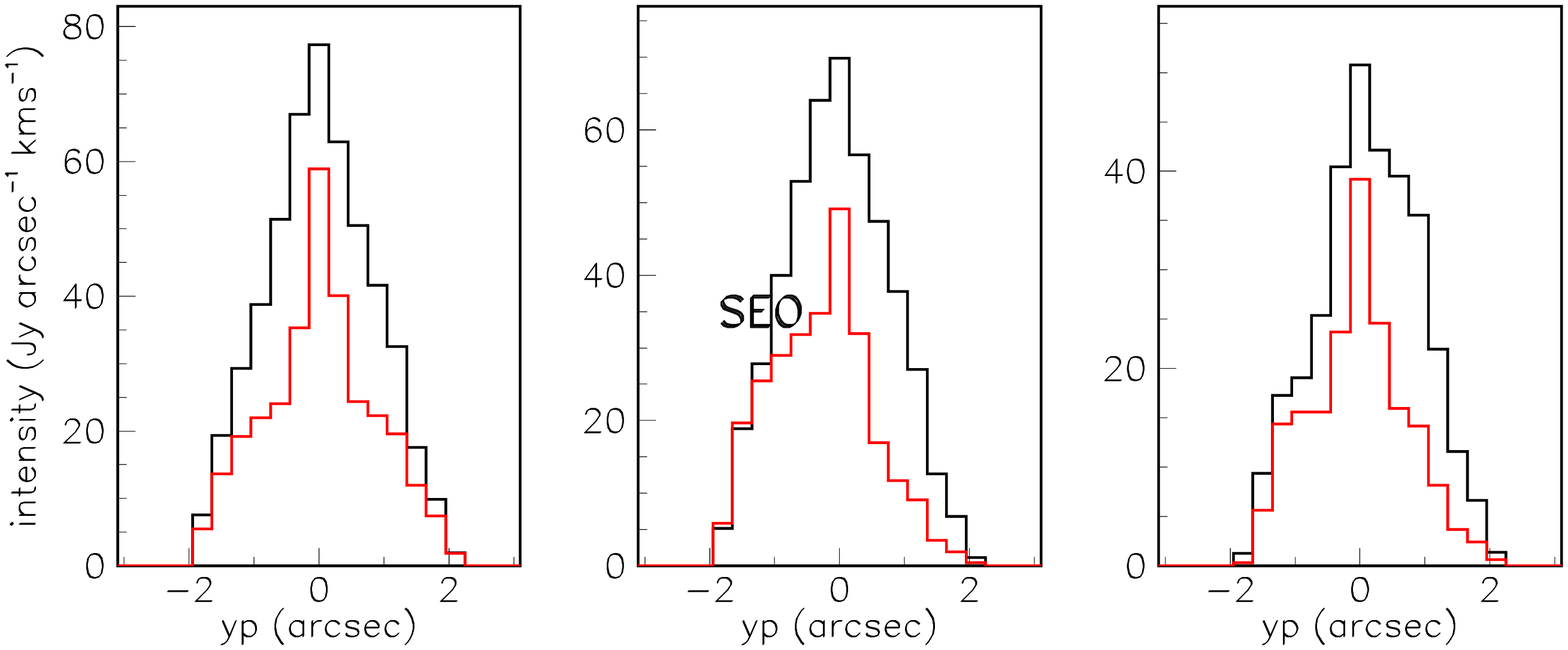}
  \caption{       Dependence on $x'$ (upper panels) and on $y'$ (lower panels) of the flux density  of the $^{12}$CO(3-2) (black) and $^{13}$CO(3-2) (red, multiplied by 5) emissions integrated inside region C over $V_z$ and over $y'$ (upper panels) or over $x'$ (lower panels), respectively. In each panel, Doppler velocity intervals are indicated on top of each column.}
 \label{fig8}
\end{figure*}

Another advantage of this representation is to make it clear that the NEA and the SWO are nearly detached from the central emission: they are associated with masses of gas that have been emitted several decades ago. In contrast, the NEO and SWS are in continuity with the central emission and are therefore associated with on-going mass loss. Together with the stream of wind focused by Mira B (SEO), they are blowing toward the blue hemisphere. As a further illustration of this result, we display in Figure \ref{fig8} projections on the $x'$ and $y'$ axes for three separate intervals of Doppler velocity: $-$4 to $-$1 \kms,  $-$1 to 1 \kms\ and 1 to 4 \kms.  

For each of the fragments making up region C, we evaluate the $^{12}$CO(3-2) and $^{13}$CO(3-2) emissions and calculate their ratio. The result is listed in Table \ref{tab3}, which defines which data cube elements are included in the definition of each fragment; in addition to be part of region C, their coordinates $x'$, $y'$ and $V_z$ must satisfy  inequality relations listed in the table. The $^{12}$CO/$^{13}$CO emission ratios vary between 6.6 and 10.0 with mean$\pm$rms values of 7.7$\pm$1.2, in good agreement with the mean value of 7.3 quoted by HTN20; they are smaller for the SEO and the NEO, which are blue-shifted, than for the south-western outflow.  

\begin{deluxetable*}{cccccccc}
\tablenum{3}
\tablecaption{  Compared emissions of the $^{12}$CO(3-2), $^{13}$CO(3-2) and SiO(5-4) lines in the fragments of region C. The values are shown with 3$\sigma$ uncertainties (see Section 4).\label{tab3}}
\tablehead{  \colhead{} & \colhead{$x'$} & \colhead{$y'$} & \colhead{$V_z$} & \colhead{$^{12}$CO} & \colhead{$^{13}$CO} & \colhead{$^{12}$CO/$^{13}$CO} & \colhead{SiO(5-4)}\\
 \colhead{} & \colhead{(arcsec)} & \colhead{(arcsec)} & \colhead{(\kms)} & \colhead{(Jy \kms)} & \colhead{(Jy \kms)} & \colhead{ratio} &\colhead{(Jy \kms)}}
\startdata
NEA&$>$0.8&-&$>$1.0&25$\pm$2&3.4$\pm$0.8&7.4$\pm$1.7&0.07$\pm$0.10\\
SWO&$<-$1.2&-&$>-$1.0&65$\pm$3&6.5$\pm$0.9&10.0$\pm$1.4&11.2$\pm$0.1\\
NEO&$>$0.5&-&$<-$1.3&45$\pm$2&6.7$\pm$0.8&6.7$\pm$1.2&0.65$\pm$0.10\\
SWS&-&$>$0.7&$<-$1.5&29$\pm$2&3.7$\pm$0.7&7.8$\pm$1.5&4.65$\pm$0.09\\
SEO&-&$>-$1.7,$<-$0.7&$>-$3,$<$1&54$\pm$2&8.2$\pm$0.8&6.6$\pm$0.7&0.88$\pm$0.11\\
\enddata
\end{deluxetable*}

\subsection{Mass loss rate}

The observed fragmentation and variability of the Mira wind implies that some caution needs to be exerted when defining a mass loss rate. We first limit our ambition to evaluating it in the south-eastern quadrant, $y'$$<$0; we see from Figure \ref{fig7} that it is dominated by the SEO and confined to Doppler velocities between $-$3 and 1 \kms. In contrast with fragments such as the NEA and SWO, the SEO can be expected to be a permanent source of mass loss, flowing in the slowly changing direction of Mira B. We use the $^{13}$CO data to take advantage of the low line opacity: in the optically thin approximation, the total flux provides a direct measure of the number of molecules. 

We evaluate the mass loss rate associated with the SEO as its yearly radial flux. To do so we measure its mean mass $M_R$ per unit of $R$ and its radial velocity $V_R$ projected on the plane of the sky, the mass loss rate being the product of these two quantities: $\dot{M}=M_R \times V_R$.

To measure $M_R$, we use the mean emission $F_R$ of the $^{13}$CO(3-2) line per unit of $R$. To obtain the relation between $F_R$ and $M_R$ we assume a radial dependence of the temperature of the form $T$[K]$=$109/$r$[arcsec] \citep{Ryde2001}, implying that the temperature in the relevant radial range varies between $\sim$50 and $\sim$200 K. Given the values of the Einstein coefficient and of the energy of the upper level listed in Table \ref{tab2}, an emission of 1 Jy \kms\ requires the presence of (2.0$\pm$0.9)$\times$10$^{46}$ $^{13}$CO molecules. The relation between the total emission ($F$) and the number of molecules ($N$) is $F=hc/(4\pi d^2)NA_{ji}f_{pop}$ where $h$ is Planck constant, $c$ the velocity of light, $d$ the distance to the Earth, $A_{ji}$ the Einstein coefficients and $f_{pop}=(2J+1)\exp(-E_u/T)/(T/2.7)$ the population of $^{13}$CO at level $J$. Assuming a $^{12}$C/$^{13}$C isotopic ratio of 12 and an abundance relative to H$_2$ of 4$\times$10$^{-4}$ \citep{vanDishoeck1992, Khouri2018}, this implies a total hydrogen mass of 12$\times$10$^{50}$ proton masses, or 1.0$\times$10$^{-6}$ solar masses for an emission of 1 Jy \kms. 

To measure $V_R$, we use the value of the velocity of the wind blowing from Mira A to Mira B, which is what SEO is made of, evaluated in HTN20 as 3.9$\pm$1.3 \kms. We recall that about a century ago, Mira B was 70 to 80 au south of Mira A in the sky plane. Over the past century, it moved in the blue-shifted direction on an orbit inclined by $\sim$60\dego\ with respect to the plane of the sky, and is presently nearly east of Mira A \citep{Prieur2002, Vlemmings2015, Planesas2016}. The wind blowing from Mira A to Mira B projects on the plane of the sky as $\sim$3.9$\times$cos50\dego$=$2.5 \kms\ at the time of the observation and as $\sim$3.9 \kms\ a century ago. On average over the SEO, the mean projected radial velocity is therefore $\sim$3.0$\pm$0.5 \kms\ or $\sim$(6.3$\pm$1.0)$\times$10$^{-3}$ arcsec yr$^{-1}$. 

\begin{figure*}
  \centering
  \includegraphics[height=4.5cm,trim=1.2cm 1.5cm 2cm 1cm,clip]{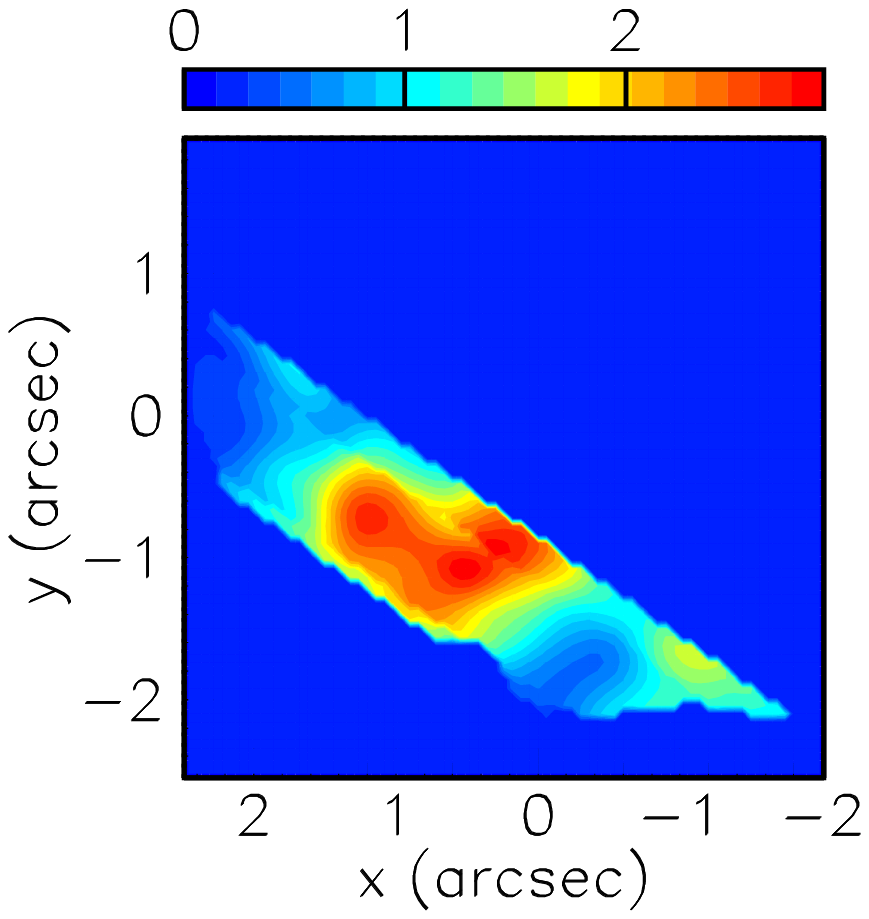}
  \includegraphics[height=4.5cm,trim=1.2cm 1.5cm 2cm 1cm,clip]{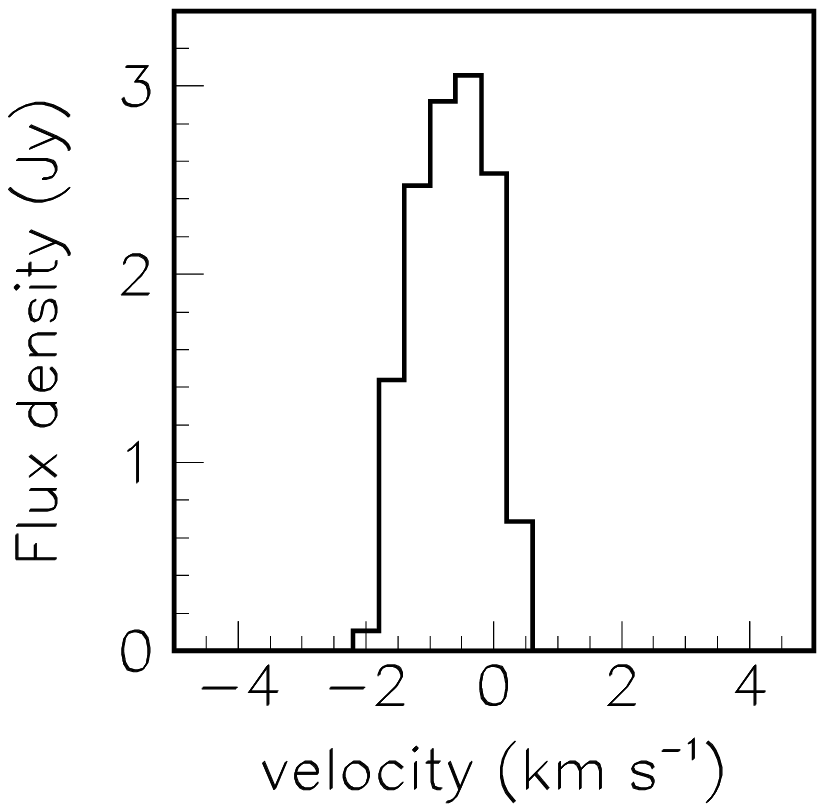}
  \includegraphics[height=4.5cm,trim=1.2cm 1.5cm 2cm 1cm,clip]{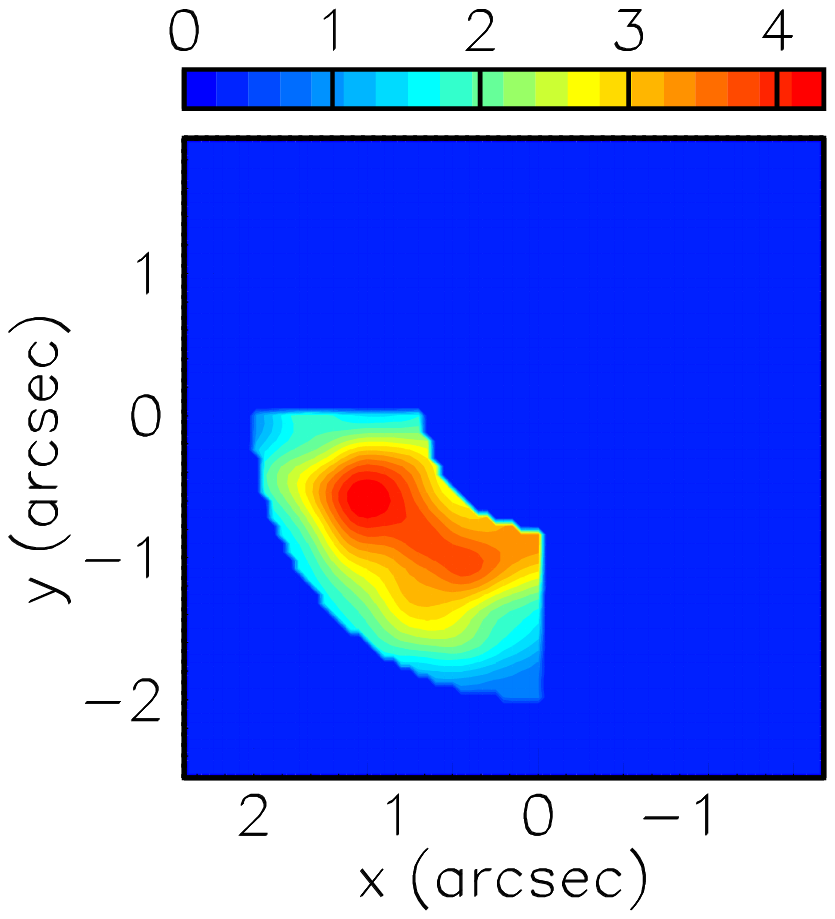}
  \includegraphics[height=4.5cm,trim=1.2cm 1.5cm 2cm 1cm,clip]{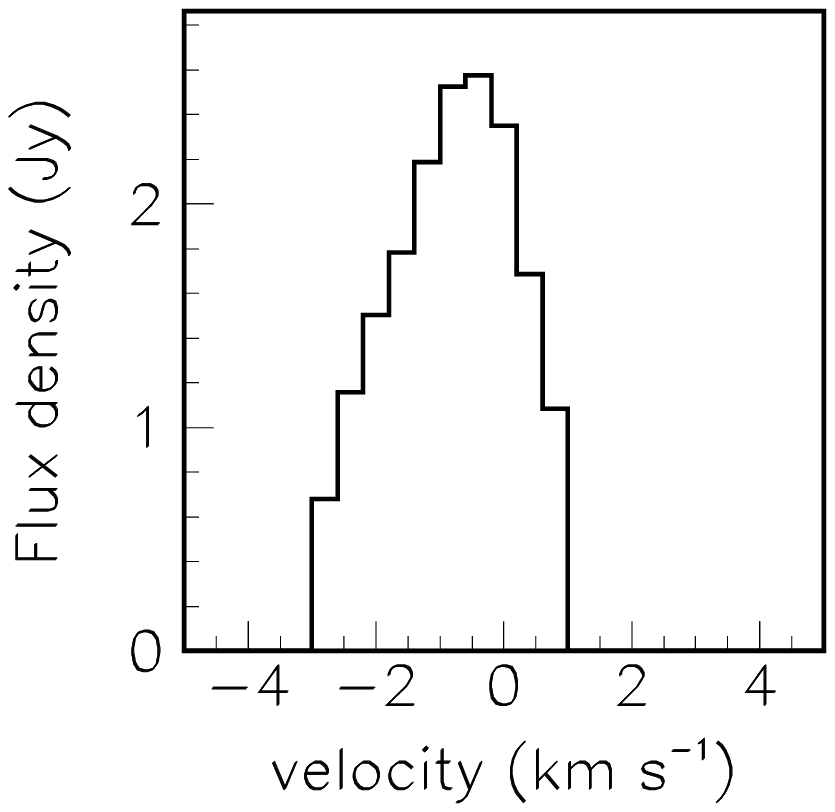}
  \caption{Intensity maps and Doppler velocity spectra of the $^{13}$CO(3-2) line emission for the two selections used to evaluate the mass loss rate in the south-eastern quadrant (see text). The color scales are in units of Jy arcsec$^{-2}$\kms.}
 \label{fig9}
\end{figure*}

In order to illustrate the evaluation of $F_R$, we present here two sets of selection criteria, which give total fluxes of 5.3 and 6.9 Jy \kms, respectively, each covering a radial range of $\sim$1.0$\pm$0.2 arcsec. Figure \ref{fig9} displays intensity maps and Doppler velocity spectra for both selections. On average, $F_R$ is (6.2$\pm$1.4)$\times$10$^{-6}$ solar masses per arcsec, giving a mass loss rate of (6.2$\pm$1.4)$\times$10$^{-6}$$\times$(6.3$\pm$1.0)$\times$10$^{-3}$$=$(3.9$\pm$1.7)$\times$10$^{-8}$ \msun yr$^{-1}$.   

The other main component of the current mass loss, the NEO, has a total flux of 6.7 Jy \kms\ (Table \ref{tab3}).  We estimate the values of its projected radial velocity and of its radial extension (HTN20) to be 2.9$\pm$0.8 \kms\ and 1.5$\pm$0.3 arcsec, respectively, giving a mass loss rate of (2.7$\pm$1.0)$\times$10$^{-8}$ \msun yr$^{-1}$. 

Adding together the contributions of the SEO and NEO, we obtain a mass loss rate of (6.6$\pm$2.0)$\times$10$^{-8}$ \msun yr$^{-1}$,  accounting for $\sim$30\% of the mass loss rate obtained from single dish observations \citep{Ryde2001, Heras2005}. But the main interest of the present analysis is to illustrate that  the concept of mass loss rate can only be defined on average when the mass loss proceeds by episodic and fragmented ejections, as has been shown to be the case for Mira Ceti. What can be defined precisely in such a case is the mass carried away by each fragment, and the radial velocity at which it escapes the star. The PV maps displayed in Figure \ref{fig7} have shown clearly that the SWO and NEA are now detached fragments and, therefore, do not contribute to the present mass loss rate evaluated close to the star. But averaged over a long period, they do. The present mass loss rate receives some contribution from the SWS, in addition to the SEO and NEO, increasing its value to the scale of 10$^{-7}$ \msun yr$^{-1}$. While very crude, this evaluation is the result of a direct measurement and is therefore a valuable addition to the indirect estimates usually quoted, which are typically twice as large. 
  
The episodic nature of the mass loss process, at the scale of decades, and its anisotropy, are not understood. They seem unrelated to the interaction between Mira A and Mira B, which has important consequences on the accretion by Mira B of part of the Mira A wind \citep{Wood2006, Sokoloski2010} but is nearly irrelevant to the genesis of the Mira A wind \citep[e.g.][and references therein]{Khouri2018}. The X-ray outburst observed in 2003 is believed to be associated with a magnetic flare followed by mass ejection, \citep{Karovska2005} but seems also unrelated to the general mass loss process. The variability and lumpiness observed at visible and infrared wavelengths \citep{Chandler2007, Kaminski2016, Wittkowski2016, Khouri2018} and at mm continuum emission \citep{Matthews2015, Vlemmings2015, Planesas2016} is at the scale of weeks and the variability of the mass loss process, at the scale of decades, seems again to be unrelated to it. A factor of relevance may be the difficulty to escape the gravity of Mira A+B ($\sim$2.7 solar masses): at 15 au from the center of the star, the escape velocity from Mira A is 15.4 \kms, larger than the boost that the shocks induced by pulsations and convective cell ejections usually produce. Beyond this distance, these shocks are no longer playing a role in the generation of the nascent wind \citep{Freytag2019}. The presence of a wind, with matter escaping the gravity of Mira A+B, relies therefore fully on whatever acceleration mechanism is active to follow up on the original boost. The complexity of the mechanisms which we know about, essentially collisions  of the gas molecules with dust grains to which the UV radiation of the star transfers momentum, is such that we cannot hope for accurate predictions. As was emphasized by \citet{Woitke2006}, oxygen-rich stars, in strong contrast with carbon-rich stars, are not good at producing dust grains that can efficiently accelerate the wind. The physico-chemistry at stake depends on the size and nature of the dust grains, on the nature and the rate of their formation, on their temperature, on how refractory and transparent they are, all of which can take a broad range of values. Qualitatively, these considerations, even if quite vague, may play a role in causing both the confinement of a large gas density at short distance from the star and the episodic and anisotropic nature of the mass loss: to a first approximation, there is simply no mass loss and gas accumulates around the star; to a second approximation mass loss occurs wherever and whenever there is a weak point through which matter can leak, causing episodic and anisotropic mass loss. Quantitatively, however, a credible picture of the physics at stake in the variability and anisotropy of the mass loss process is still lacking.

\section{ Peculiar morphology of the SiO(5-4) emission }
\begin{figure*}
  \centering
  \includegraphics[height=4.cm,trim=0.cm 0cm 0.cm 0.9cm,clip]{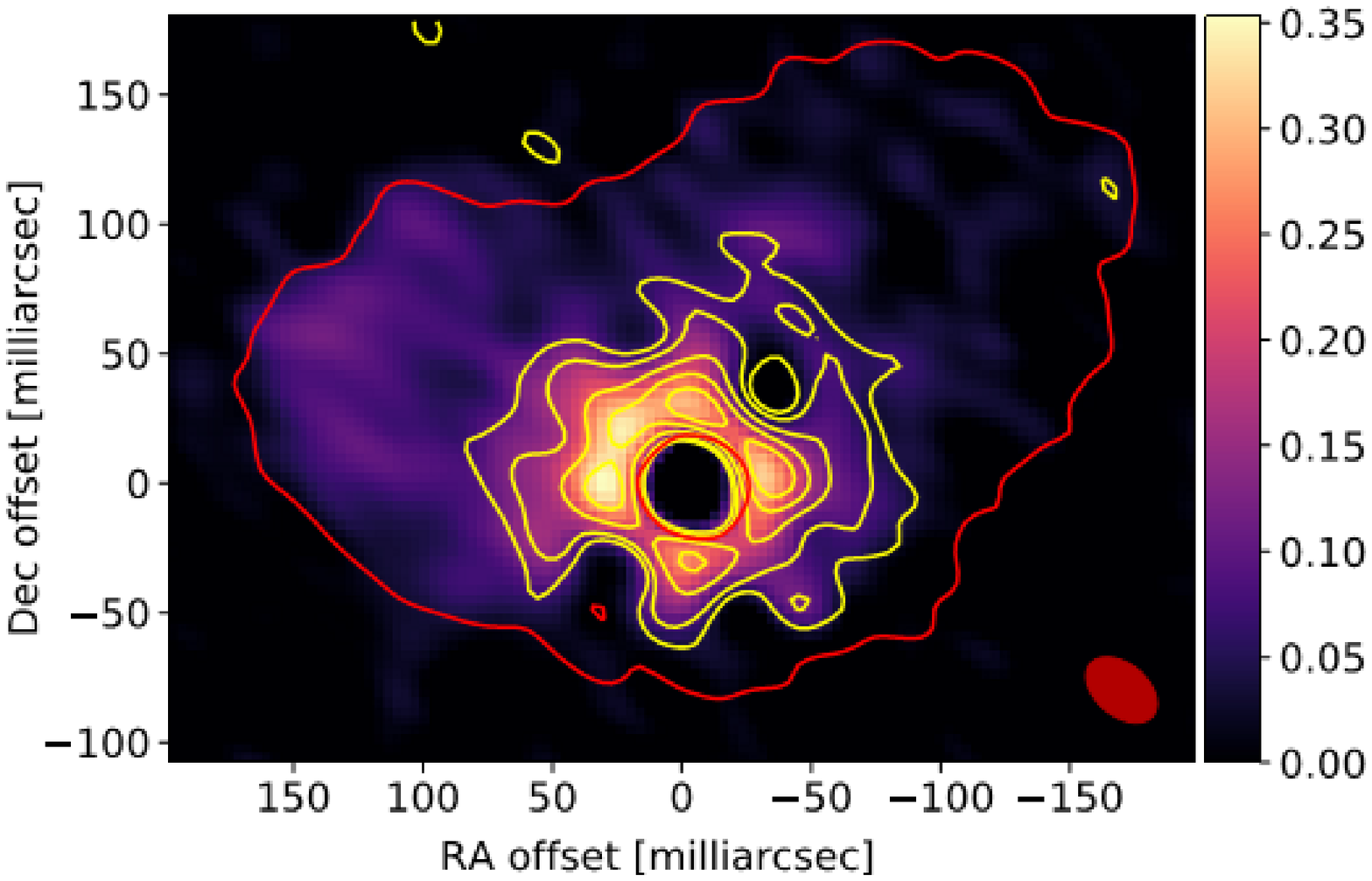}
  \includegraphics[height=4.3cm,trim=0cm 1.5cm 0cm 1.5cm,clip]{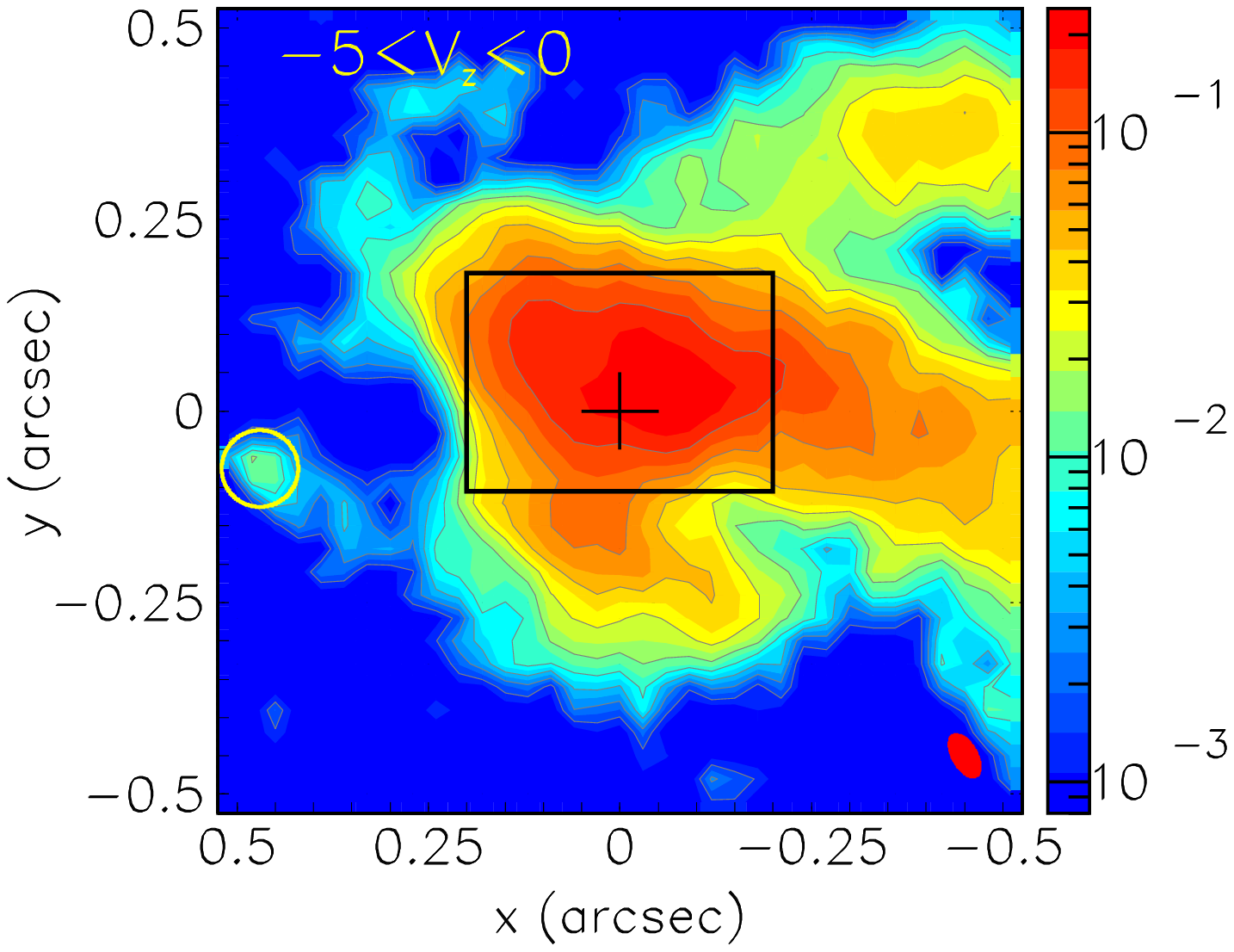}
  \includegraphics[height=4.3cm,trim=0cm 1.5cm 0cm 1.5cm,clip]{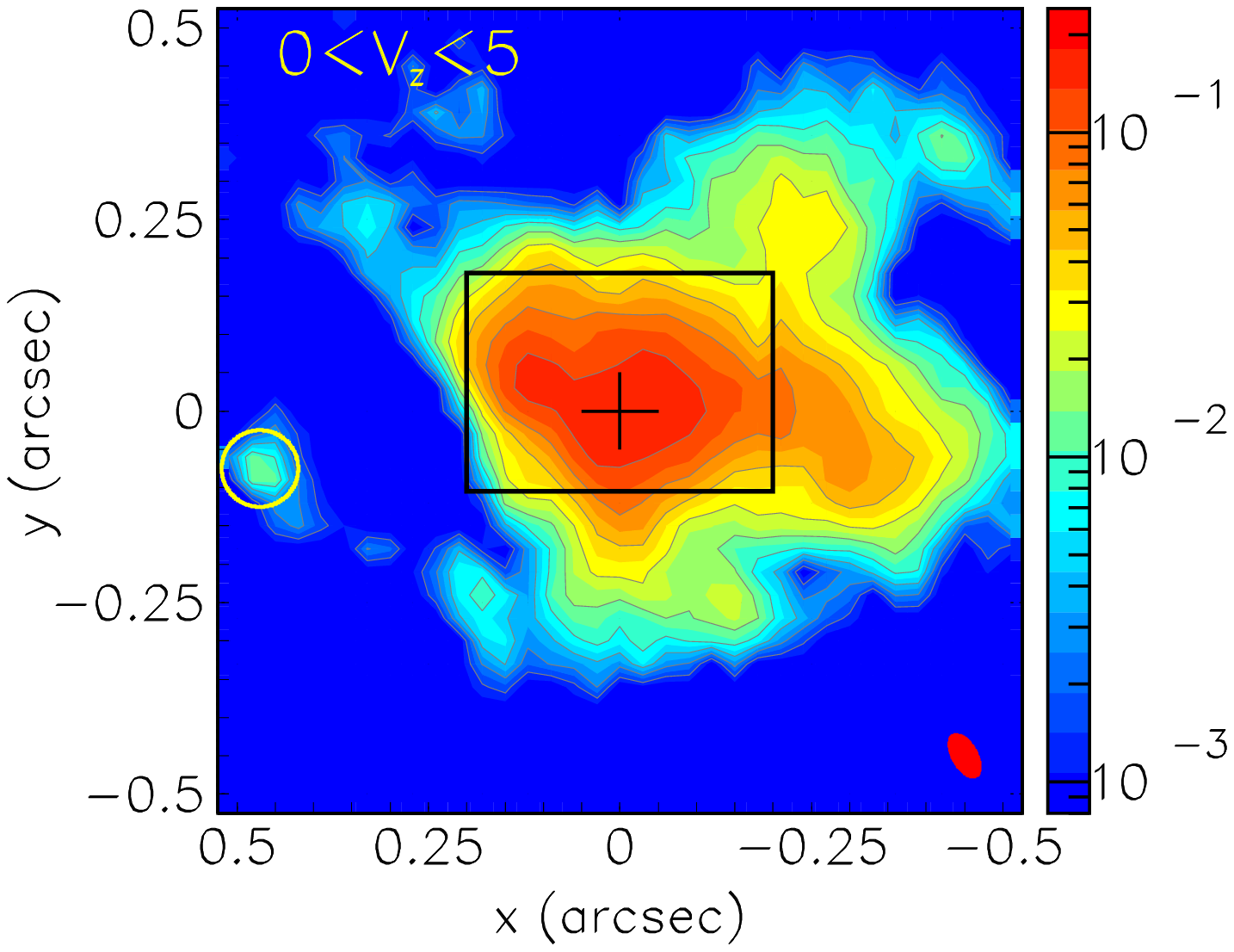}
  \caption{Left: Figure 8 of  \citet{Khouri2018}, comparing the emissions of AlO($N$=9-8)(color map, the color scale is in Jy \kms\ beam$^{-1}$), SO($N_J=8_8-7_7$) (red contour) and CO($\nu$=1,3-2) (yellow contours). Dust covers the northern part of the SO emission region, beyond some 80 mas north of the star. Center and right: intensity maps of the SiO(5-4) emission integrated in the blue-shifed (center) and red-shifted (right) hemispheres respectively. Mira B is clearly revealed by its continuum emission that has been studied by \citet{Planesas2016}. Its location is indicated by a yellow circle. The color scales are in units of Jy beam$^{-1}$\kms.}
 \label{fig10}
\end{figure*}

\citet{Wong2016} studied the Doppler velocity spectrum of the emission of several SiO lines observed over the stellar disc of Mira Ceti. In contrast to other similar stars, this spectrum displays no blue-shifted spike signaling the terminal wind velocity and radiative transfer modeling reveals an abrupt decline of SiO density beyond some 60-80 mas from the center of Mira A. As illustrated in the left panel of Figure \ref{fig10}, \citet{Khouri2018, Khouri2019} have shown that similar confinement affects the emissions of the SO($N_J$$=$8$_8$-7$_7$), $^{13}$CO(3-2), CO($\nu$=1,3-2) and AlO($N$=9-8), as well as other molecular line emissions; they give evidence for dust clustering on the northern and north-western edges of the region of SO emission, beyond some 80 mas projected distance from the star center. In the preceding section (Figure \ref{fig4}), we have shown that the present observations of the emission of the CO(3-2) line was consistent with such confinement for both isotopologues, however with low angular resolution. The SiO line emission studied in Section 3 is instead seen with high angular resolution that allows for a meaningful comparison (Figure \ref{fig10}) with the observations of \citet{Khouri2018}; it displays a similar confinement, within 100-200 mas projected distance from the star, as seen for the SO line. In the present section, we use these observations of the SiO line emission with the aim to shed light on the nature of the observed confinement. Indeed, these observations are challenging in two respects: the absence of SiO emission over most of the sky plane is at variance with what we know of other oxygen-rich AGB stars; its presence in a small south-western region needs therefore to be explained, paying particular attention to significant differences displayed there by the CO and SiO emissions.

\subsection{Confinement at short distances of the SiO emissions outside the south-western quadrant}

\begin{figure*}
  \centering
  \includegraphics[height=4.4cm,trim=1.cm 1.5cm 1.5cm 0.5cm,clip]{fig7b-13co-xpvzmap-cregion.eps}
  \includegraphics[height=4.4cm,trim=1.cm 1.5cm 1.5cm 0.5cm,clip]{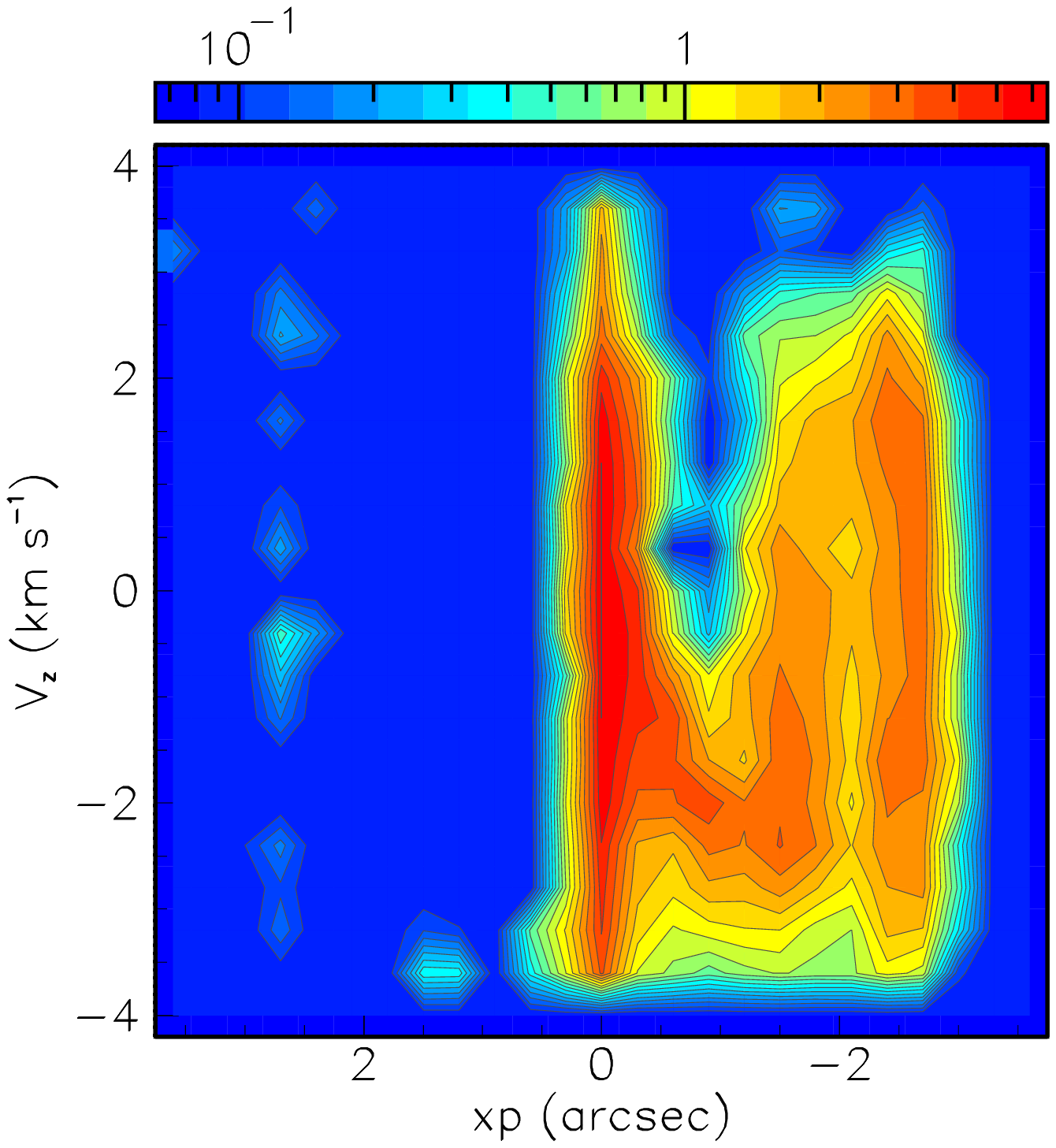}
  \includegraphics[height=4.4cm,trim=1.cm 1.5cm 1.5cm 0.5cm,clip]{fig7e-13co-ypvzmap-cregion.eps}
  \includegraphics[height=4.4cm,trim=1.cm 1.5cm 1.5cm 0.5cm,clip]{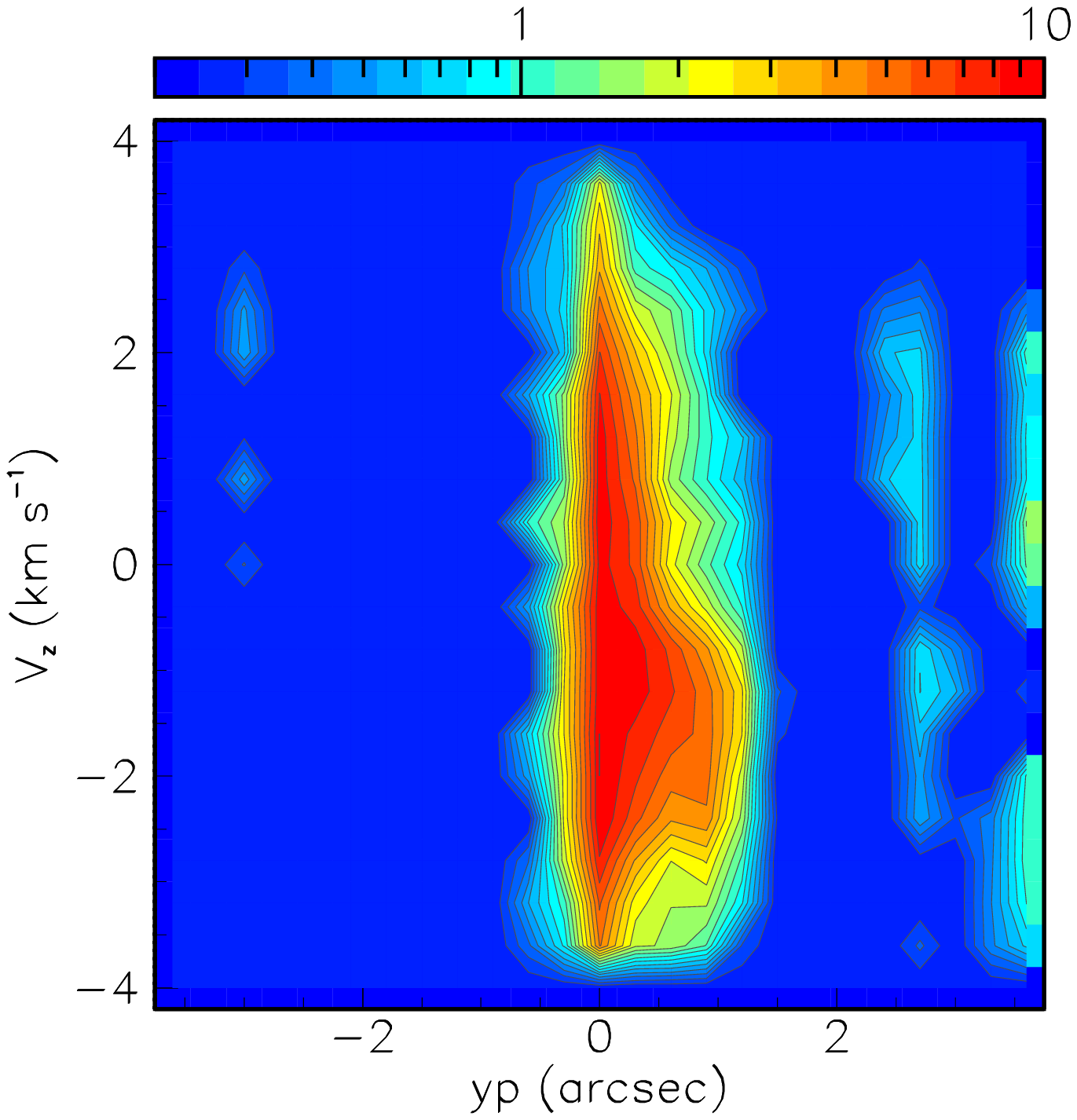}
  \caption{ Comparing $^{13}$CO(3-2) and SiO emissions: PV maps $V_z$ vs $x'$ (left pair) and vs $y'$ (right pair).  In each pair of panels $^{13}$CO(3-2) is left and SiO(5-4) is right. The labels are the same as defined in Figure \ref{fig7}. CO emission is from region C only but SiO emission is not restricted to a particular region. The color scales are in units of Jy arcsec$^{-1}$.}
 \label{fig12}
\end{figure*}

Figure \ref{fig10} shows that Mira A is currently ejecting SiO gas molecules mostly toward the blue-shifted hemisphere, as it does for CO molecules. However, closer inspection reveals major differences between the two emissions. This is illustrated in Figure \ref{fig12}, which compares PV maps of $V_z$ vs $x'$ and $y'$ for the emissions of $^{13}$CO and SiO molecules as was done in Figure \ref{fig7} for the two CO isotopologues. The comparison is made with $^{13}$CO rather than $^{12}$CO in order to minimize effects of opacity. SiO is essentially absent from the SEO, the NEA and the NEO but is present in the SWO and the SWS. As the SiO observations were obtained with a very different and much more extended antenna configuration than the CO observations, we checked that imaging was well-behaved up to al least 3 arcsec projected distance from the star. We also made sure that the arguments developed in the present section are not significantly affected by the very different angular resolutions of the CO and SiO observations.  

\begin{figure*}
  \centering
  \includegraphics[height=4.8cm,trim=1.cm 1.3cm 1.2cm 0.5cm,clip]{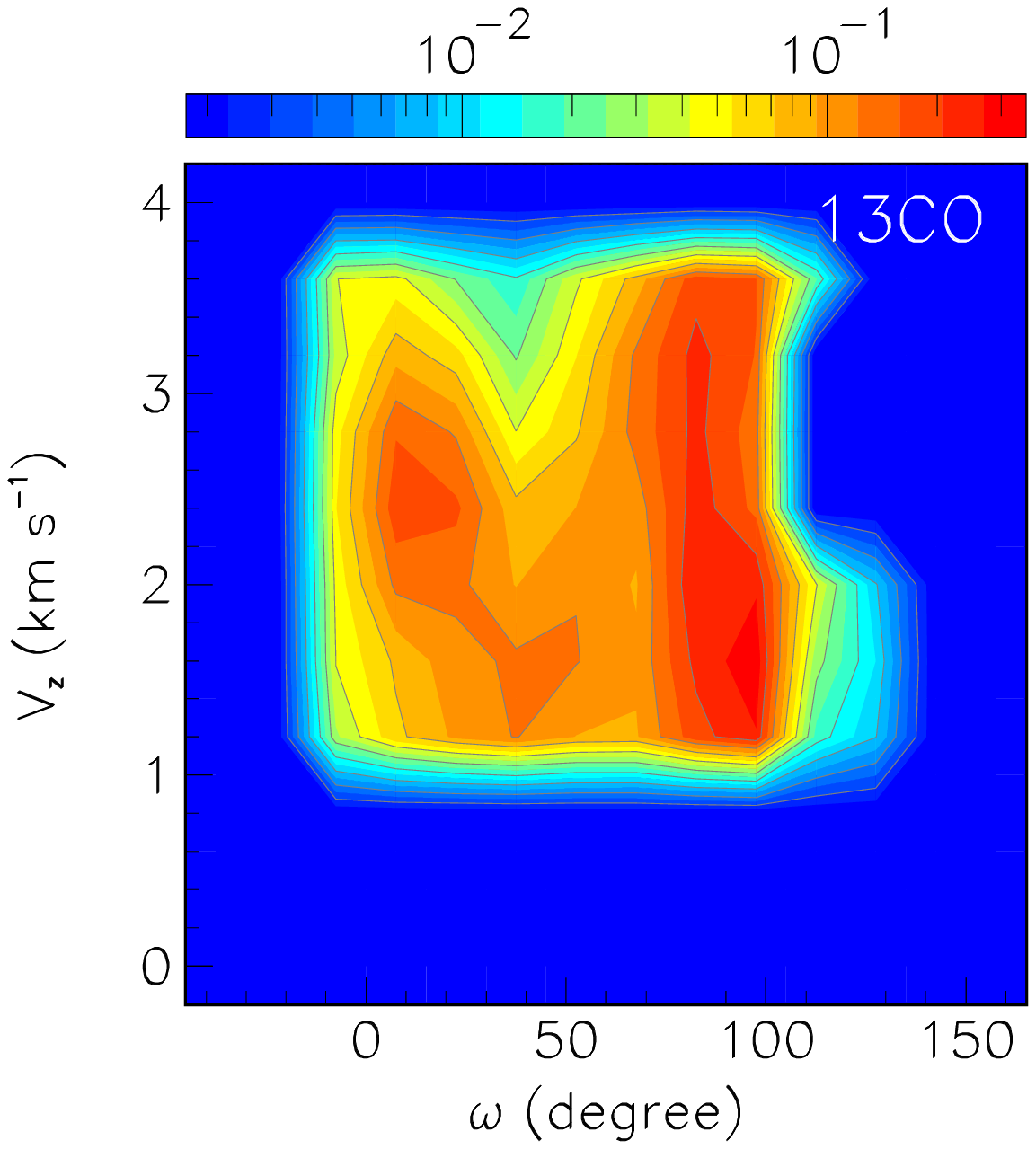}
  \includegraphics[height=4.8cm,trim=1.cm 1.3cm 1.2cm 0.5cm,clip]{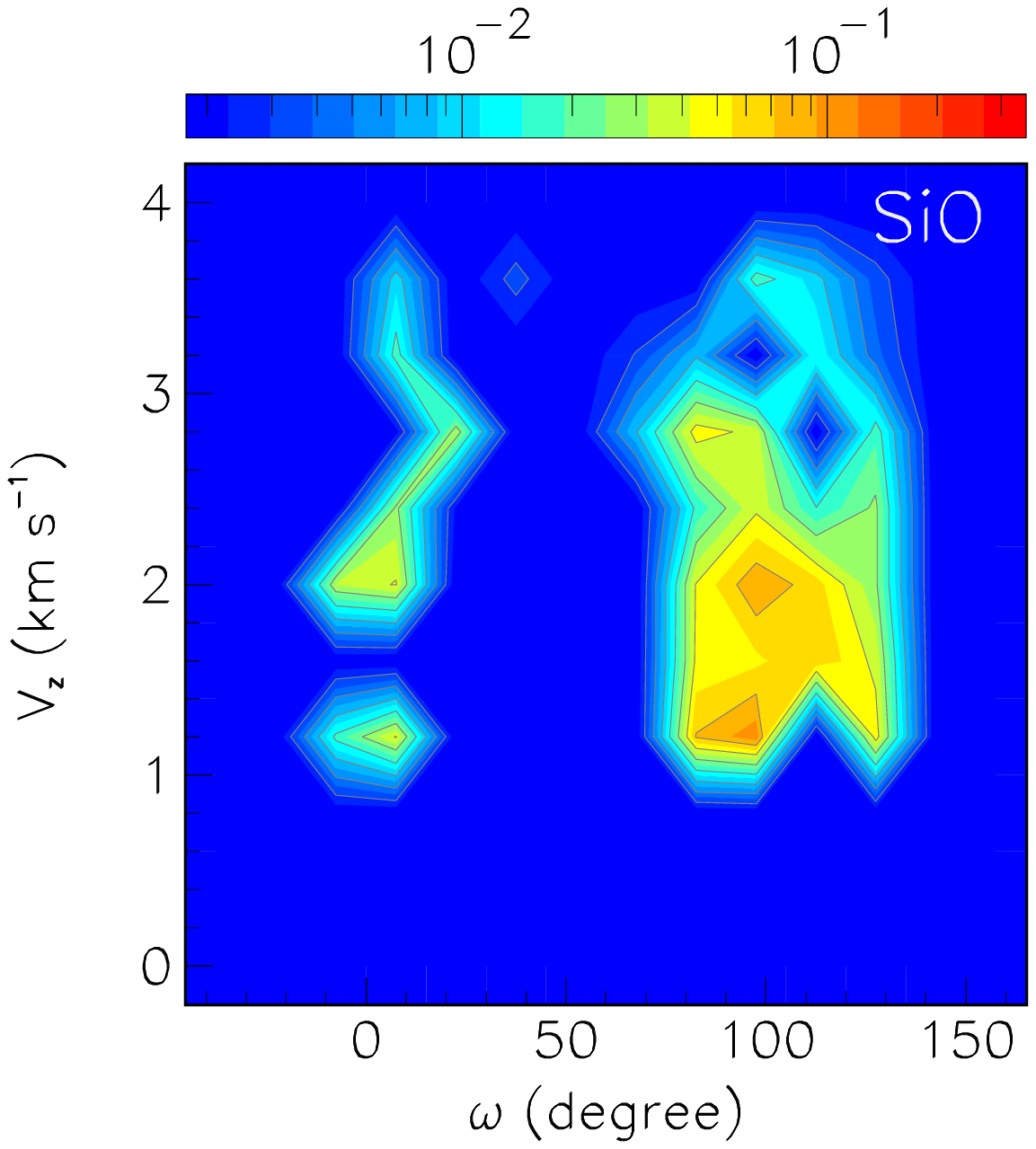}
  \includegraphics[height=4.8cm,trim=.7cm 1.3cm 1.2cm 0.5cm,clip]{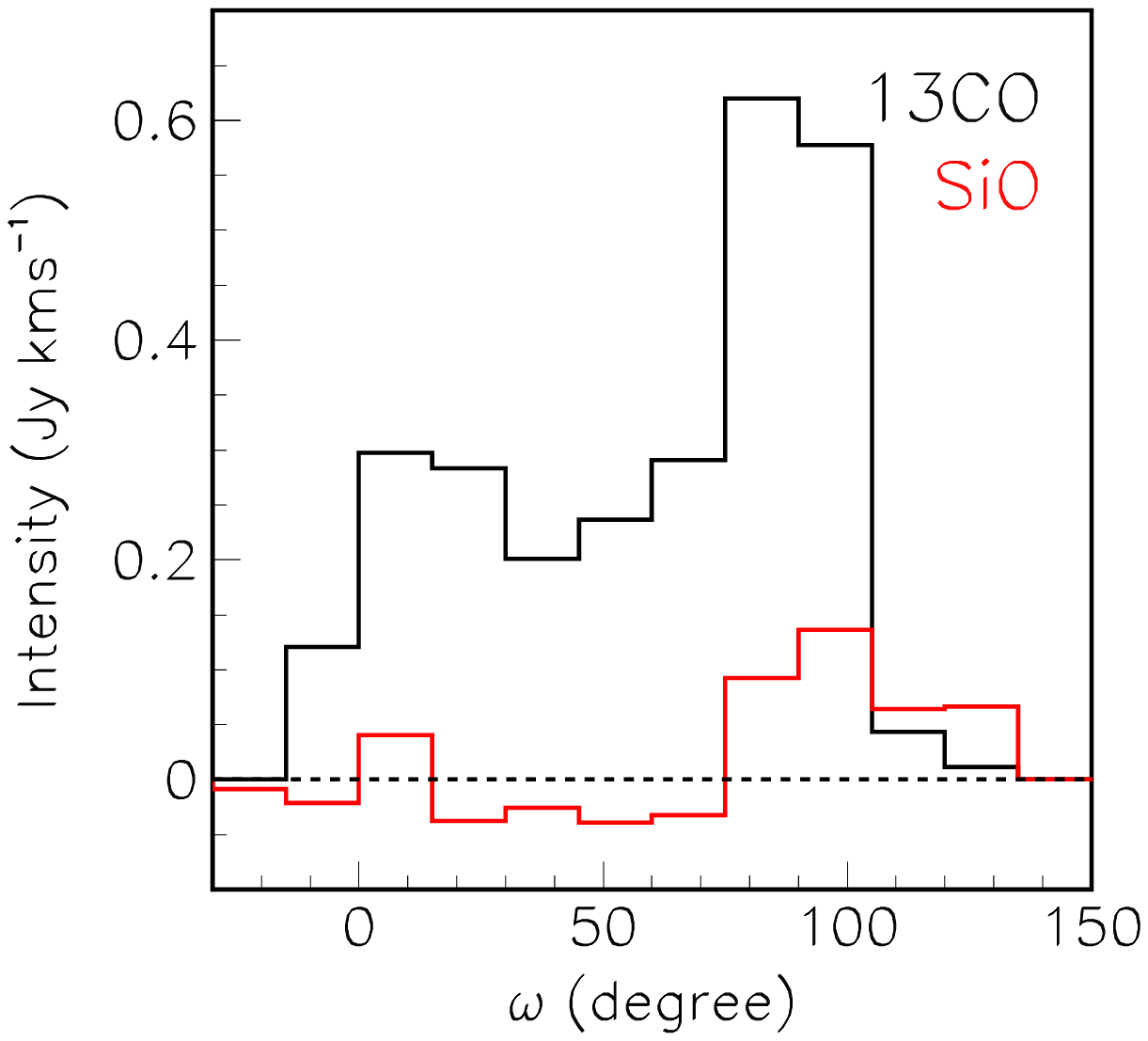}
  \caption{Comparison of the NEA intensities (Table \ref{tab3}) observed in $^{13}$CO(3-2) and SiO(5-4) emissions. PV maps are shown in the left ($^{13}$CO) and central (SiO) panels. The $\omega$ distributions are shown in the right panel. The color scales are in units of Jy.}
 \label{fig13}
\end{figure*}

The last column of Table \ref{tab3} lists the values of the measured SiO(5-4) intensities; they are a bit larger than the $^{13}$CO(3-2) intensities for SWO and SWS but an order of magnitude smaller for NEA, NEO and SEO. To get deeper insight into the lack of SiO emission in most of the sky plane, we exclude the south-western quadrant from the remaining of the present section and keep its study for the next section. For the values of SiO emission listed in Table \ref{tab3} to be meaningful, the identification of each fragment in terms of intervals of $x'$, $y'$ and $V_z$ must properly match both CO and SiO observations: is SiO completely absent outside the south-western quadrant or is it present in each CO fragment, but at a much lower level? We explored this issue in considerable details and concluded in favor of the latter. We illustrate it in the case of the NEA in Figure \ref{fig13} with PV maps of $V_z$ vs $\omega$ for each of the two lines.  We choose NEA because it is most clearly identified in CO emission and because the value quoted in Table \ref{tab3} for the SiO line is compatible with noise. While being an order of magnitude smaller than the CO intensity (0.23 vs 2.68 Jy \kms) the SiO intensity displays an $\omega$ distribution clearly correlated with that of the CO intensity. 

In the optically thin approximation, the ratio between the SiO and $^{13}$CO column densities associated with a same emissivity is nearly temperature independent because of the similarity between the upper energy levels and is dominated by the ratio of the Einstein coefficients multiplied by the ($2J+1$) and partition function factors; its value is $\sim$7.5 10$^{-3}$. Using a typical SiO(5-4)/$^{13}$CO(3-2) intensity ratio of $\sim$0.1 outside the south-western quadrant and a $^{12}$CO/$^{13}$CO abundance ratio of 12 gives an SiO/$^{12}$CO abundance ratio of 6.3 10$^{-5}$. Excluding the south-western quadrant, the episodic mass loss of Mira A produces therefore fragments containing both SiO and $^{12}$CO molecules, the latter being typically 1 to 2$\times$10$^4$ times more abundant than the former. 

\begin{figure*}
  \centering
  \includegraphics[height=4.5cm,trim=.5cm 1.5cm 1.5cm 1.cm,clip]{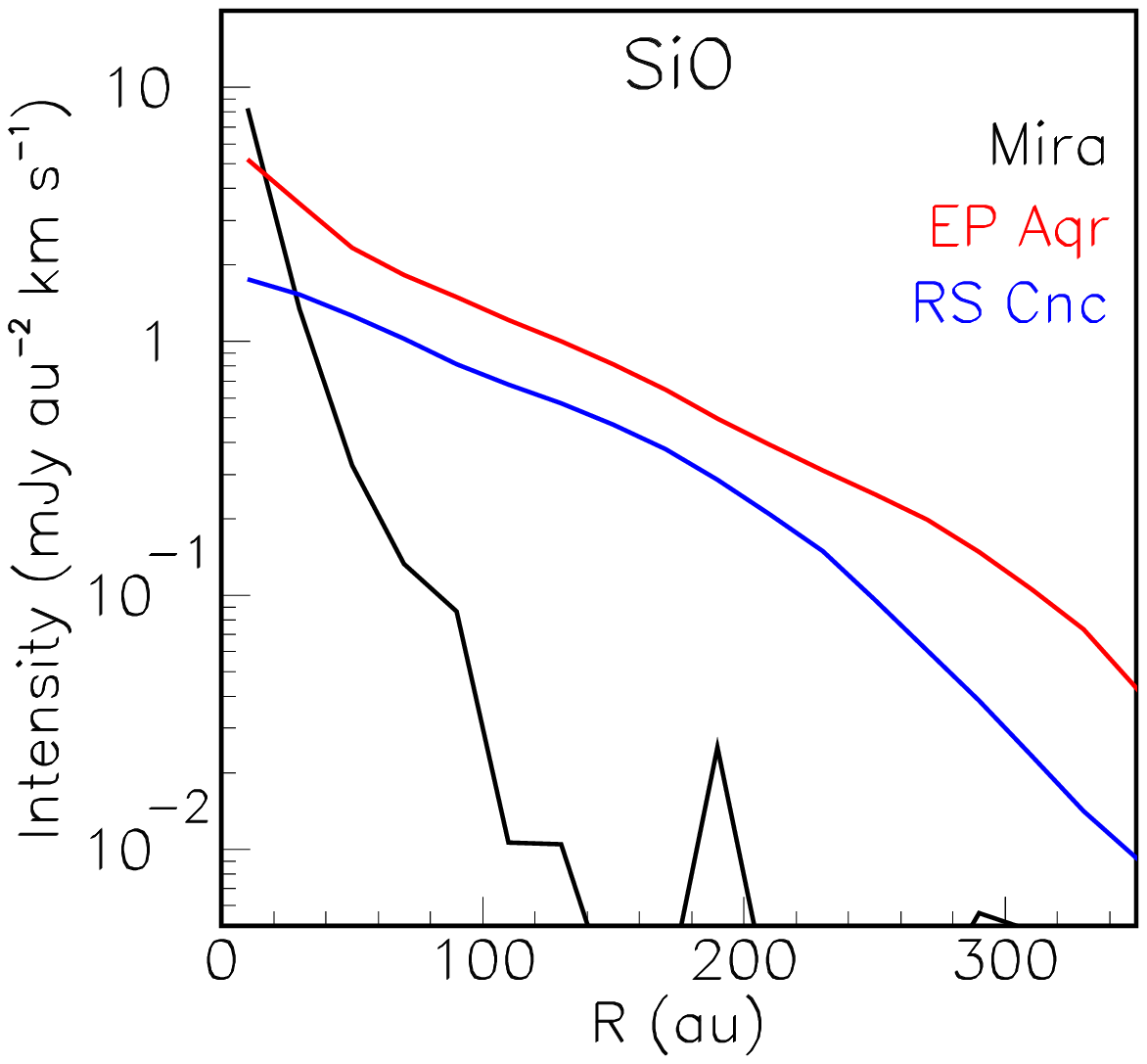}
  \includegraphics[height=4.5cm,trim=.5cm 1.5cm 1.5cm 1.cm,clip]{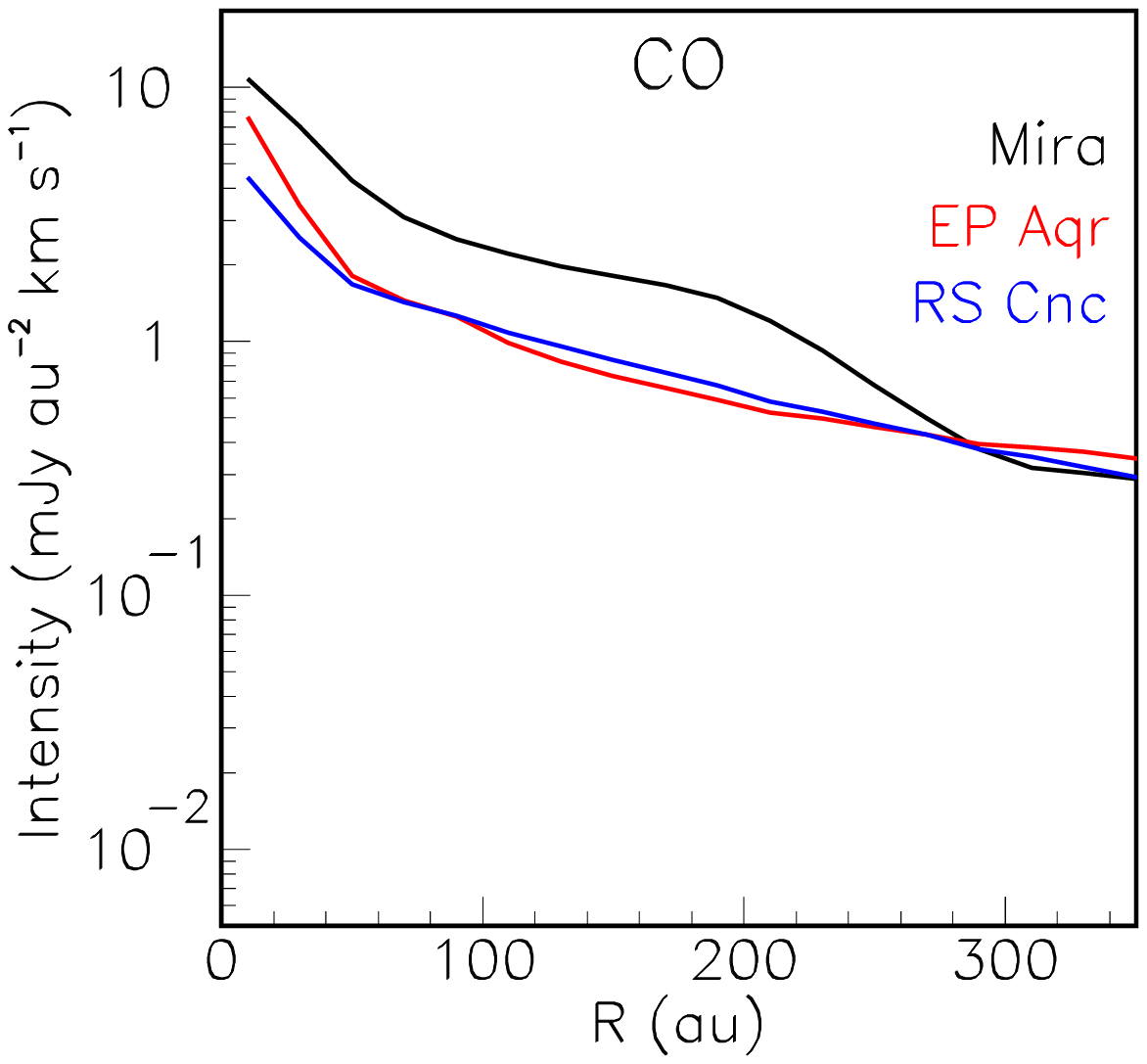}
  \includegraphics[height=4.5cm,trim=.5cm 1.5cm 1.5cm 1.cm,clip]{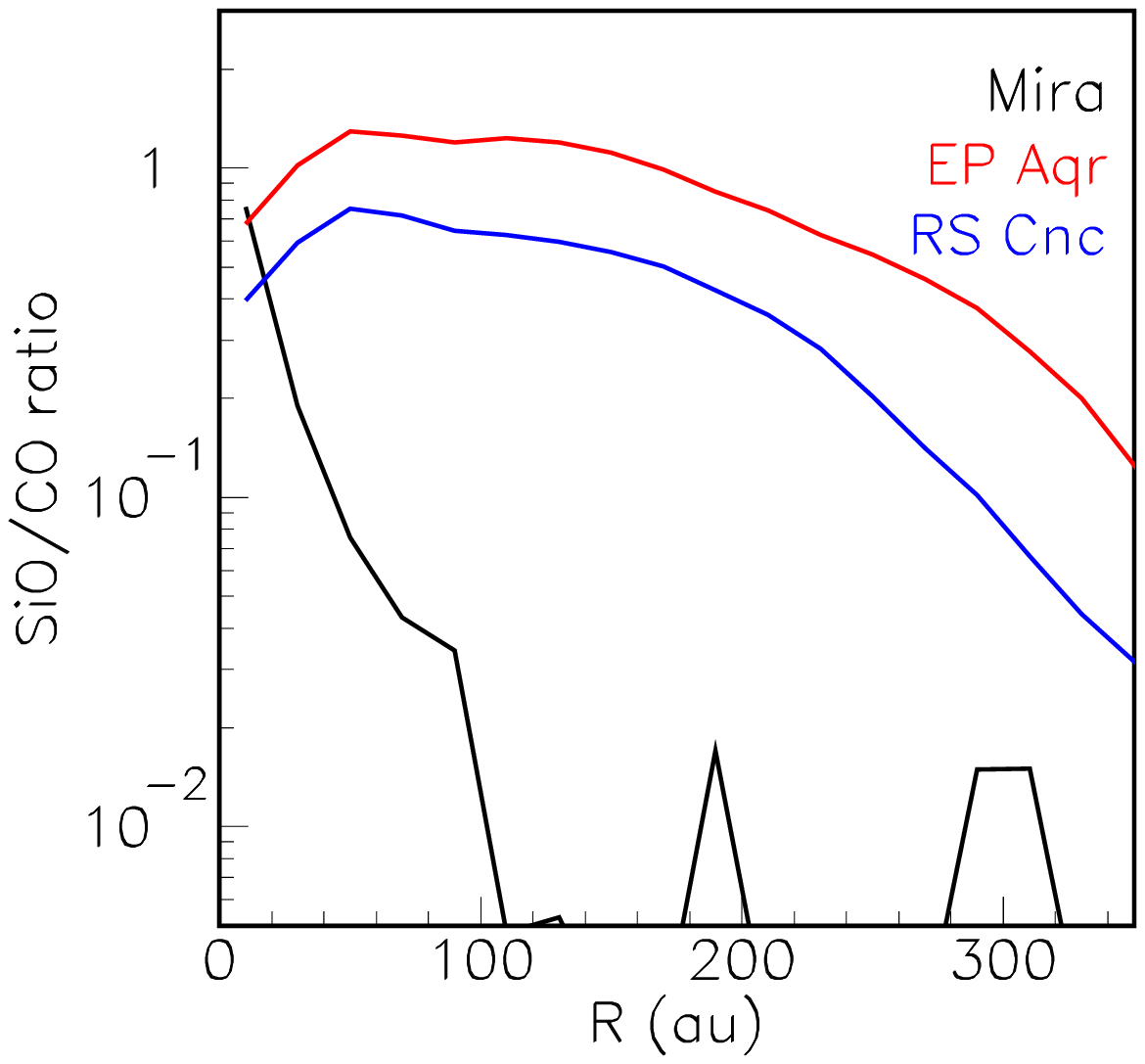}
  \caption{Comparison between the radial distributions of the $^{12}$CO and $^{28}$SiO emissions of Mira Ceti (this work, excluding the wedge 200\dego$<$$\omega$$<$315\dego), EP Aqr \citep{Hoai2019} and RS Cnc \citep{Winters2021}. Left : SiO(5-4) intensity. Center: CO(3-2) intensity. In the case of EP Aqr and RS Cnc these are obtained from CO(2-1) data (see text). Right: ratio between the SiO and CO intensities.   }
 \label{fig14}
\end{figure*}

According to our current understanding of the formation of the wind in oxygen-rich AGB stars \citep{Hofner2018, Freytag2019} one expects transparent dust grains to form first, within some two stellar radii from the star, and SiO dust grains to form at larger distances, where the temperature is low enough. The SiO gas phase is then progressively depleted over some 100 au before being photo-dissociated by interstellar UV radiation \citep{Gonzalez2003}. The radial distribution of the SiO emission, when interpreted in such terms, is therefore expected to provide essential information to the comparison between different stars and to the confrontation of observations with the predictions of the standard model. Figure \ref{fig14} compares Mira, excluding the south-western quadrant (200\dego$<\omega<$315\dego), with RS Cnc and EP Aqr, two oxygen-rich AGB stars having mass loss rates of $\sim$3$\times$10$^{-7}$ \citep{Winters2021} and $\sim$1.6$\times$10$^{-7}$ \citep{Hoai2019} \msun yr$^{-1}$, respectively, and for which the SiO(5-4) emission has been observed over a wide radial range. The left panel displays the radial dependence of the SiO(5-4) emission, illustrating the spectacular difference between Mira and the other two stars. However, as these data are prone to important opacity within some 100 au from the star, a reliable quantitative description would require a radiative transfer modeling accounting for the different morphologies and temperature gradients of the three stars. The central panel displays the radial dependence of the CO(3-2) emission. In the case of EP Aqr and RS Cnc, available observations are of the CO(2-1), rather than CO(3-2), emission. For the comparison with Mira to be meaningful, we have therefore multiplied the CO(2-1) brightness by ($A_{32}/A_{21}$)(7/5)[$\exp(-E_{u32})/T$]/[$\exp(-E_{u21})/T$] where $A_{21}=6.91\times10^{-7}$ s$^{-1}$ and $A_{32}=2.50\times10^{-6}$ s$^{-1}$ are the Einstein coefficients, $E_{u21}=16.6$ K and $E_{u32}=33.2$ K are the energies of the upper level and $T$ is the temperature. The temperature dependence is taken from the comparison between CO(2-1) and CO(1-0) emissions presented in \citet{Hoai2019}; we checked that it gives very similar results to a form $T$[K]$=$109/$r$[arcsec] \citep{Ryde2001}, which is valid in the relevant radial range. The CO data show the importance of the central reservoir in all three stars. The right panel compares the radial dependence of the SiO(5-4)/CO(3-2) emission ratio. Beyond 100 au (1 arcsec) from Mira, excluding the south-western quadrant, the SiO(5-4) intensity is at the 0.01 mJy au$^{-2}$ \kms\ level, while the $^{12}$CO(3-2) intensity is $\sim$2 mJy au$^{-2}$ \kms. The ratio between the SiO(5-4) and $^{12}$CO(3-2) intensities beyond 100 au and excluding the south-western quadrant is therefore $\sim$0.005, consistent with our 0.1 estimate for the SiO(5-4)/$^{13}$CO(3-2) ratio from Table \ref{tab3}. In comparison, at similar distances, the SiO(5-4)/$^{12}$CO(3-2) ratio is $\sim$0.6 for RS Cnc and $\sim$1.2 for EP Aqr. The comparison between the three stars reveals therefore the much lower SiO/CO abundance ratio observed for Mira.

We checked that this result is consistent with observations in some other oxygen-rich AGB stars; however, the comparison is often complicated by the lack of data on the same lines, the limitation of the $R$ range having been explored or important opacity. This is in particular the case for R Dor where the emission of the line for which data are available, SiO(8-7), is very opaque, requiring radiative transfer modeling to reach a meaningful conclusion \citep{Nhung2021}. 

The global confinement of a gas reservoir within some 100 au from the center of the star had been reported earlier, in particular by \citet{Wong2016} and \citet{Khouri2018}, the latter authors arguing that such confinement is evidence for a steep density decline at the edge of the reservoir; they considered that the presence of a pulsation-induced shock producing a density contrast at the shock front and causing dust formation to happen efficiently in the post-shocked gas was an unlikely scenario, such shocks being expected to have dissipated well below a few 10 au from the star. They favor instead a scenario in which the mass-loss rate of Mira has suddenly increased in 2003, in association with the emission of the soft X-ray burst reported by \citet{Karovska2005}.  However, none of these two scenarios can explain the second specificity of Mira A: the very low SiO/CO ratio.   

Both depletion and photo-dissociation of the SiO gas phase of the CSE of oxygen-rich AGB stars have been studied, in particular, by \citet{Gonzalez2003} and \citet{Schoier2004}; the latter authors considered specifically the case of L$_2$ Pup, which has a low mass-loss rate of 2.7$\times$10$^{-8}$ \msun yr$^{-1}$ and a low terminal velocity of 2.1 \kms, both similar to, but smaller than that measured for the Mira SEO in Section 4.2. They assume an unshielded photo-dissociation rate of 2.5$\times$10$^{-10}$ s$^{-1}$ and find a photo-dissociation radius of $\sim$60-70 au, only twice the condensation radius of $\sim$30 au associated with the depletion of the SiO gas phase by adsorption on dust grains. Note that shielding from UV radiation is provided by dust, not by gas and even a very small flow of gas will survive as long as it is sufficiently shielded by the dust. Their result suggests therefore that, for densities and wind velocities comparable to those present in the different fragments of the Mira CSE, the SiO gas phase, within 0.5 arcsec from Mira A, could have been mostly depleted on dust grains and whatever survived would have been rapidly photo-dissociated by the interstellar UV radiation. In contrast, the photo-dissociation radius obtained by \citet{Schoier2004} for R Dor is three times as large as for L$_2$ Pup. Moreover, in the present case of Mira Ceti, the UV radiation emitted by Mira B \citep{Wood2006, Sokoloski2010} must also contribute to photo-dissociation, in particular in the SEO. Validation of such interpretation requires a detailed modeling that is well beyond the scope of the present article; if confirmed, it would explain the low SiO/CO ratio as a simple effect of depletion and photo-dissociation, in line with current understanding. Moreover, it would encourage a systematic study, for oxygen-rich AGB stars, of the correlation between mass loss rate and radial extensions of the SiO gas phase and of the SiO/CO emission ratio.

\begin{figure}
  \centering
  \includegraphics[height=6cm,trim=1.cm 1.5cm 1.2cm 0.5cm,clip]{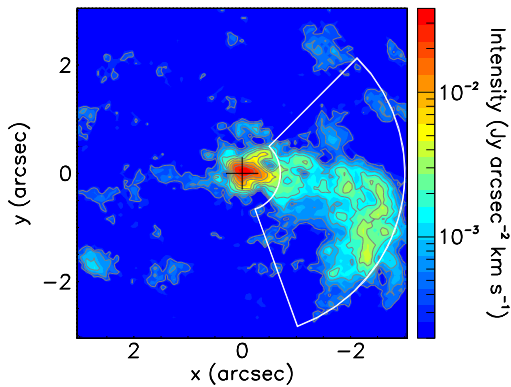}
  \caption{Intensity map of SiO(5-4) emission integrated in the $|V_z|$$<$ 4 \kms\ interval. The position of Mira A is marked with a black cross. The white lines define the south-western region used in Figure \ref{fig16}.  }
 \label{fig15}
\end{figure}

\begin{deluxetable*}{ccccccc}
\tablenum{4}
\tablecaption{Parameters of relevance to the comparison presented in Figure \ref{fig14}.\label{tab4}}
\tablehead{ &\multicolumn{2}{c}{Mira}&\multicolumn{2}{c}{RS Cnc} &\multicolumn{2}{c}{EP Aqr}\\
    &\multicolumn{2}{c}{(this work)}&\multicolumn{2}{c}{\citep{Winters2021} } &\multicolumn{2}{c}{\citep{TuanAnh2019}}}
\startdata
D (pc)&
\multicolumn{2}{c}{100} &
\multicolumn{2}{c}{150} &
\multicolumn{2}{c}{114}\\
Line&
CO(3-2)&
SiO(5-4)&
CO(2-1) &
SiO(5-4) &
CO(2-1) &
SiO(5-4)\\
Beam (arcsec$^2$)&
0.39$\times$0.36 &
0.06$\times$0.03 &
0.48$\times$0.30 &
0.51$\times$0.32 &
0.33$\times$0.30&
0.29$\times$0.25\\
Noise ($\sigma$, mJy beam$^{-1}$)&
14 &
0.66 &
2.9 &
3.4 &
6 &
3.9\\
Frequency (GHz)$^a$&
345.8&
217.1&
230.5&
217.1&
230.5&
217.1\\
Einstein coef. $A_{ji}$ (s$^{-1}$)$^a$&
2.50$\times$10$^{-6}$ &
5.2$\times$10$^{-4}$&
6.91$\times$10$^{-7}$&
5.2$\times$10$^{-4}$&
6.91$\times$10$^{-7}$&
5.2$\times$10$^{-4}$\\
Upper level $E_u$ (K)$^a$&
33.2&
31.3&
33.2 &
75.0 &
16.6&
31.3\\
\enddata
\tablecomments{$^a$ The values are taken from Leiden Atomic and Molecular Database (LAMDA; \citet{Schoier2005}).}
\end{deluxetable*}

\subsection{Different emissions of the CO and SiO lines in the south-western quadrant}

\begin{figure*}
  \centering
  \includegraphics[height=4.5cm,trim=1.cm 1.5cm 1.5cm 0.5cm,clip]{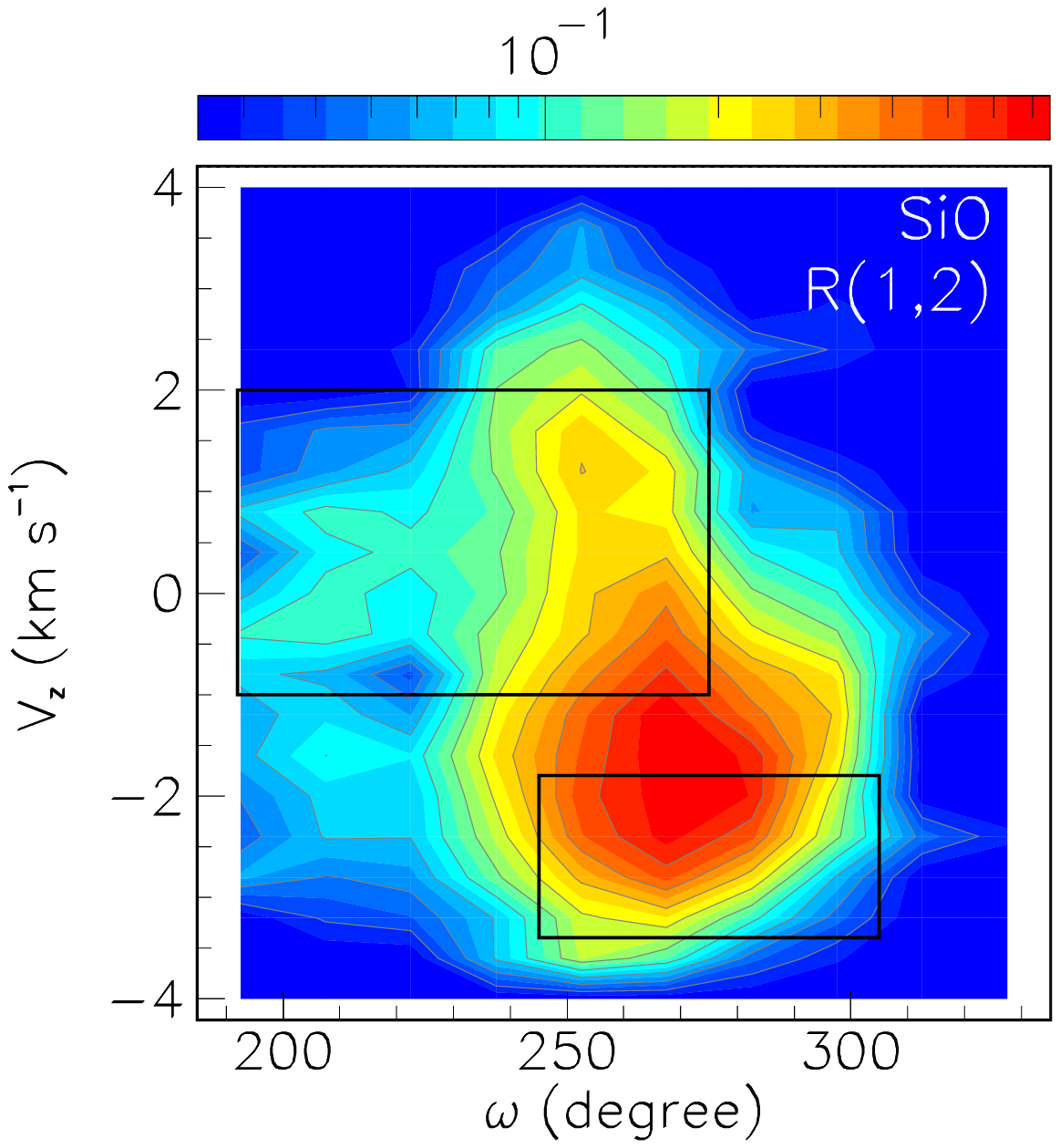}
  \includegraphics[height=4.5cm,trim=1.cm 1.5cm 1.5cm 0.5cm,clip]{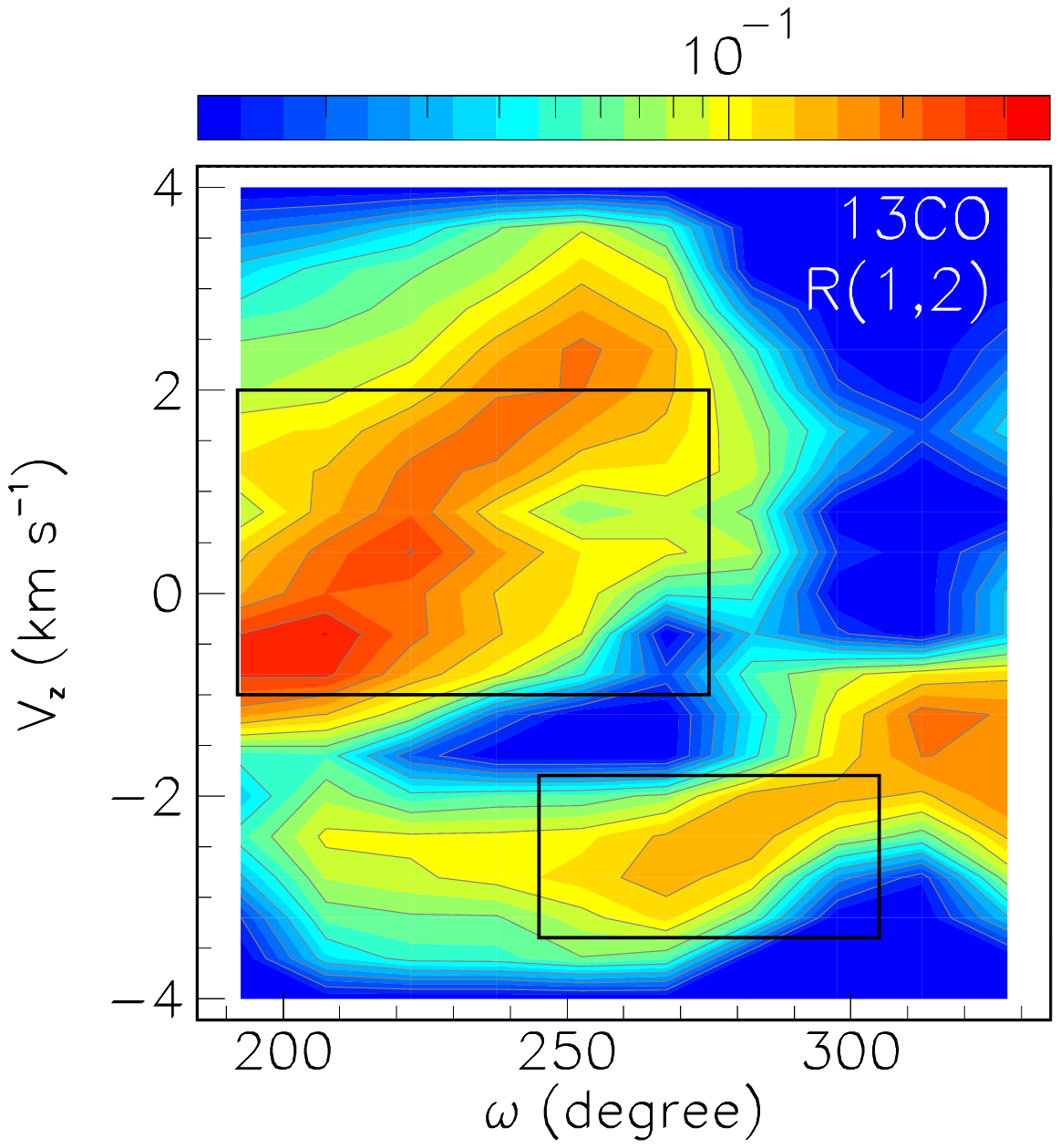}
  \includegraphics[height=4.5cm,trim=1.cm 1.5cm 1.5cm 0.5cm,clip]{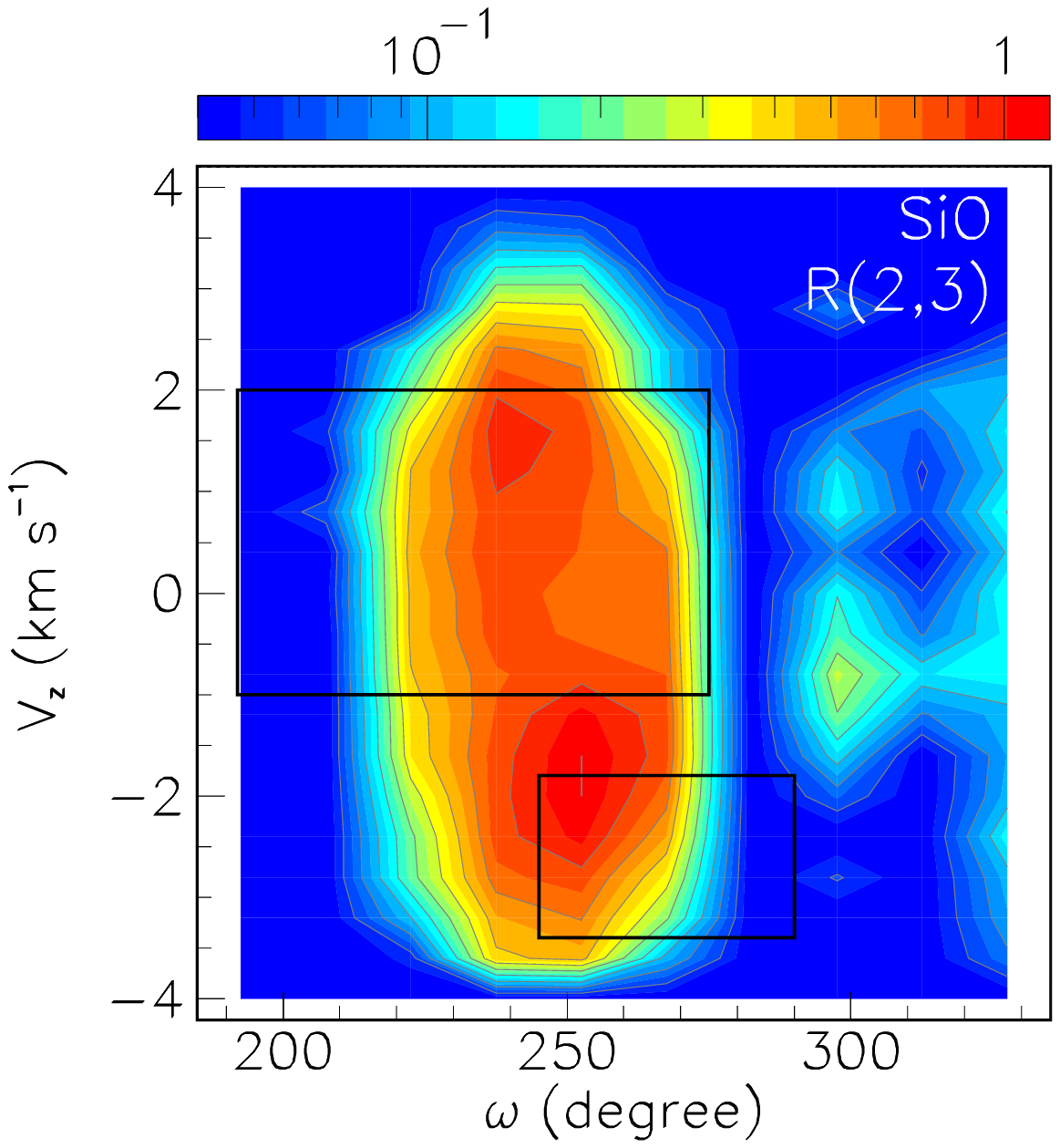}
  \includegraphics[height=4.5cm,trim=1.cm 1.5cm 1.5cm 0.5cm,clip]{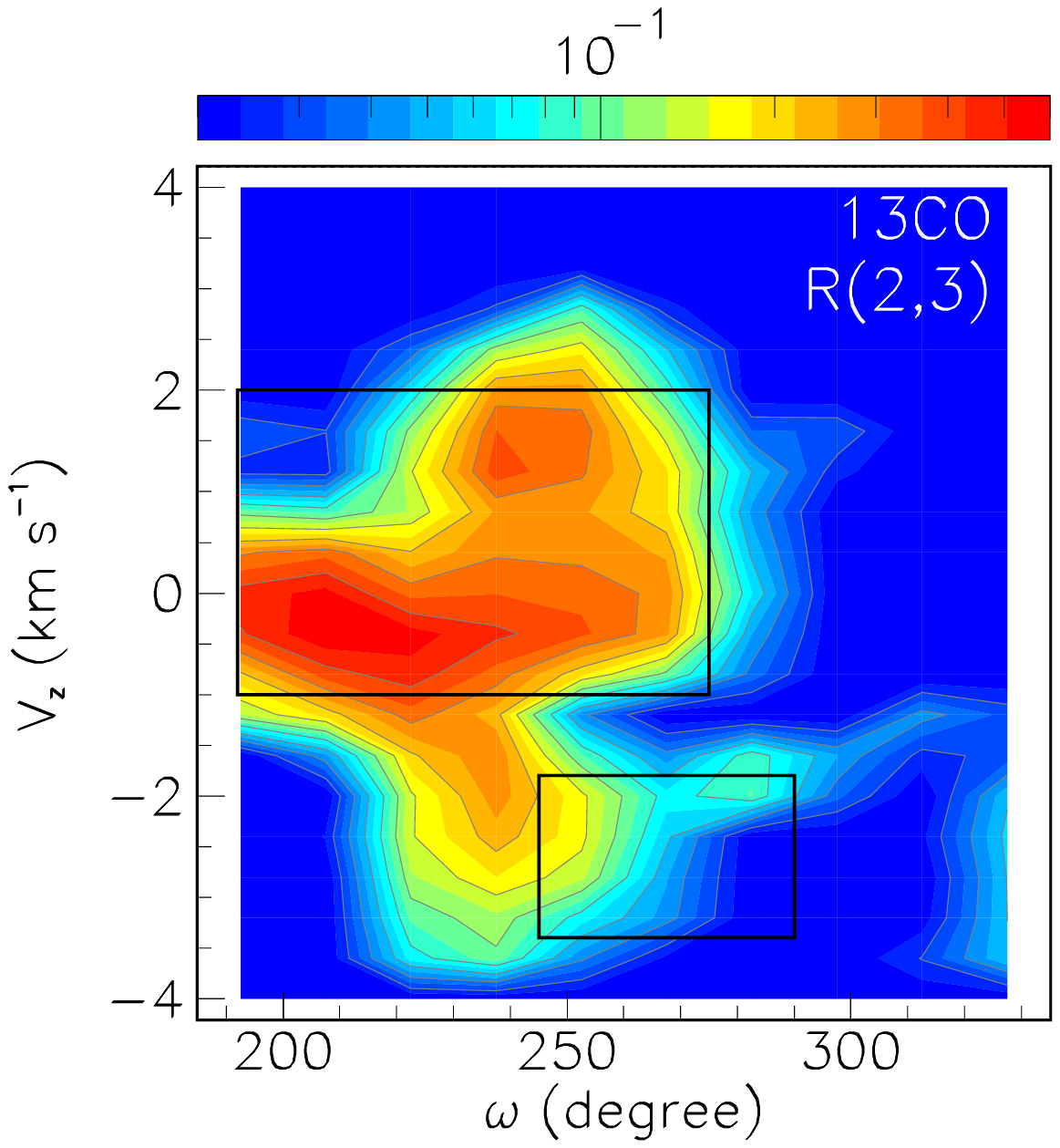}
  \caption{ PV maps ($V_z$ vs $\omega$) of SiO(5-4) and $^{13}$CO(3-2) emissions in the south-western region. The first two panels are integrated over 1$<$$R$$<$2 ($R$(1,2)) arcsec, the last two over 2$<$$R$$<$3 ($R$(2,3)) arcsec. In each pair, SiO is left and $^{13}$CO is right. The large rectangles indicate the location of the SWO fragment and the small rectangles of the SWS fragment as defined in Table \ref{tab3}. The color scales are in units of Jy. }
 \label{fig16}
\end{figure*}

HTN20 presented a detailed study of the south-western region that includes the fragments SWO and SWS of the present study. In the present section we extend the analysis to a comparison between the SiO(5-4) (Figure \ref{fig15}) and $^{13}$CO(3-2) emissions, which is more free of opacity effects than the SiO(5-4) vs $^{12}$CO(3-2) comparison presented in HTN20. Figure \ref{fig16} displays PV maps ($V_z$ vs $\omega$) in two intervals of $R$ separately: 1$<$$R$$<$2 arcsec and 2$<$$R$$<$3 arcsec. At first glance, CO and SiO emissions seem to be completely unrelated. Essentially, SiO emission reveals a flow covering $V_z$=$-$1.7$\pm$0.7 \kms\ and $\omega$=265\dego$\pm$15\dego\ for $R$ between 1 and 2 arcsec and a broad area covering $|V_z|$$<$4 \kms\ and $\omega$=250\dego $\pm$30\dego\ for $R$ between 2 and 3 arcsec. HTN20 refers to these as first and second components of SiO emission, respectively. CO emission covers the whole south-western quadrant but is strongly depressed toward the west at $V_z$$\sim$$-$1.5$\pm$0.5 \kms.
In HTN20, we speculated that a mass ejection associated with the 2003 X-ray burst, probably caused by a magnetic flare, \citep{Wood2006} and centered on the first SiO component has been punching a hole through the CO volume at a velocity of some 130 \kms\ to reach  $R\sim$3 arcsec  in 2014. We note that the dates of observation were only 4.5 months apart: 12-15 June for CO and 1st November for SiO, during which time the mass ejection and associated shock wave would have covered $\sim$10 au (100 mas). The mass ejection left accordingly a low density wake behind it, and is assumed to have been accompanied by a shock front that caused the SiO molecules trapped in dust grains to outgas \citep[][ and references therein]{Gusdorf2008} and/or UV-dissociated SiO molecules to recombine..

\begin{figure*}
  \centering
  \includegraphics[height=4.5cm,trim=1.cm 1.5cm 0cm 1.5cm,clip]{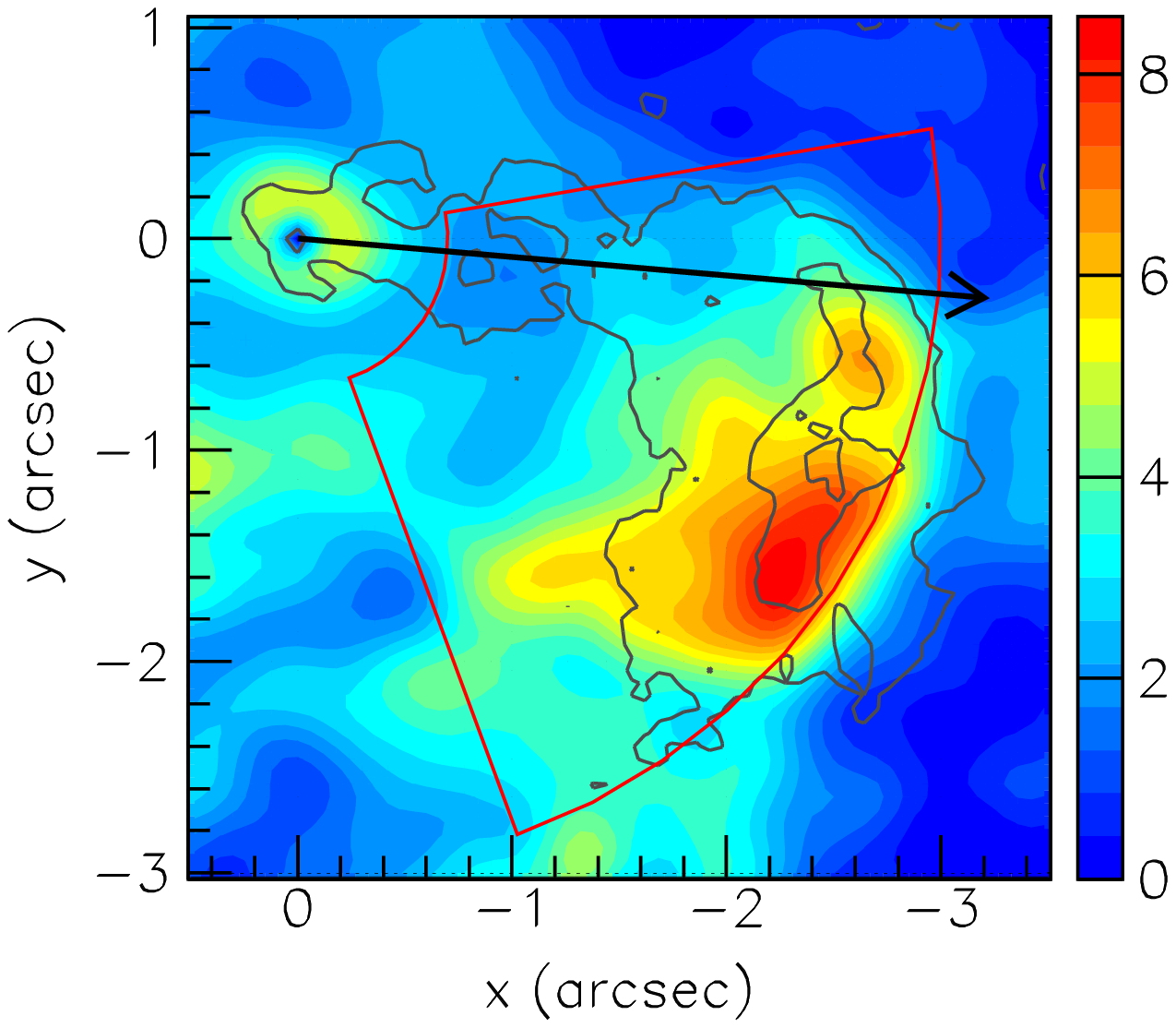}
  \includegraphics[height=4.5cm,trim=1.cm 1.5cm 0cm 1.5cm,clip]{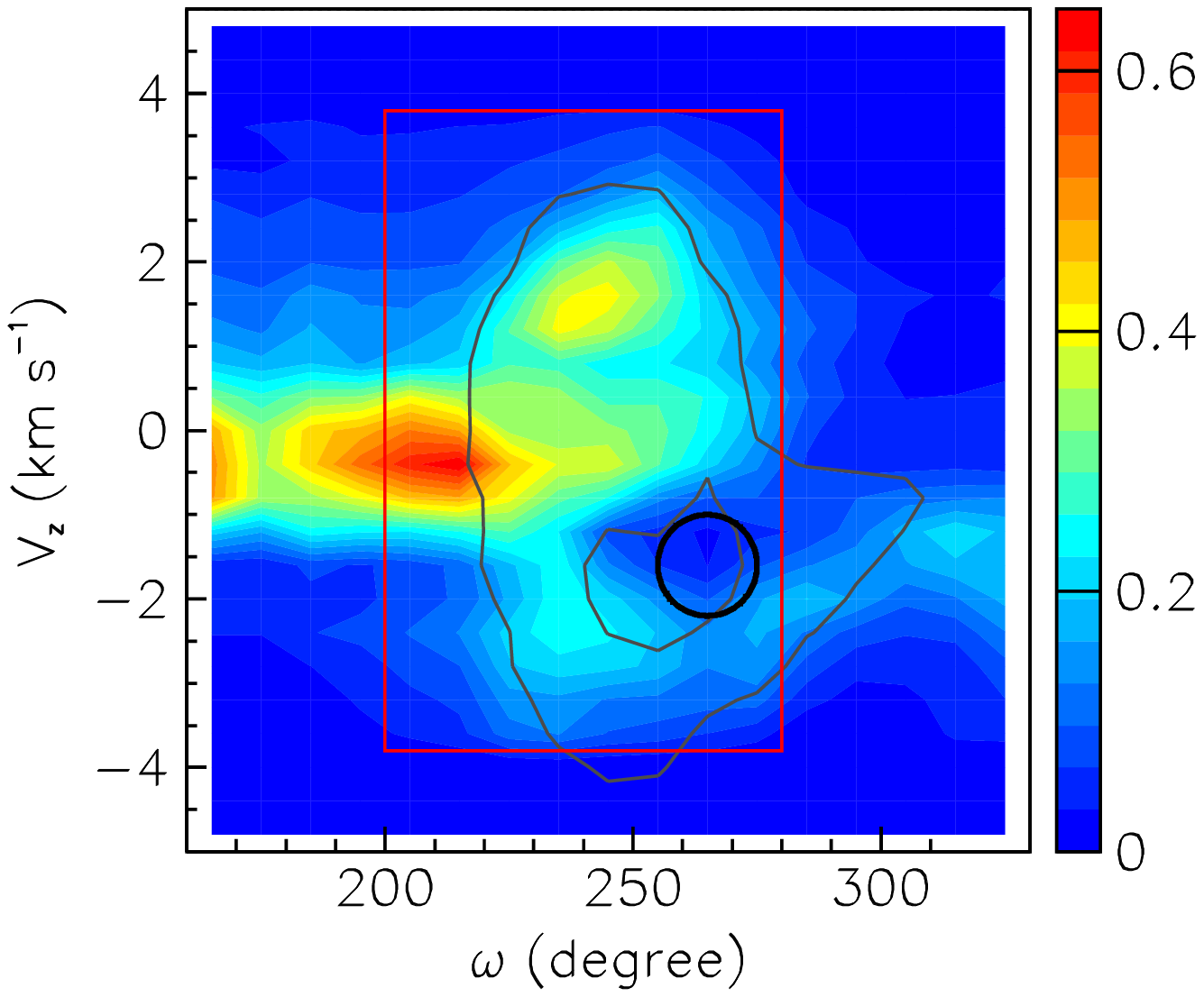}
  \includegraphics[height=4.5cm,trim=1.cm 1.5cm 0cm 1.5cm,clip]{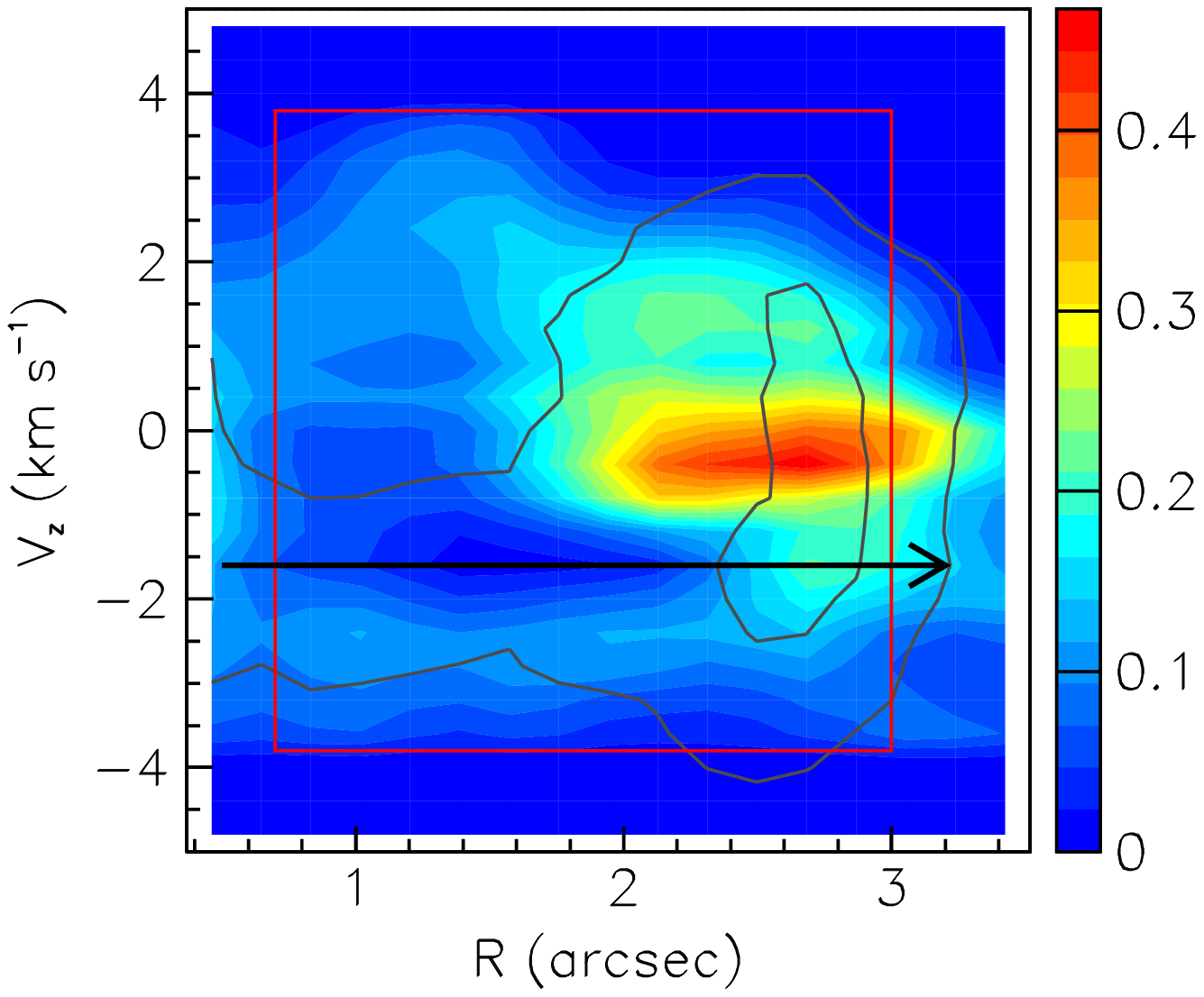}\\
  \includegraphics[height=4.5cm,trim=1.cm 1.5cm 0cm 1.5cm,clip]{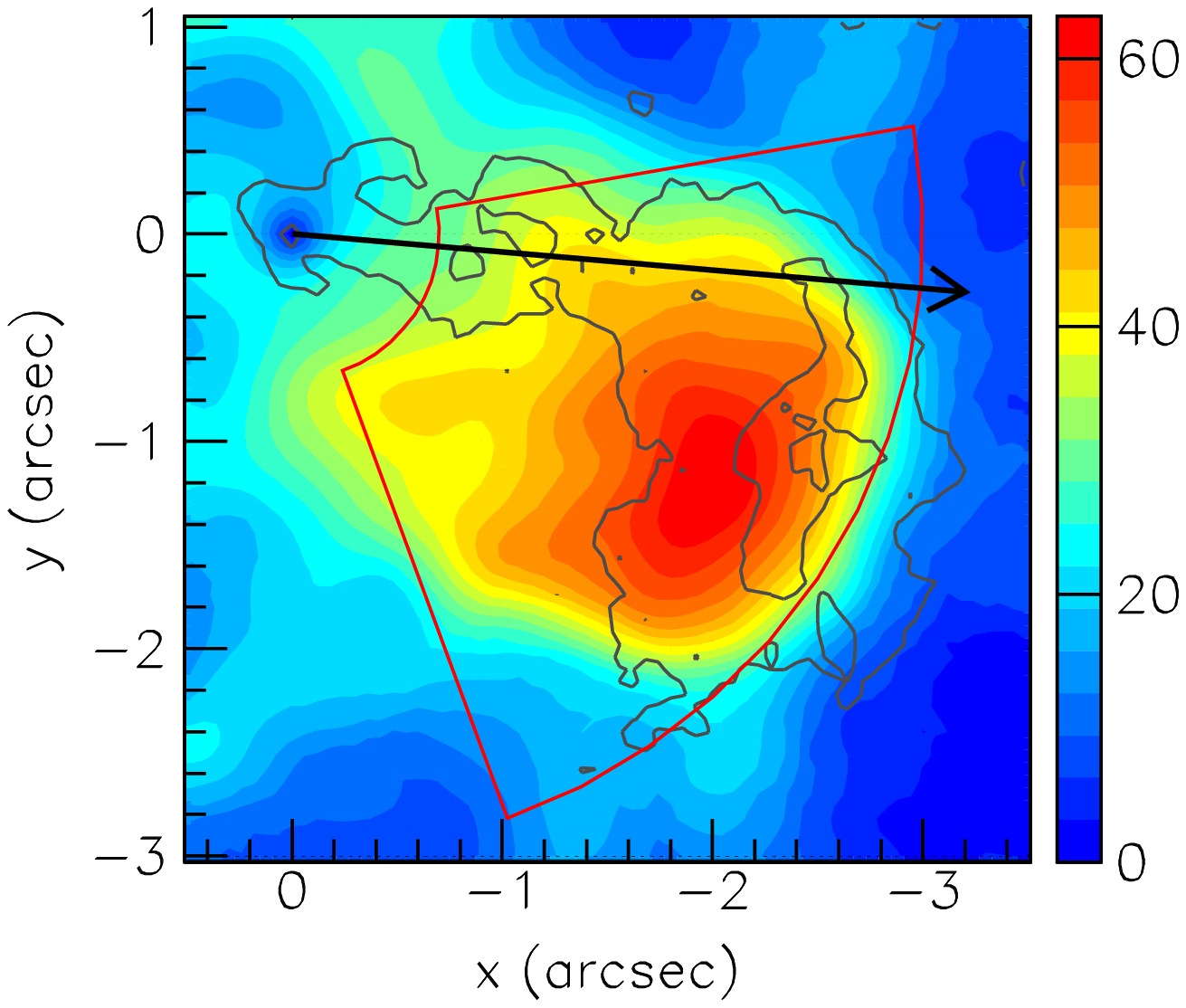}
  \includegraphics[height=4.5cm,trim=1.cm 1.5cm 0cm 1.5cm,clip]{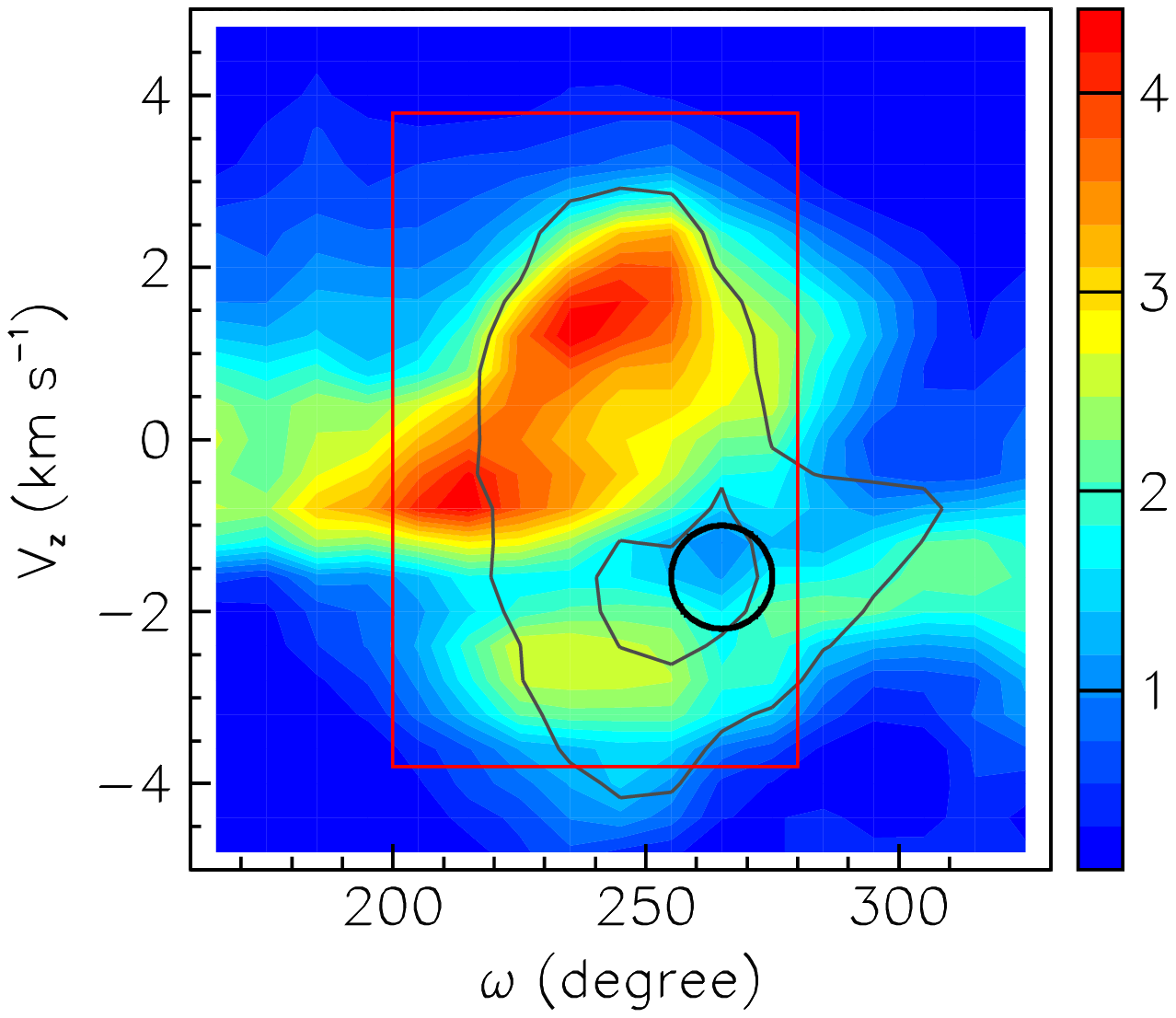}
  \includegraphics[height=4.5cm,trim=1.cm 1.5cm 0cm 1.5cm,clip]{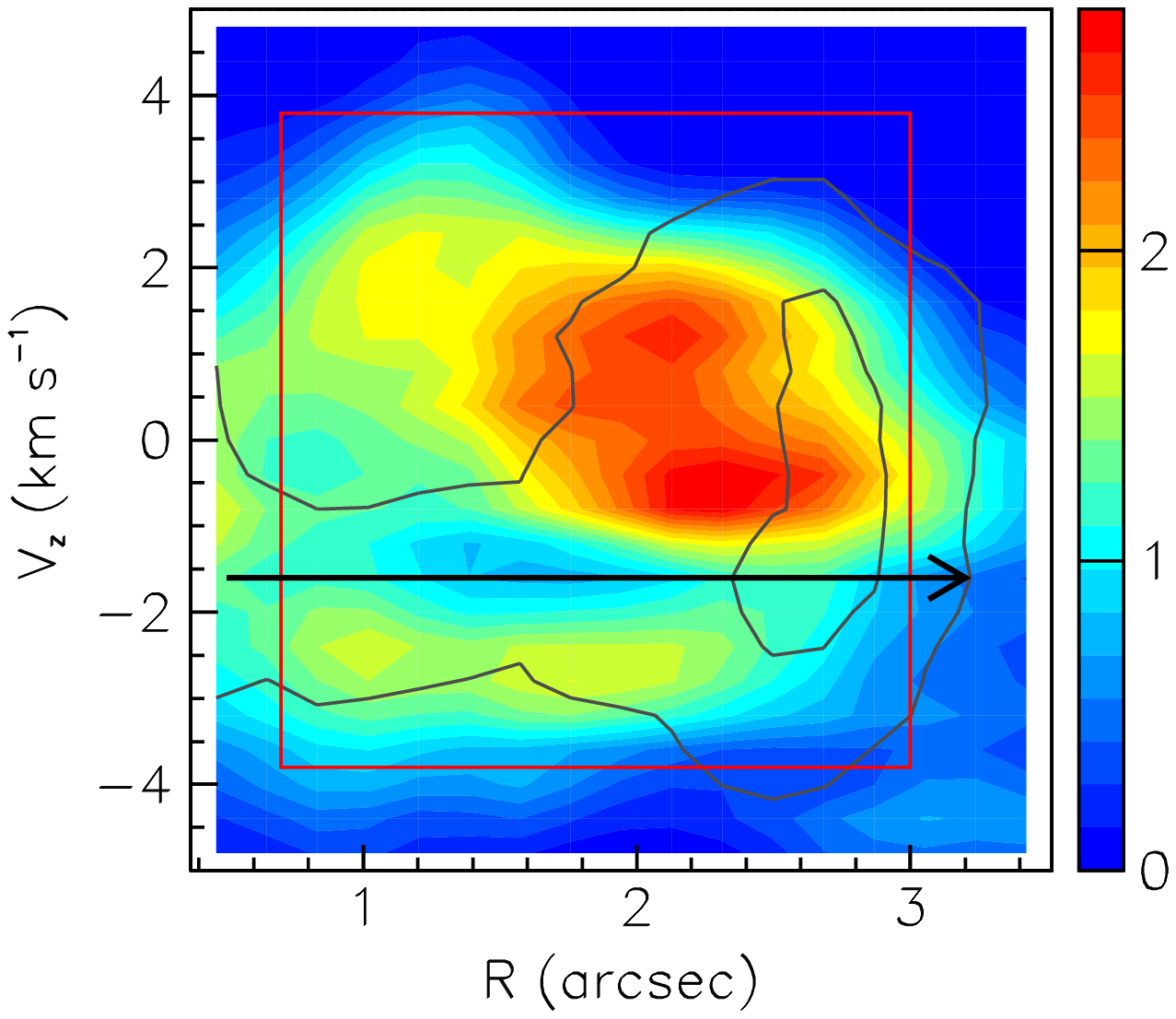}\\
  \includegraphics[height=4.5cm,trim=1.cm 1.5cm 0cm 1.5cm,clip]{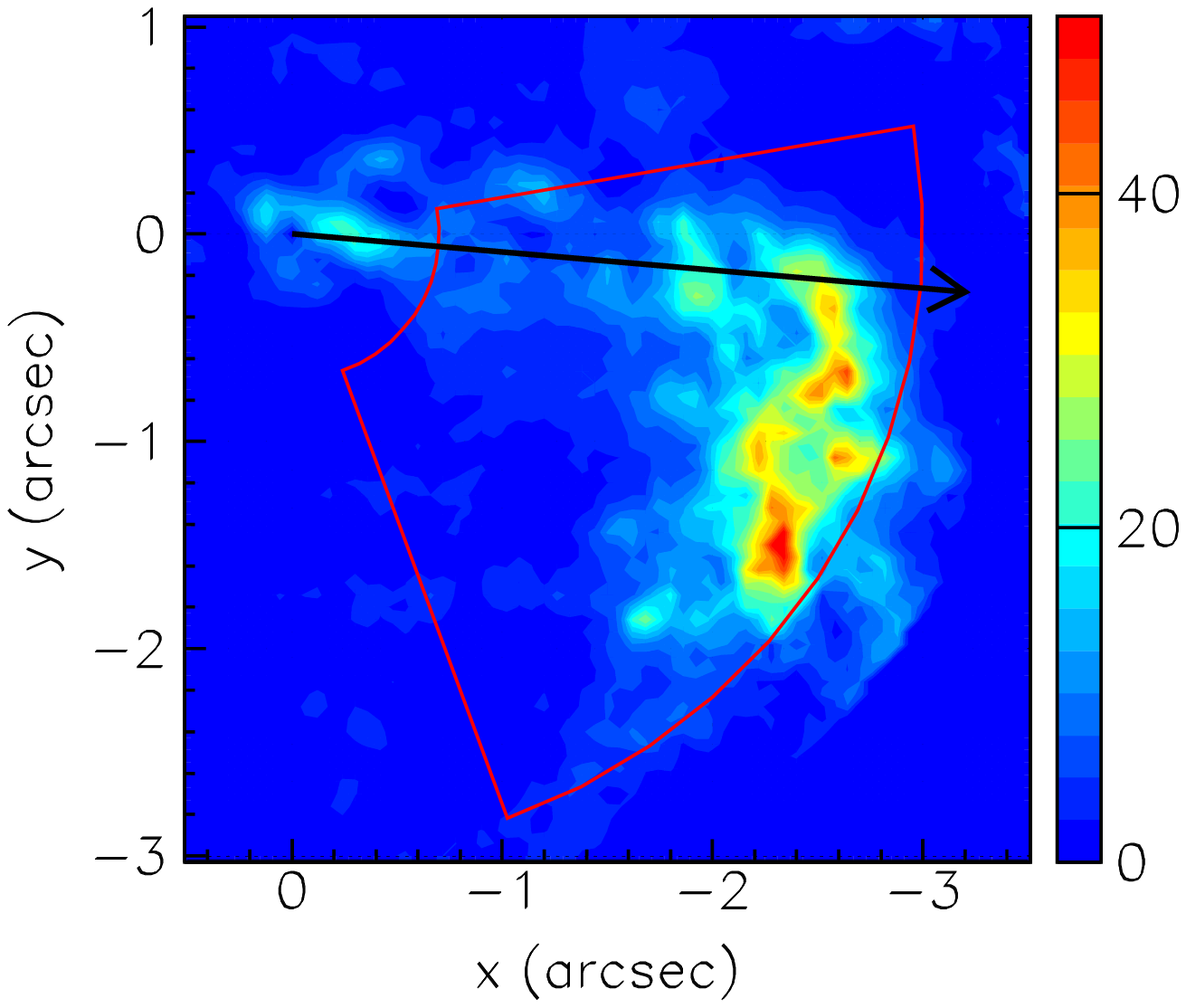}
  \includegraphics[height=4.5cm,trim=1.cm 1.5cm 0cm 1.5cm,clip]{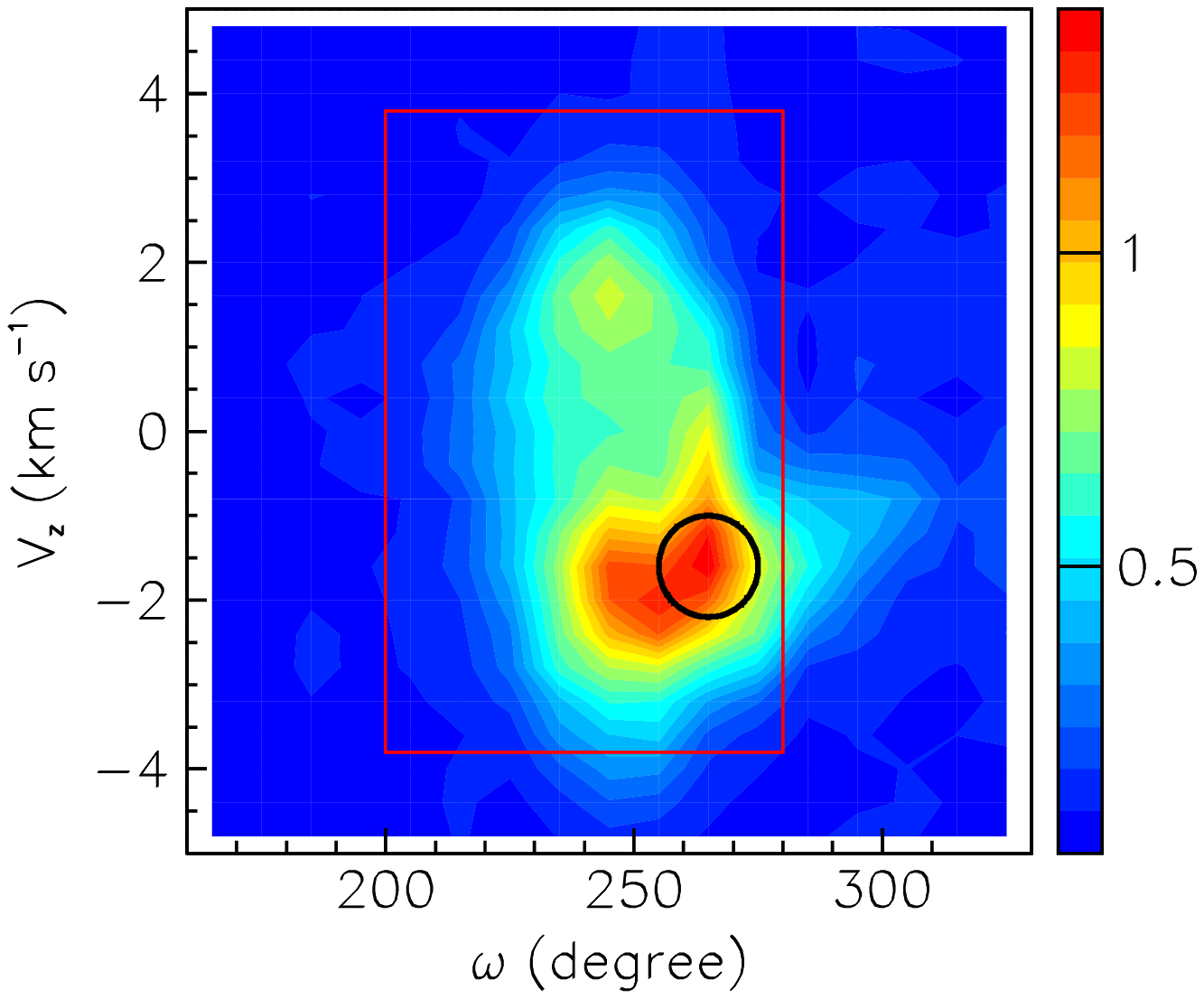}
  \includegraphics[height=4.5cm,trim=1.cm 1.5cm 0cm 1.5cm,clip]{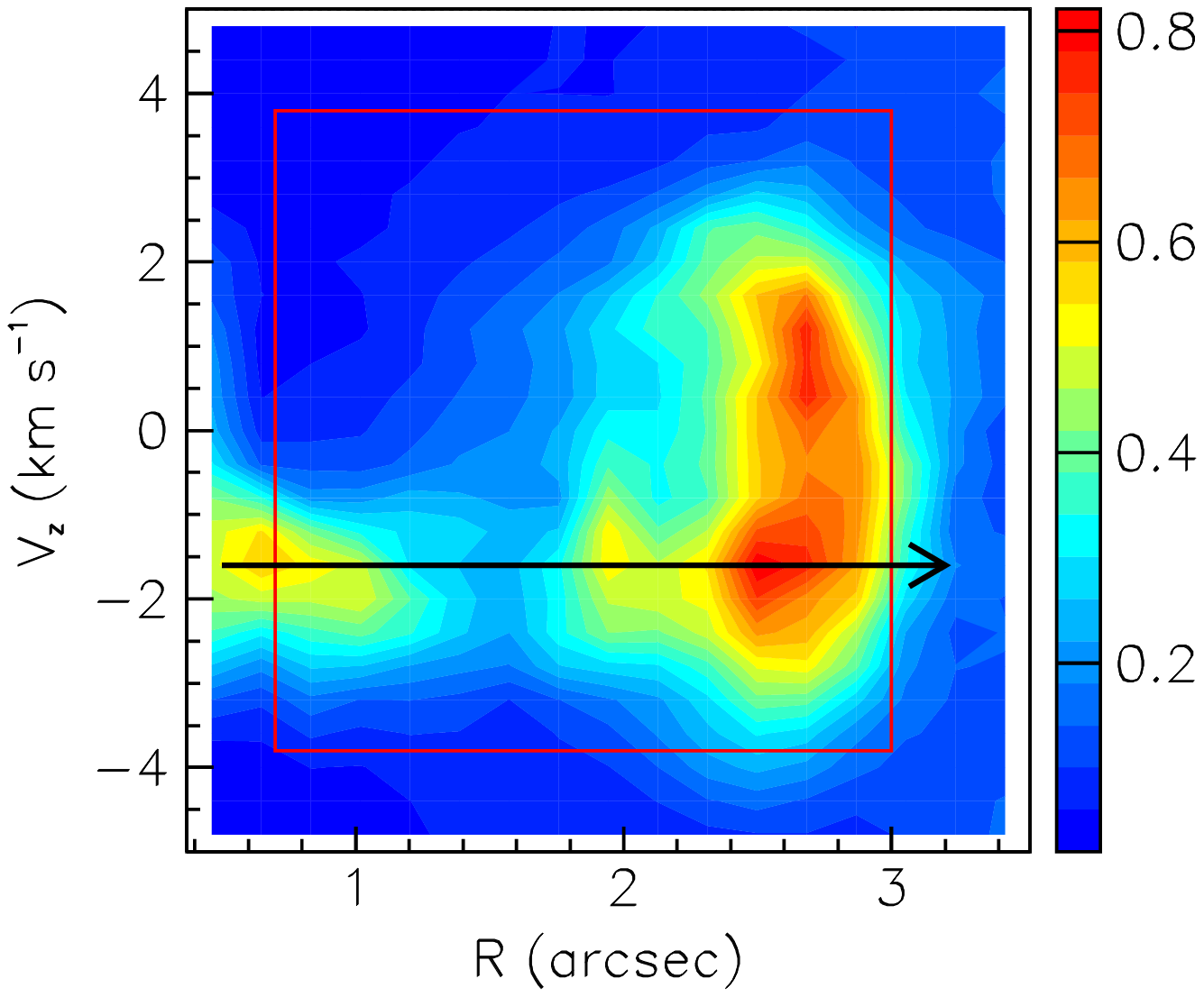}
  \caption{Projections of the south-western region of the data cube delimited by red lines, on the ($x,y$) plane (left, multiplied by $R$), the ($V_z,\omega$) plane (center) and the ($V_z,R$) plane (right). Units for the left panels are Jy arcsec$^{-1}$ \kms\ and for the other panels Jy. From up down, $^{13}$CO, $^{12}$CO and SiO emissions. SiO contours are superimposed on the CO panels. The arrows and circles indicate the supposed trajectory of the 2003 mass ejection. The $^{12}$CO and SiO panels were reproduced from Figure 7 of HTN20 with permission of the RAS.}
 \label{fig17}
\end{figure*}

\begin{figure*}
  \centering
  \includegraphics[height=4.5cm,trim=.5cm 1.5cm 1.5cm 1.5cm,clip]{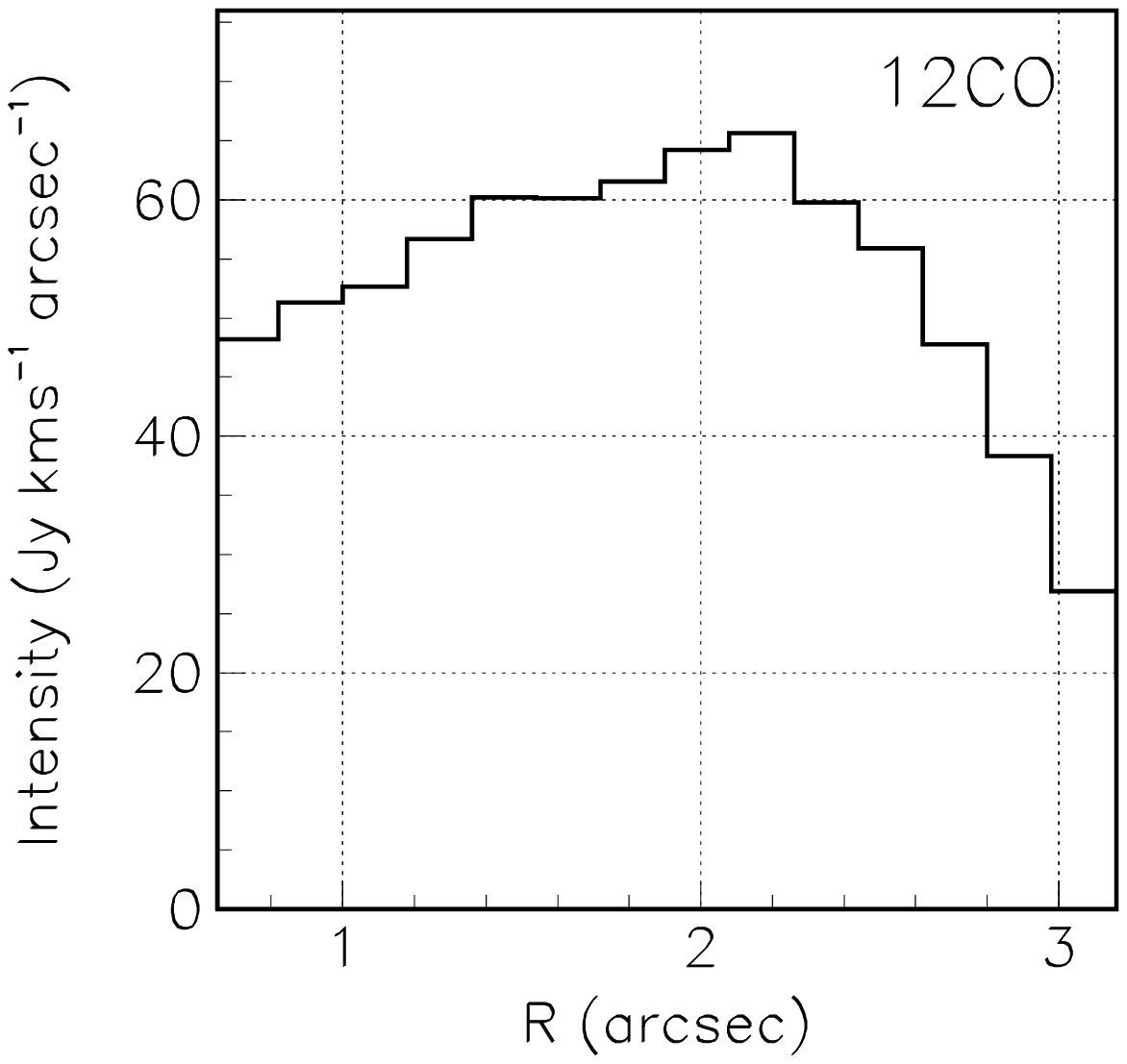}
  \includegraphics[height=4.5cm,trim=.5cm 1.5cm 1.5cm 1.5cm,clip]{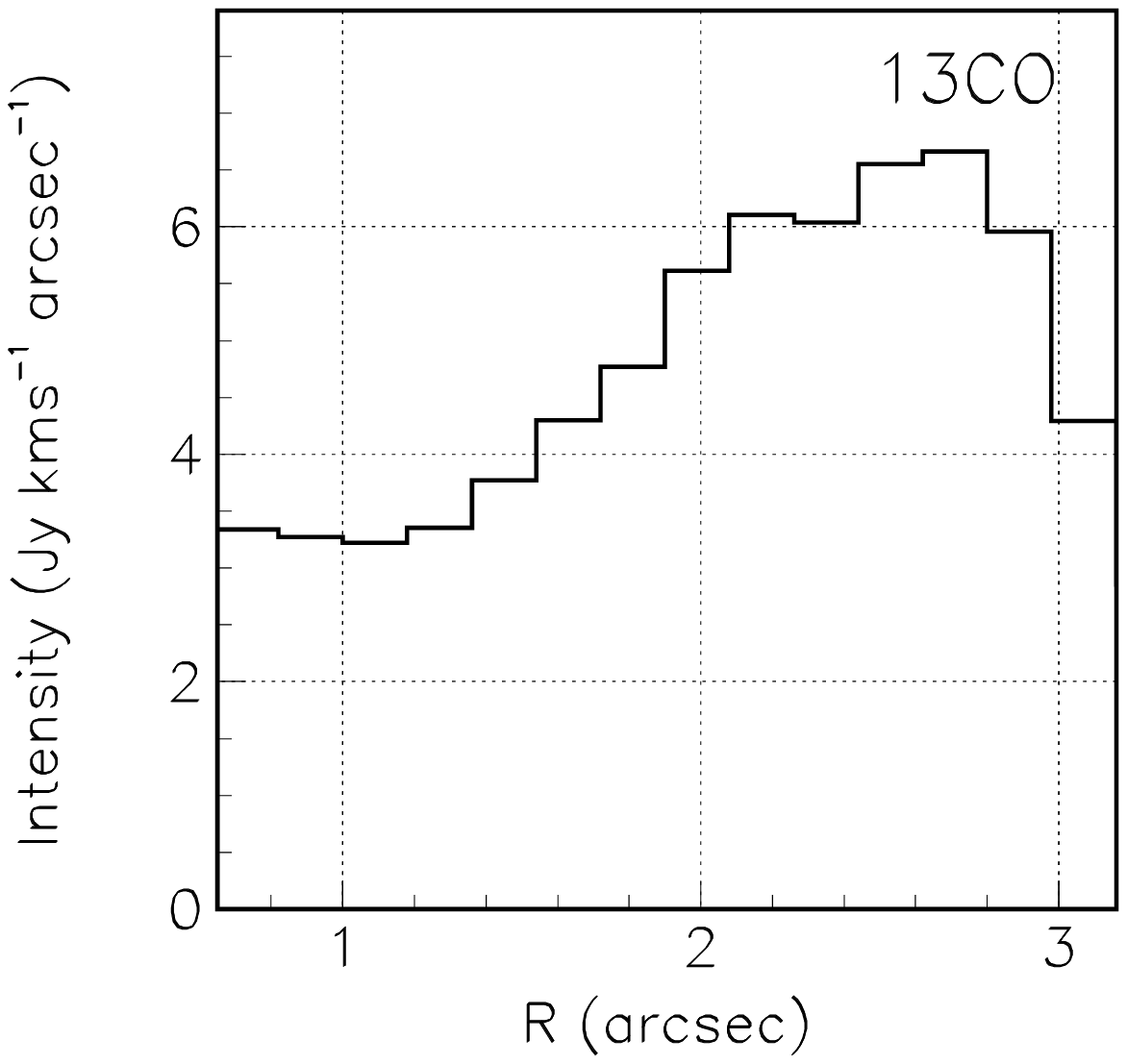}
  \includegraphics[height=4.5cm,trim=.5cm 1.5cm 1.5cm 1.5cm,clip]{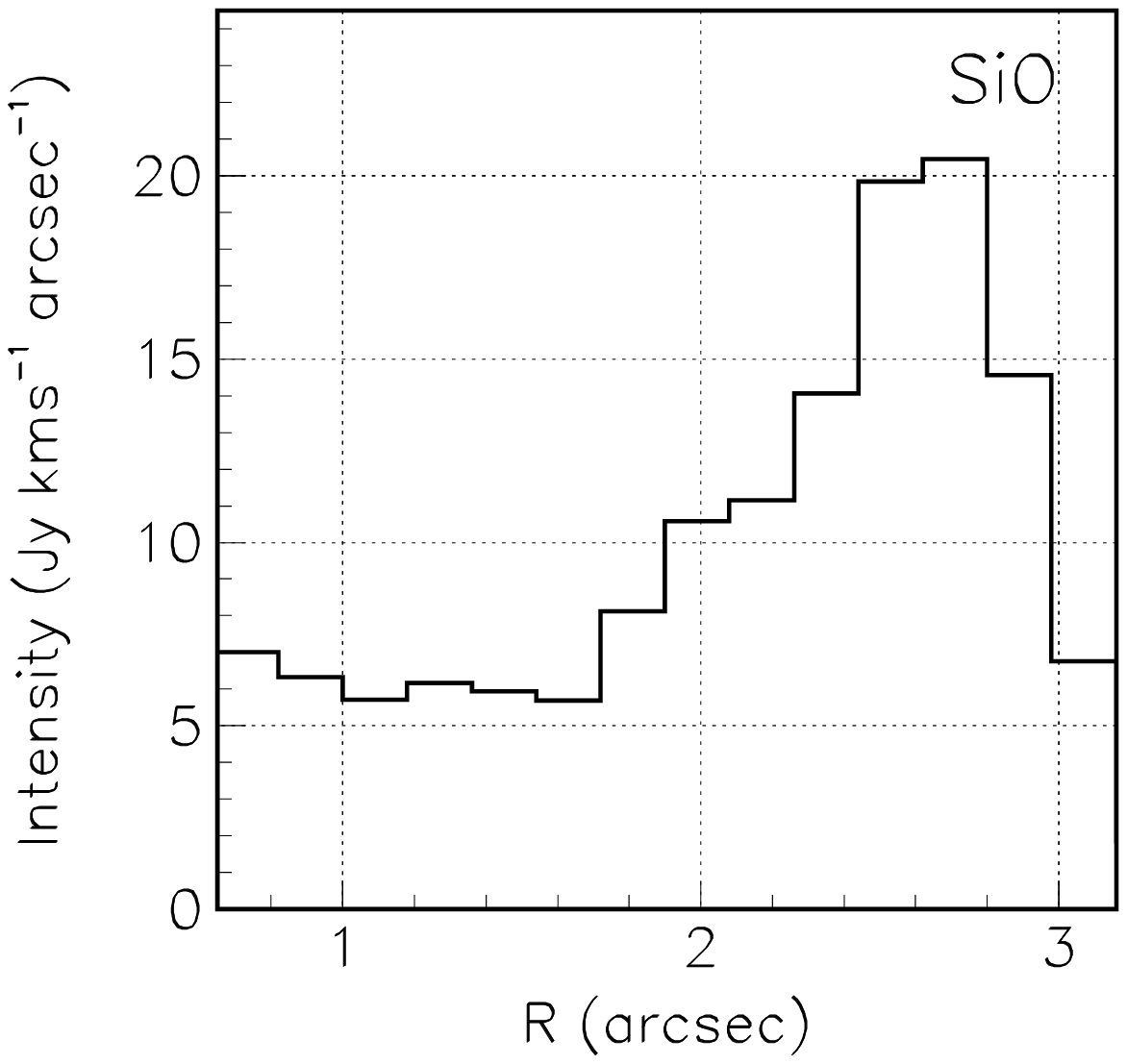}
  \caption{Radial profiles of SiO, $^{13}$CO  and $^{12}$CO emissions integrated in the region defined in Figure \ref{fig17}. }
 \label{fig18}
\end{figure*}

We refer the reader to HTN20 for  the details of the argumentation but in spite of its speculative nature this scenario accounts well, at least qualitatively, for the main observed features of both SiO and CO emissions. Figure \ref{fig17} projects the south-western region of the data cube on the ($x,y$), ($V_z$,$\omega$) and ($V_z$,$R$) planes for each of the $^{13}$CO(3-2), $^{12}$CO(3-2) and SiO(5-4) line emissions  and Figure \ref{fig18} displays the radial profiles; together with Figure \ref{fig16}, they illustrate well the relevant features of the morpho-kinematics. Accepting this interpretation would require a quantitative modeling of the event, which is well beyond the scope of the present article. We retain however that such a scenario has the advantage of decoupling the mechanisms of the disturbance caused by the mass ejection from that of generation of the nascent wind, in which we are interested. It is consistent with the latter being described as episodic mass loss in the form of fragments strongly depleted of SiO molecules, with a SiO/CO abundance ratio at the level of 0.5-1.0$\times$10$^{-4}$: the whole SiO emission is interpreted as due to SiO molecules, which recombined or outgassed, eleven years before the observation, from dust grains present in the pre-existing SWO.
We studied in considerable details the anti-correlation displayed by the CO(3-2) and SiO(5-4) emissions in the south-western quadrant but failed to find another sensible interpretation than proposed in HTN20. If true, this interpretation has the merit of having significantly unraveled the complexity of the observed morpho-kinematics: models aiming at a description of the genesis of the wind of Mira A should simply ignore the south-western quadrant. 

\section{Summary and conclusion}

We have revisited, and extended to $^{13}$CO(3-2) emission, a number of analyses of the morpho-kinematics of the nascent wind of Mira A presented in HTN20, in the light of current knowledge of its main features.

Some results have strengthened and confirmed earlier results. These include:

- the broad line width displayed by the SiO(5-4) emission at distances from the star between 5 and 15 au, evidence for the importance of shocks induced by pulsations and convective cell ejections as abundantly revealed by observations at shorter wavelengths;

- the confinement of a high density gas volume around the star, again  evidenced by observations at shorter wavelengths such as by \citet{Khouri2018};

- increased support to the scenario presented in HTN20 to explain the presence of SiO emission in the south-western quadrant.

Some have presented new results. These include:

- the possible presence of rotation about an axis projecting 40\dego$\pm$15\dego\ east of north with a projected velocity of 0.7$\pm$0.2 \kms, in agreement with results obtained from the observation of SiO masers;

- the presence of two fragments, SWO and NEA, emitted a few decades ago and nearly detached from the central gas reservoir, giving additional evidence for the episodic nature of the mass loss;

- the current mass loss proceeding toward the blue-shifted hemisphere along three favored directions: the SEO, dominated by the wind focused by Mira B, the NEO and the SWS;

- a detailed exploration of the $^{13}$CO(3-2) emission, optically thinner than the $^{12}$CO(3-2) emission and essentially confirming the results presented in HTN20;

- an evaluation of the mass loss rate associated with the main fragments and a comparison with that obtained from single dish observations;

- a detailed study of the suppression of SiO abundance beyond the central high density gas volume, and a comparison with other AGB stars;

- an evaluation of the SiO/CO abundance ratio at the level of 0.5-1$\times$10$^{-4}$ when excluding the south-western quadrant, at least two orders of magnitude smaller than other typical low mass-loss-rate oxygen-rich AGB stars, and a possible interpretation in terms of the combined effect of depletion on dust grains and photo-dissociation by interstellar UV radiation.

The main contribution of the present work to our understanding of the wind of Mira Ceti has been to reduce the problem to explaining why and how the mass loss is episodic and anisotropic. Here episodic should  be understood as modulations of the mass loss rate over periods of several decades with peak-to-valley ratios of a few, no more than an order of magnitude. Moreover, we note that the north-east/south-west direction seems to be a preferred direction for the outflows. A probably permanent component of the wind is the part focused by Mira B, but the direction of the flow evolves slowly, at the pace of the orbital period, some 500 years. We have underscored the absence of sensible explanation of the variability and anisotropy of the mass loss process and have commented about arguments of possible relevance. Qualitatively, it looks as if small perturbations to the zero-order picture where there is simply no mass loss are sufficient to trigger episodes of fragmentary loss. Quantitatively, however, understanding the underlying physics requires a detailed modeling of the dust-gas chemistry at stake, probably well beyond what current models can address.

The present results cannot be ignored by hydro-dynamical and physico-chemical models attempting a description of the nascent wind. Indeed, if Mira A is an archetype in terms of its variability, it is rather atypical in terms of the generation of its wind. Such models can be expected to find it very difficult to account for its peculiar features, small variations in the parameters deciding when and where mass loss can proceed significantly. 


\begin{acknowledgments} 
  We thank Professors J. Alcolea, B. Freytag, T. Khouri, M. Karovska, G. Perrin, and K.T. Wong for having helped us to better understand Mira by answering and/or clarifying questions we had concerning both their work and the general picture. We express our deep gratitude to the anonymous referee for the pertinence of his/her comments and suggestions, which helped us with improving significantly the quality of the presentation. This paper uses ALMA data ADS/JAO.ALMA\#2011.0.00014.SV and ADS/JAO.ALMA\#2013.1.00047.S. ALMA is a partnership of ESO (representing its member states), NSF (USA) and NINS (Japan), together with NRC (Canada), MOST and ASIAA (Taiwan), and KASI (Republic of Korea), in cooperation with the Republic of Chile. The Joint ALMA Observatory is operated by ESO, AUI/NRAO and NAOJ. The data are retrieved from the JVO/NAOJ portal. We are deeply indebted to the ALMA partnership, whose open access policy means invaluable support and encouragement for Vietnamese astrophysics. Financial support from the World Laboratory, the Odon Vallet Foundation and VNSC is gratefully acknowledged. This research is funded by the Vietnam National Foundation for Science and Technology Development (NAFOSTED) under grant number 103.99-2019.368.
\end{acknowledgments}

\bibliography{filted_mira2021apj}{}
\bibliographystyle{aasjournal}

\begin{figure*}
  \centering
  \includegraphics[width=15cm]{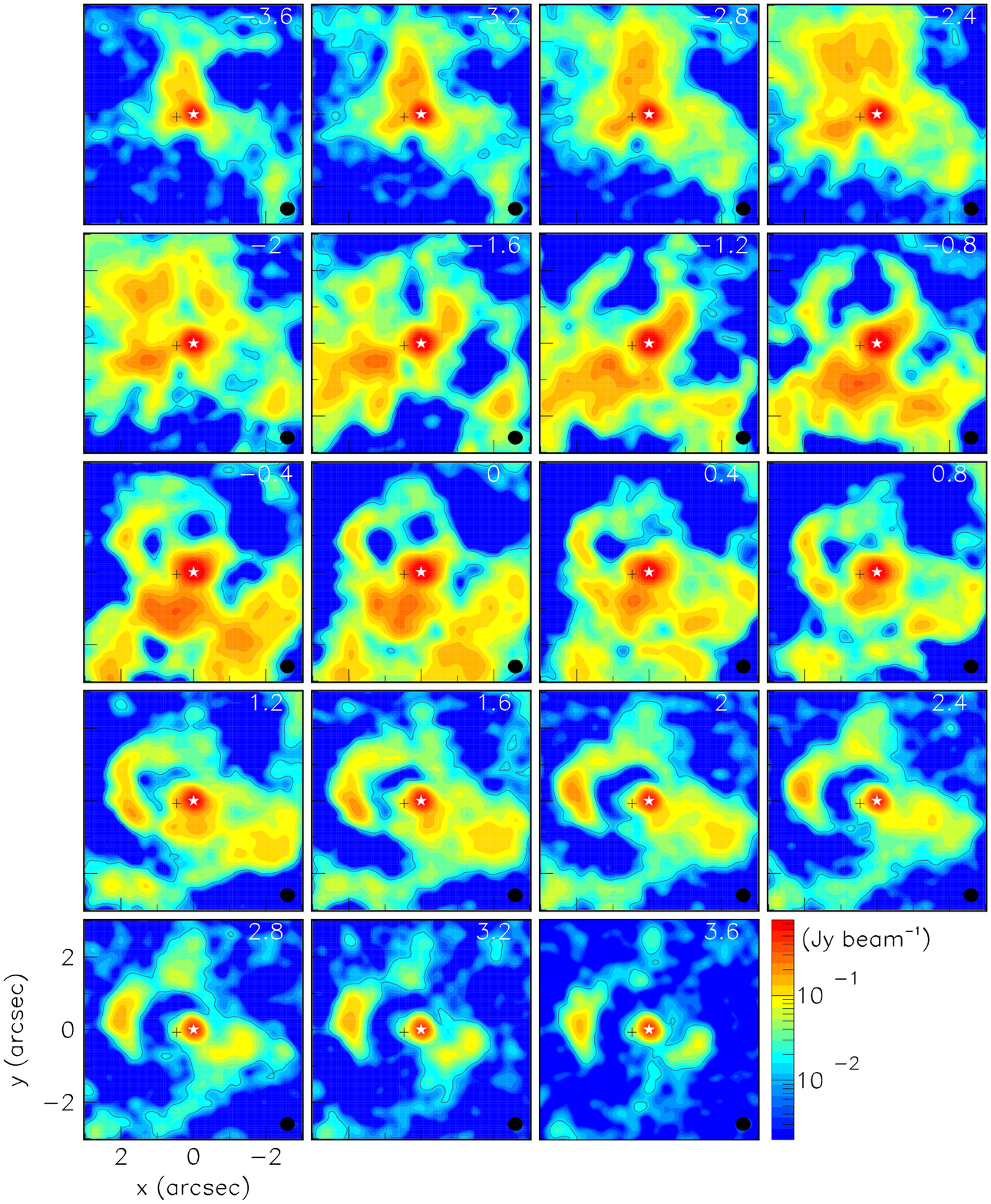}
  \caption{Channel maps of the $^{13}$CO(3-2) emission. In each panel, the Doppler velocity with respect to the systemic velocity is indicated in the upper-right corner, the beam size is shown in the lower-right corner, the position of Mira A is marked with a  white star and that of Mira B with a black cross. The contours show the emission at 3-$\sigma$ noise. }
  \label{figA1}
\end{figure*}

\appendix
Channel maps of the $^{13}$CO(3-2) emission are displayed in Figure \ref{figA1}.


\end{document}